\documentclass[aps, prl, twocolumn, superscriptaddress, longbibliography]{revtex4-1}
\usepackage[colorlinks, linkcolor=blue, anchorcolor=blue, citecolor=blue]{hyperref}
\usepackage{amsmath}
\usepackage{graphicx}
\usepackage{braket}
\usepackage{multirow}
\usepackage{color}
\usepackage{soul}
\usepackage{diagbox}
\usepackage{rotating}

\usepackage{ulem}

\begin{document}

\title{Enhanced Spin-polarization via Partial Ge1-dimerization as the Driving Force of the 2$\times$2$\times$2 CDW in FeGe}
\author{Yilin Wang}\email{yilinwang@ustc.edu.cn}    
\affiliation{School of Future Technology, University of Science and Technology of China, Hefei, Anhui 230026, China} 

\date{\today}

\begin{abstract}
    A $2\times2\times2$ charge density wave (CDW) was recently observed deep inside the antiferromagnetic phase of a Kagome metal FeGe. A key question is whether the CDW in FeGe is driven by its electronic correlation and magnetism. Here, we address this problem using density functional theory and its combination with $U$ as well as dynamical mean-field theory. Our calculations show that large dimerization ($\sim$1.3 \AA) of Ge1-sites along $c$-axis will enhance electronic correlation of the Fe-$3d$ orbitals and, as a result, it enhances the spin-polarization and saves more magnetic exchange energies. We find that the balance between magnetic energy saving and structural energy cost via partially dimerizing Ge1-sites in an enlarged superstructure, could induce a new local minimum in total energies. The response to the large partial Ge1-dimerization will induce additional small modulations ($<$0.05 \AA) of other sites in the Kagome and honeycomb layers, which further reduces the total energy and leads to a stable $2\times 2\times 2$ CDW ground state in FeGe. Our results are in good agreement with the existing experiments and reveal a novel CDW mechanism driven by the interplay of structure, electronic correlation and magnetism.
\end{abstract}

\maketitle

\textit{Introduction.}---Charge density wave (CDW), the static modulations of small amplitude in the electron density distribution accompanied by a periodic lattice distortion~\cite{gruner:1988,gruner:2019}, is one of the most important phases of matter in condensed matter physics. Common driving force for a CDW includes Fermi surface nesting~\cite{Peierls:2001} and electron-phonon couplings~\cite{McMillan:1977,Varma:1983,Mazin:2008,Jiandong:2015}. More exotic mechanism driven by strong electron-electron interactions is also suggested, for example, in copper oxides~\cite{Tranquada:1995,Hoffman:2002,Kivelson:2003,Wise:2008,Reznik:2012,Blackburn:2013,Comin:2014,LeTacon:2014,Fujita:2014} and nickel oxides~\cite{Tranquada:1994,Lee:1997}. Recently, an exotic chiral $2\times 2\times 2$ CDW that breaks time-reversal symmetry was observed~\cite{Jiang:2021,Mielke:2022} in a Kagome metal AV$_3$Sb$_5$ (A=K, Rb, Cs)~\cite{Ortiz:2019,Stephen:2020,Mazhar:2020}, which was suggested to be triggered by van Hove singularities (VHSs)~\cite{Jiangping:2021,Thomale:2013,Qianghua:2013,Binghai:2021,Miao:2021,Denner:2021,Rahul:2021,Balents:2021,Rafael:2021,Zhao:2021,JiangKun:2022,Nie:2022}. Since Kagome metals can simultaneously feature geometry frustration~\cite{syozi:1951,Sachdev:1992,Norman:2016}, flat-bands-induced electronic correlation~\cite{wenxg:2011,Yin:2018,Yin:2020,Liu:2020,Kang:2020,Yilin:2023} and magnetism~\cite{Tasaki:1992,Zhenyu:2018,Yin:2019,Yin:2020b,Miao:2021b,Changgan:2022}, non-trivial topology~\cite{Stephen:2020,Ye:2018,Liang:2021,Ziqiang:2022}, VHSs~\cite{Qianghua:2013,Thomale:2013} as well as strong electron-phonon interactions~\cite{GangXu:2022,Pengcheng:2022,Liu:2022,Okazaki:2022}, it has become an ideal platform for exploring CDWs driven by various mechanisms. 

\begin{figure}
    \centering
    \includegraphics[width=0.5\textwidth]{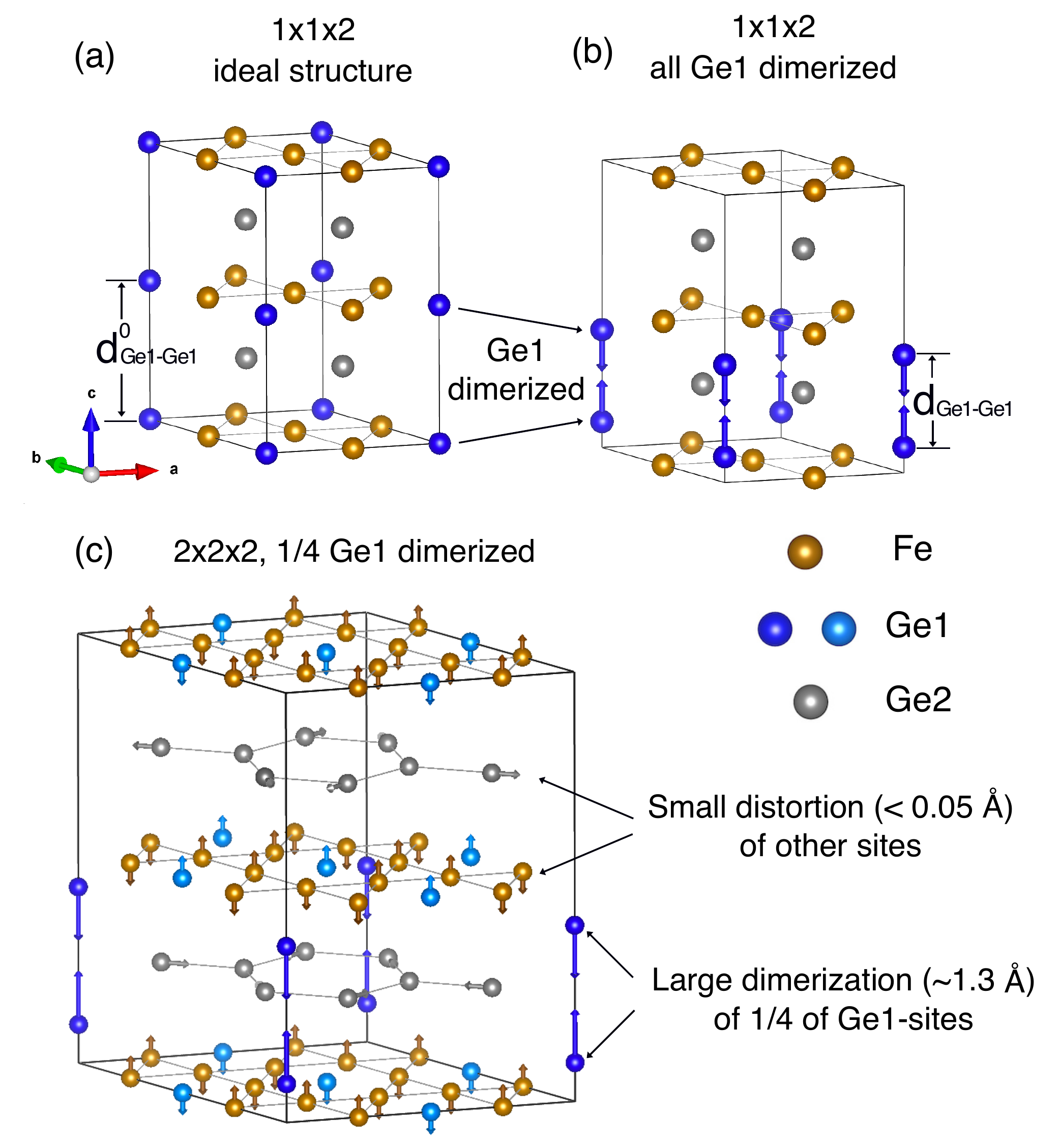}
    \caption{(a) 1$\times$1$\times$2 ideal Kagome structure of FeGe. There are two types of Ge sites. Ge1 (dark blue) is at the center of the hexagon of the Fe Kagome layer. Ge2 (grey) forms a honeycomb layer in between two Kagome layers. (b) All the Ge1 atoms deviate from the Kagome layers and form dimers along the c-axis. The strength of Ge1 dimerization is defined as $d=d_{\text{Ge1-Ge1}}^0-d_{\text{Ge1-Ge1}}$, where $d_{\text{Ge1-Ge1}}^0$ and $d_{\text{Ge1-Ge1}}$ are the bond lengths of Ge1-Ge1 before and after dimerization, respectively. (c) A 2$\times$2$\times$2 superstructure of FeGe with space group P6/mmm. It consists of a large dimerization ($|d|\sim1.3$ \AA) of 1/4 of the Ge1-sites along c-axis and small distortion ($<0.05$ \AA, indicated by short arrows) of other sites.}
    \label{fig:struct}
\end{figure}

Very recently, another $2\times 2\times 2$ CDW was observed around $T_{\text{CDW}}=100$ K, deep inside the A-type antiferromagnetic (AFM) phase ($T_{\text{N}}\sim 410$ K) of a magnetic Kagome metal FeGe~\cite{Teng:2022, YinJiaXin:2022}. Both neutron~\cite{Teng:2022} and x-ray~\cite{Miao:2022} scattering experiments indicate a first-order nature of this CDW transition. Possible anomalous Hall effect~\cite{Teng:2022} and topological edge modes~\cite{YinJiaXin:2022} were observed in its CDW phase, which are similar to those observed in AV$_3$Sb$_5$~\cite{Mazhar:2020,Yingjianjun:2021}. However, there are more differences between the CDWs in FeGe and AV$_3$Sb$_5$. (i) The ordered magnetic moments were found to be enhanced in the CDW phase of FeGe~\cite{Teng:2022}. (ii) A sharp superlattice peak, $Q=(0,0,2.5)$, that probes lattice distortion along $c$-axis, was observed in FeGe by the x-ray scattering experiment with the same onset temperature as CDW~\cite{Miao:2022}, which is absent in AV$_3$Sb$_5$. (iii) First-principle calculations find only little phonon softening around the three $L$-points, (0.5, 0, 0.5), (0, 0.5, 0.5), (-0.5, 0.5, 0.5), but the phonon frequencies never become negative in FeGe in the parameter regime that produces the correct ordered magnetic moment~\cite{Miao:2022}. Such behavior in phonon spectrum is very different from AV$_3$Sb$_5$~\cite{Binghai:2021,GangXu:2022}, while it shows some similarities to the electronic-correlation-driven CDWs in cooper and nickel oxides. These indicate a different origin of the CDW in FeGe, in sharp contrast to AV$_3$Sb$_5$. A natural question thus arises: whether the CDW in FeGe is driven by its electronic correlation and magnetism? Despite several investigations by experiment~\cite{Teng:2022,YinJiaXin:2022,Teng:2023,Miao:2022} and theoretical calculations~\cite{Setty:2022,Guoqing:2022, Miao:2022, Xiangang:2022, Xiangang:2023,Jianpeng:2023}, the driving force of the CDW in FeGe still remains an open question. 

In a previous work~\cite{Miao:2022}, guided by the soft phonon mode at $L$-points and the first-order nature of the CDW transition in FeGe, we have obtained a stable $2\times 2\times 2$ superstructure of FeGe with lower energy than its ideal structure, via structural relaxation using density functional theory (DFT). A key finding is that there is a large dimerization ($\sim$1.3 \AA) of 1/4 of Ge1-sites along the $c$-axis in the $2\times 2\times 2$ superstructure. Motivated by this observation, in this work, we carry out more calculations to identify the driving force for the CDW in FeGe, using DFT and its combination with $U$ (DFT+$U$) as well as dynamical mean-field theory (DFT+DMFT)~\cite{lichtenstein:2001,kotliar:2006}. Our calculations show that large dimerization of Ge1-sites along $c$-axis will enhance electronic correlation of the Fe-$3d$ orbitals and, as a result, it enhances the spin-polarization and saves more magnetic exchange energies. We find that the balance between magnetic energy saving and structural energy cost via partially dimerizing Ge1-sites in an enlarged superstructure, could induce a new local minimum in total energies. The response to the large partial Ge1-dimerization will induce additional small modulations ($<$0.05 \AA) of other sites in the Kagome and honeycomb layers, which further reduces the total energy and leads to a stable $2\times 2\times 2$ CDW ground state in FeGe. Our results thus reveal a novel CDW mechanism driven by the interplay of structure, electronic correlation and magnetism, which is in good agreement with the existing experiments.

\textit{Methods.}---As shown in Fig.~\ref{fig:struct}(a), hexagonal FeGe is consisting of a Kagome layer of Fe$_3$Ge and a honeycomb layer of Ge$_2$. There are two non-equivalent Ge-sites: Ge1 (blue) in the center of the hexagon of the Fe Kagome structure and Ge2 (grey) in the honeycomb layer. The DFT calculations are performed using the VASP package~\cite{kresse:1996,blochl:1994}, with exchange-correlation functional of generalized gradient approximation (GGA)~\cite{perdew:1996}. Although FeGe is a strongly correlated magnet, the DFT calculations without Hubbard $U$ correction have already correctly produced the ordered magnetic moments of its AFM phase (around 1.5 $\mu_B$/Fe), observed by neutron scattering experiment~\cite{Teng:2022}. Therefore, DFT calculations are applicable to the AFM phase of FeGe. The experimental lattice parameters of FeGe, $a=4.985$ \AA\: and $c=4.048$ \AA\:~\cite{Brian:2020,Teng:2022}, are used in calculations.

To better capture the strong electronic correlations of FeGe, we also perform fully charge self-consistent single-site DFT+DMFT calculations for its paramagnetic (PM) and AFM phases, using the EDMFTF code developed by Haule \textit{et al.}~\cite{Haule:2010,Haule:2015free} based on the WIEN2K package~\cite{Blaha:2020}. For AFM calculation, a non-magnetic calculation is performed in the DFT part, while the AFM spin-polarization is considered in the DMFT part by breaking the spin degeneracy of the local self-energy. The Hubbard $U$ and Hund's coupling $J_H$ are chosen to be 3.9 eV and 0.85 eV, respectively, to reproduce the ordered magnetic moments observed experimentally. More computational details are presented in the Supplementary Materials~\cite{suppl}. 

\begin{figure*}
    \centering
    \includegraphics[width=0.98\textwidth]{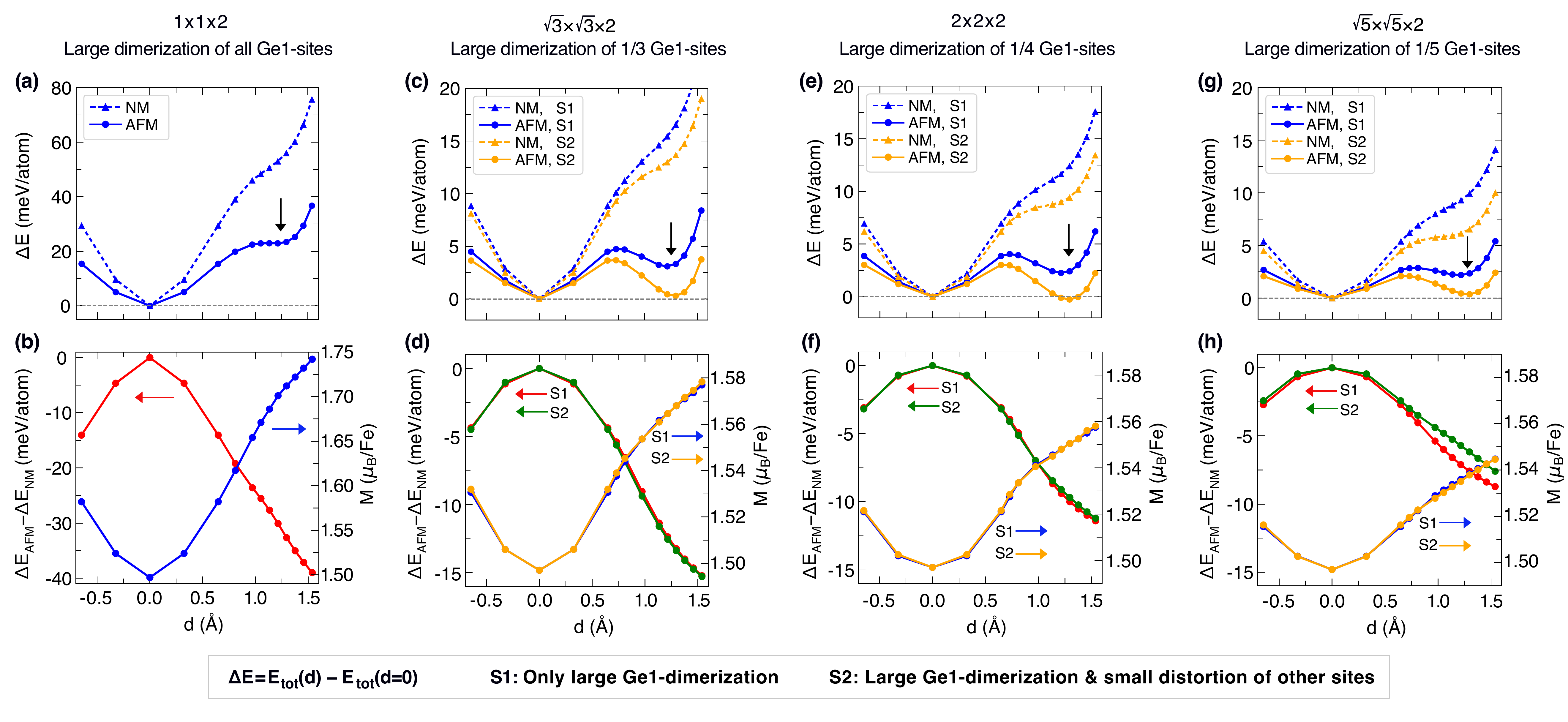}
    \caption{Enhanced spin-polarization via partial Ge1-dimerization calculated by DFT. (a) The difference of total energy between the distorted and ideal Kagome structure, $\Delta E=E_{\text{tot}}(d)-E_{\text{tot}}(d=0)$, as functions of the strength of Ge1-dimerization $d$. Dashed and solid curves are for non-magnetic and anti-ferromagnetic states, respectively. The black arrow labels a local energy minimum around $d=1.3$ \AA. (b) The magnetic exchange energy, $\Delta E_{\text{AFM}}-\Delta E_{\text{NM}}$, and the average of ordered magnetic moments per Fe in the AFM state as functions of $d$, are shown in the left and right $y$-axis, respectively. (a) and (b) are for the 1$\times$1$\times$2 structure with large dimerization of all Ge1-sites. (c)-(h) are in analogous to (a) and (b), but for enlarged superstructures and two different cases labeled by ``S1'' and ``S2'' (see the main text). (c) and (d) are for the $\sqrt{3}\times\sqrt{3}\times2$ superstructure (P6/mmm) with large dimerization of $1/3$ of Ge1-sites. (e) and (f) are for the $2\times2\times2$ superstructure (P6/mmm) with large dimerization of $1/4$ of Ge1-sites. (g) and (h) are for the $\sqrt{5}\times\sqrt{5}\times2$ superstructure (Cmmm) with large dimerization of $1/5$ of Ge1-sites.}
    \label{fig:energy}
\end{figure*}

\textit{Results.}---Fig.~\ref{fig:energy} show the results calculated by DFT. Fig.~\ref{fig:energy}(a) shows the total energy difference of the $1\times 1\times 2$ superstructure of FeGe between the one with large dimerization of all the Ge1-sites along $c$-axis and the one without dimerization, $\Delta E=E_{\text{tot}}(d)-E_{\text{tot}}(d=0)$, as functions of the Ge1-dimerization strength $d$ (see Fig.~\ref{fig:struct}). $\Delta E$ for the non-magnetic (NM) state keeps increasing with $|d|$, indicating that it has to pay for energies for structure distortions. However, the increasing rate slows down when the dimerization strength $d$ is around 1.0 \AA$\sim$1.3 \AA. Although $\Delta E$ also increases with $|d|$ for AFM state, its magnitude is much smaller than that of NM state, indicating that the spin-polarization are enhanced and more magnetic exchange energies are saved by Ge1-dimerization. This is shown in Fig.~\ref{fig:energy}(b), where the magnetic exchange energies, $\Delta E_{\text{AFM}}-\Delta E_{\text{NM}}$, are negative and their magnitude keep increasing with $|d|$ (left $y$-axis), and the ordered magnetic moments also increase with $|d|$ (right $y$-axis). This is a main finding of the present work. The competition between magnetic energy saving from enhanced spin-polarization and structural energy cost from Ge1-dimerization may induce a new local minimum in the total energies of the AFM state, as illustrated by the black arrow around $d=1.3$ \AA\: in Fig.~\ref{fig:energy}(a). It may further become a global energy minimum if the magnetic energy wins, and drive a first-order phase transition. This has not achieved in the $1\times 1\times 2$ superstructure because it costs too much structural energies to dimerize all the Ge1-sites. Therefore, a straightforward strategy to balance the magnetic energy saving and structural energy cost is to enlarge the structure along the $a$ and $b$ crystal axis, but dimerize partial Ge1-sites. 

Along this line, we construct three superstructures: $\sqrt{3}\times \sqrt{3}\times 2$ with large dimerization of 1/3 of Ge1-sites [Fig.~\ref{fig:energy}(c) and (d)], $2\times 2\times 2$ with large dimerization of 1/4 of Ge1-sites [Fig.~\ref{fig:energy}(e) and (f)], and $\sqrt{5}\times \sqrt{5}\times 2$ with large dimerization of 1/5 of Ge1-sites [Fig.~\ref{fig:energy}(g) and (h)]. We then perform calculations for two different cases. (S1) The superstructure with only fixed large dimerization of partial Ge1-sites but other sites are not relaxed; (S2) The superstructure with fixed large dimerization of partial Ge1-sites and all the other sites are relaxed until the force of each of those atoms are less than 1 meV/\AA. 

As shown by the solid blue curves in Fig.~\ref{fig:energy}(c), (e) and (g), well-defined local energy minimums have already formed around $d=1.3$ \AA\: in all the three superstructures with only large partial Ge1-dimerization (S1). The energy minimums are only 2$\sim$3 meV/atom higher than that of the non-distorted structures ($d=0$). The magnetic exchange energies and the ordered magnetic moments shown in Fig.~\ref{fig:energy}(d), (f) and (h) follow the same behaviors as the $1\times 1\times 2$ superstructure. Large dimerization of partial Ge1-sites will exert extra potential on other sites, such that they will slightly relax their positions to further reduce the total energy, which are shown by the solid orange curves in Fig.~\ref{fig:energy}(c), (e) and (g). The energy minimums around $d=1.3$ \AA\: of all the three superstructures get much closer to that of the ideal structure, but it only becomes a global minimum in the $2\times 2\times 2$ superstructure, consistent with the experimentally observed $2\times 2\times 2$ CDW in FeGe. This is because the 1/3 Ge1-dimerization in the $\sqrt{3}\times \sqrt{3}\times 2$ superstructure pays for more structural distortion energy and the 1/5 Ge1-dimerization in the $\sqrt{5}\times \sqrt{5}\times 2$ superstructure cannot save enough magnetic energy, comparing to the 1/4 Ge1-dimerization in the $2\times 2\times 2$ superstructure. Therefore, we can conclude that the form of $2\times 2\times 2$ CDW superstructure in FeGe results from a subtle balance between magnetic exchange energy saving and structure energy cost via large structural dimerization of 1/4 of Ge1-sites. 

The distortions of other sites are very small ($<$0.05 \AA). The arrows in Fig.~\ref{fig:struct}(c) illustrates such distortions in the $2\times 2\times 2$ superstructure. The atoms mainly move out-of-plane along $c$-axis in the Kagome layers, and move in-plane in the honeycomb layers to form a Kekul\'{e}-type distortion~\cite{Herrera:2020,Guoqing:2022}. We note that the magnetic exchange energies and ordered magnetic moments are not further enhanced by those additional distortions [green curves in Fig.~\ref{fig:energy}(d), (f) and (h)], indicating that the driving force for the enhanced spin-polarization is primarily from the large dimerization of Ge1-sites. The small $2\times 2$ charge modulations in Kagome and honeycomb layers which are observed by the STM experiments~\cite{YinJiaXin:2022,Teng:2022,Ziyuan:2023}, are induced as a consequence to respond to the extra potential induced by the large dimerization of partial Ge1-sites.

\begin{figure}
    \centering
    \includegraphics[width=0.5\textwidth]{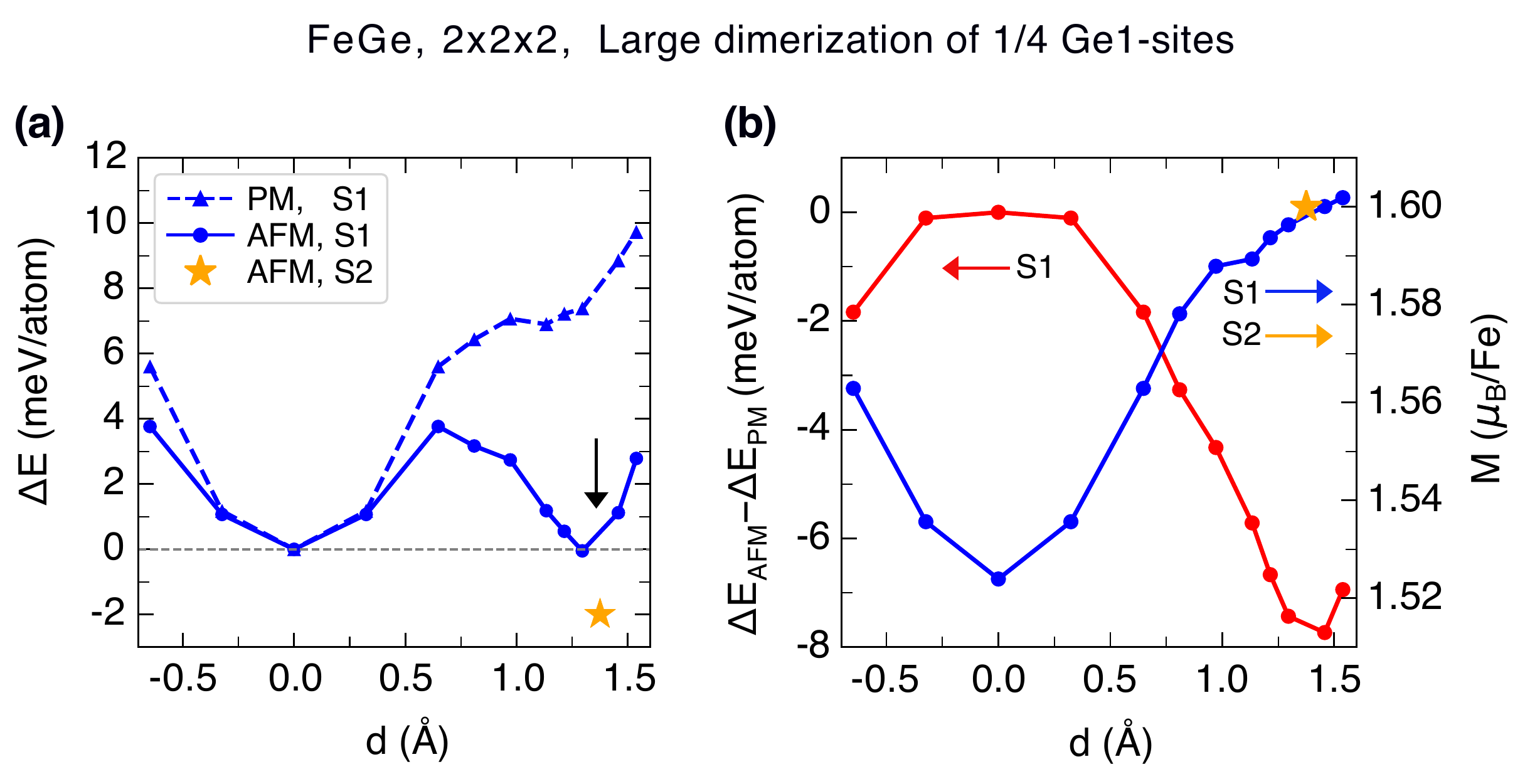}
    \caption{Analogous to Fig.~\ref{fig:energy}, but calculated by DFT+DMFT for paramagnetic (PM) and AFM phases of the $2\times 2\times 2$ superstructure, at $T=290$ K, $U=3.9$ eV and $J_H=0.85$ eV. Here, $E$ stands for total free energy. The orange star in (a) and (b) stands for the total free energy and ordered magnetic moments, respectively, of the fully relaxed $2\times 2\times 2$ superstructure.}
    \label{fig:dmft}
\end{figure}

In order to understand how Ge1-dimerization will enhance the spin-polarization, we compare the electronic correlations of the paramagnetic state of the $1\times 1\times 2$ superstructure with and without Ge1-dimerization by DFT+DMFT calculations. The calculated mass-enhancement $m^{\star}/m^{\text{DFT}}$ of Fe-$3d$ orbitals at $T=290$ K, $U=3.9$ eV and $J_H=0.85$ eV are tabulated in Table.~\ref{tab:mass}. The mass-enhancement, in particular, of $d_{x^2-y^2}$, $d_{xz}$ and $d_{yz}$ orbitals increase substantially from the ideal Kagome structure to the structure with a Ge1-dimerization of $d=1.3$ \AA. It indicates that the height of Ge1 atoms relative to the Fe Kagome plane will affect the electronic correlations of Fe-$3d$ orbitals. This is similar to the iron-based superconductors, where the electronic correlations of Fe-$3d$ orbitals are found to be very sensitive to the heights of As or Se sites~\cite{YinZP:2011,Ieki:2014,Haule:2016force}. Stronger electronic correlations will enhance the spin-polarization of the AFM state of FeGe. Analogous to Fig.~\ref{fig:energy}, we also perform DFT+DMFT calculations for the $2\times 2\times 2$ superstructure, which are shown in Fig.~\ref{fig:dmft}. Similar profiles of energy and ordered magnetic moments as functions of $d$ are produced, and a global energy minimum is also found around $d=1.3$ \AA, which validates the findings from the simple DFT calculations. 

\begin{table}
    \centering
    \caption{The mass enhancement, $m^*/m^{\text{DFT}}=1/Z$, of Fe-$3d$ orbitals from DFT+DMFT calculations of the paramagnetic phase of the $1\times1\times2$ superstructure with ($d=1.3$ \AA) and without ($d=0$ \AA) Ge1-dimerization, at $T=290$ K, $U=3.9$ eV and $J_H=$ 0.85 eV.}
    \begin{ruledtabular}
    \begin{tabular}{c|ccccc}

                     &$d_{3z^2-r^2} $ &  $d_{x^2-y^2}$ & $d_{xz}$ & $d_{yz}$ & $d_{xy}$ \\
         \hline
         $d=0$ \AA & 2.146 & 3.232 & 3.342 & 3.239 & 2.055 \\
         \hline
        $d=1.3$ \AA & $2.287$ & $3.971$ & $4.171$ & $3.642$ & $2.136$\\
    \end{tabular}
\end{ruledtabular}
    \label{tab:mass}
\end{table}

\begin{figure}
        \centering
        \includegraphics[width=0.5\textwidth]{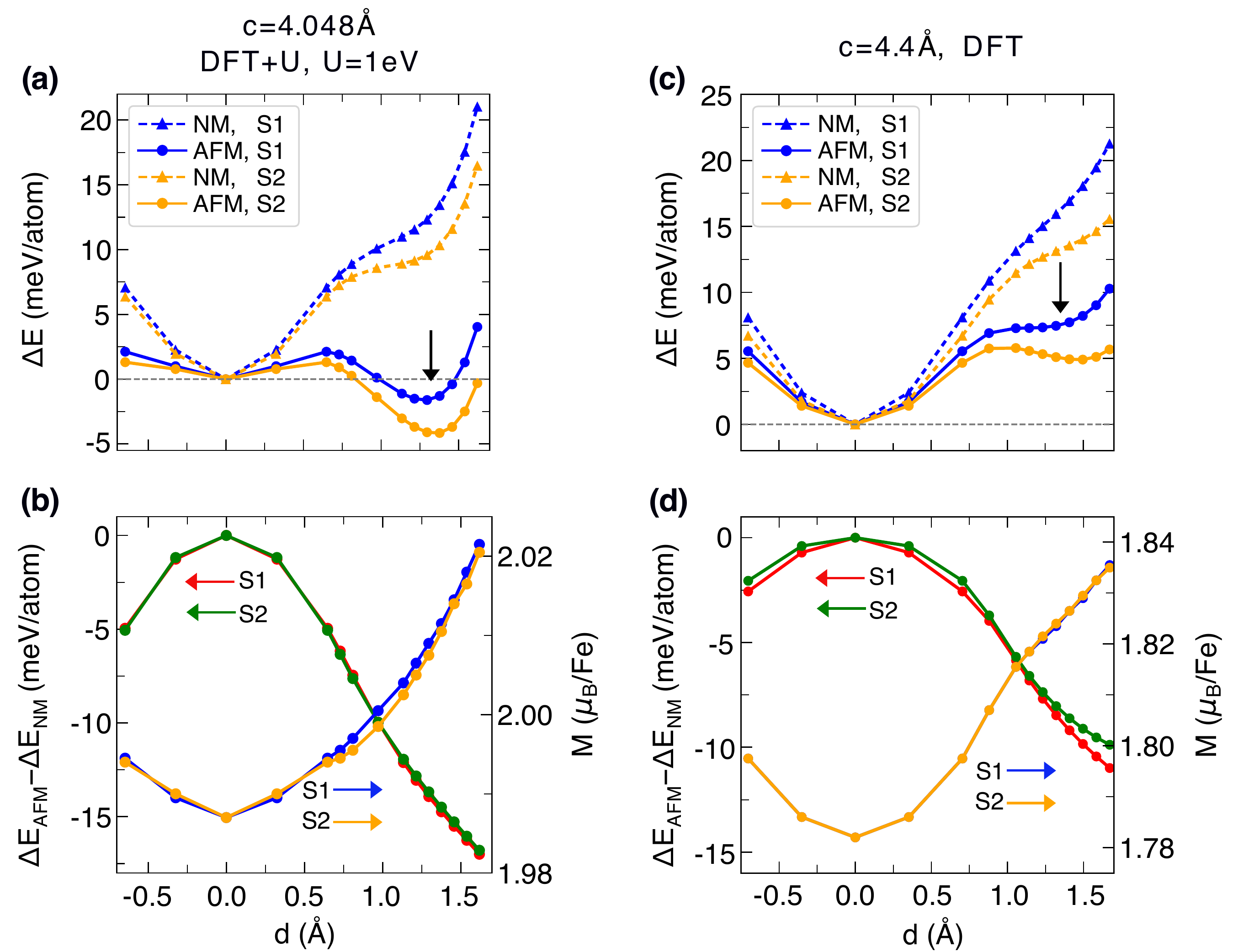}
        \caption{Analogous to Fig.~\ref{fig:energy} for $2\times 2\times 2$ superstructure of FeGe. (a) and (b) are calculated by DFT+$U$ at $U=1$ eV with the experimental lattice parameters. (c) and (d) are calculated by DFT, but with a larger lattice parameter of $c=4.4$ \AA.}
        \label{fig4}
    \end{figure}

The subtle competition between magnetic and structural energy indicates that the $2\times 2\times 2$ CDW in FeGe is sensitive to the variation of its magnetism and crystal structure. Thus, on the one hand, increasing Hubbard $U$ in DFT+$U$ calculations is expected to further save the magnetic energy and stabilize the CDW state. The DFT+$U$ results at $U=1$ eV for the $2\times 2\times 2$ superstructure are shown in Fig.~\ref{fig4}(a) and (b). The ordered magnetic moment is enhanced to about 2 $\mu_B/\text{Fe}$, much larger than the experimental value, and the energy of the CDW ground state is further reduced. We note that the energy minimum around $d=1.3$ \AA\: becomes a global energy minimum even in the superstructure with only large dimerization of 1/4 of Ge1-sites (solid blue curve), which further confirms that the large dimerization of partial Ge1-sites is the driving force. We also find that the $\sqrt{3}\times\sqrt{3}\times 2$ superstructure becomes the ground state instead of the $2\times 2\times 2$ superstructure at $U=1$ eV. This is also expected within our theory, since the magnetic energy saving becomes even greater in the superstructure with more dimerized Ge1-sites at larger $U$, such that the $\sqrt{3}\times\sqrt{3}\times 2$ superstructure saves the most energies among the superstructures at $U=1$ eV. This result further supports our conclusion that the CDW in FeGe results from a subtle competition between magnetic energy saving and structural energy cost.

On the other hand, elongating the crystal $c$-axis of FeGe may cause it to pay for more structural distortion energy when dimerizing Ge1-sites. We show this in Fig.~\ref{fig4}(c) and (d) by increasing $c$ to 4.4 \AA. Indeed, the $2\times 2\times 2$ CDW state is not favored anymore.

\textit{Conclusion and Discussion.}---To summarize, by DFT, DFT+$U$ and DFT+DMFT calculations, we show that the driving force of the $2\times 2\times 2$ CDW in FeGe is an enhanced spin-polarization via large dimerization of partial Ge1-sites in an enlarged superstructure. We reveal that the enhancement of spin-polarization is due to stronger electronic correlations of Fe-$3d$ orbitals caused by large dimerization of Ge1-sites. The small $2\times 2$ charge modulations in Kagome and honeycomb layers observed by STM~\cite{YinJiaXin:2022,Teng:2022,Ziyuan:2023}, are thus induced as a consequence to respond the large dimerization of partial Ge1-sites. 

Our results thus indicate a first-order CDW transition in FeGe, which is consistent with both neutron~\cite{Teng:2022} and x-ray~\cite{Miao:2022} scattering experiments. The large Ge1-dimerization along the $c$-axis is consistent with the observation that an additional sharp super-lattice peak $Q=(0,0,2.5)$ was found in the elastic x-ray scattering experiment with the same onset temperature as the CDW transition~\cite{Miao:2022}, since this peak probes the structural distortion along $c$-axis. The enhancement of spin-polarization in the CDW phase is consistent with the observation from neutron scattering experiment~\cite{Teng:2022}.

Similar to AV$_3$Sb$_5$, there are also VHSs at $M$ point near the Fermi level in the electronic structure of FeGe~\cite{Teng:2022,Teng:2023,Miao:2022}, so VHSs may also play roles in driving the CDW in FeGe. However, Wan \textit{et al.}~\cite{Xiangang:2023} have computed the nesting functions of FeGe and found that the maximum of nesting function is at $K$ point instead of $M$ point. This excludes the possibility of a VHSs-induced CDW in FeGe.

We note that the Kagome FeSn, the sister compound of FeGe, shows similar electronic structures~\cite{Zhenyu:2020} and antiferromagnetic orders~\cite{Brian:2019} to FeGe, but no CDW was observed experimentally. We also perform the same calculations for FeSn and find a similar local energy minimum induced by large Sn1-dimerization (see Fig. S1 in~\cite{suppl}), but it is far from becoming a global energy minimum in the regime of reasonable Hubbard $U$ that produces the correct ordered magnetic moments ($\sim$ 1.85 $\mu_B$/Fe)~\cite{Brian:2019}. It may originate from that Sn has a much larger atomic radius and, as a result, much larger lattice parameters of FeSn~\cite{Brian:2020}, such that it has to pay for much more structural energies than FeGe by dimerizing the Sn1-sites.

\textit{Acknowledgement.}---We thank Hu Miao, Zhida Song and Yajun Yan for very helpful discussion. This project was supported by the National Natural Science Foundation of China (No. 12174365). All the calculations were preformed on TianHe-1(A), the National Supercomputer Center in Tianjin, China.

\bibliography{main}

\begin{thebibliography}{87}%
\makeatletter
\providecommand \@ifxundefined [1]{%
 \@ifx{#1\undefined}
}%
\providecommand \@ifnum [1]{%
 \ifnum #1\expandafter \@firstoftwo
 \else \expandafter \@secondoftwo
 \fi
}%
\providecommand \@ifx [1]{%
 \ifx #1\expandafter \@firstoftwo
 \else \expandafter \@secondoftwo
 \fi
}%
\providecommand \natexlab [1]{#1}%
\providecommand \enquote  [1]{``#1''}%
\providecommand \bibnamefont  [1]{#1}%
\providecommand \bibfnamefont [1]{#1}%
\providecommand \citenamefont [1]{#1}%
\providecommand \href@noop [0]{\@secondoftwo}%
\providecommand \href [0]{\begingroup \@sanitize@url \@href}%
\providecommand \@href[1]{\@@startlink{#1}\@@href}%
\providecommand \@@href[1]{\endgroup#1\@@endlink}%
\providecommand \@sanitize@url [0]{\catcode `\\12\catcode `\$12\catcode
  `\&12\catcode `\#12\catcode `\^12\catcode `\_12\catcode `\%12\relax}%
\providecommand \@@startlink[1]{}%
\providecommand \@@endlink[0]{}%
\providecommand \url  [0]{\begingroup\@sanitize@url \@url }%
\providecommand \@url [1]{\endgroup\@href {#1}{\urlprefix }}%
\providecommand \urlprefix  [0]{URL }%
\providecommand \Eprint [0]{\href }%
\providecommand \doibase [0]{http://dx.doi.org/}%
\providecommand \selectlanguage [0]{\@gobble}%
\providecommand \bibinfo  [0]{\@secondoftwo}%
\providecommand \bibfield  [0]{\@secondoftwo}%
\providecommand \translation [1]{[#1]}%
\providecommand \BibitemOpen [0]{}%
\providecommand \bibitemStop [0]{}%
\providecommand \bibitemNoStop [0]{.\EOS\space}%
\providecommand \EOS [0]{\spacefactor3000\relax}%
\providecommand \BibitemShut  [1]{\csname bibitem#1\endcsname}%
\let\auto@bib@innerbib\@empty
\bibitem [{\citenamefont {Gr\"uner}(1988)}]{gruner:1988}%
  \BibitemOpen
  \bibfield  {author} {\bibinfo {author} {\bibfnamefont {George}\ \bibnamefont
  {Gr\"uner}},\ }\bibfield  {title} {\enquote {\bibinfo {title} {The dynamics
  of charge-density waves},}\ }\href {\doibase 10.1103/RevModPhys.60.1129}
  {\bibfield  {journal} {\bibinfo  {journal} {Rev. Mod. Phys.}\ }\textbf
  {\bibinfo {volume} {60}},\ \bibinfo {pages} {1129--1181} (\bibinfo {year}
  {1988})}\BibitemShut {NoStop}%
\bibitem [{\citenamefont {Gr\"uner}(2019)}]{gruner:2019}%
  \BibitemOpen
  \bibfield  {author} {\bibinfo {author} {\bibfnamefont {George}\ \bibnamefont
  {Gr\"uner}},\ }\href@noop {} {\emph {\bibinfo {title} {Density waves in
  solids}}}\ (\bibinfo  {publisher} {CRC Press},\ \bibinfo {year}
  {2019})\BibitemShut {NoStop}%
\bibitem [{\citenamefont {Peierls}(2001)}]{Peierls:2001}%
  \BibitemOpen
  \bibfield  {author} {\bibinfo {author} {\bibfnamefont {R.~E.}\ \bibnamefont
  {Peierls}},\ }\href@noop {} {\emph {\bibinfo {title} {Quantum Theory of
  Solids}}}\ (\bibinfo  {publisher} {Oxford University Press},\ \bibinfo {year}
  {2001})\BibitemShut {NoStop}%
\bibitem [{\citenamefont {McMillan}(1977)}]{McMillan:1977}%
  \BibitemOpen
  \bibfield  {author} {\bibinfo {author} {\bibfnamefont {W.~L.}\ \bibnamefont
  {McMillan}},\ }\bibfield  {title} {\enquote {\bibinfo {title} {Microscopic
  model of charge-density waves in {2H}-{T}a{S}e$_2$.}}\ }\href {\doibase
  10.1103/PhysRevB.16.643} {\bibfield  {journal} {\bibinfo  {journal} {Phys.
  Rev. B}\ }\textbf {\bibinfo {volume} {16}},\ \bibinfo {pages} {643--650}
  (\bibinfo {year} {1977})}\BibitemShut {NoStop}%
\bibitem [{\citenamefont {Varma}\ and\ \citenamefont
  {Simons}(1983)}]{Varma:1983}%
  \BibitemOpen
  \bibfield  {author} {\bibinfo {author} {\bibfnamefont {C.~M.}\ \bibnamefont
  {Varma}}\ and\ \bibinfo {author} {\bibfnamefont {A.~L.}\ \bibnamefont
  {Simons}},\ }\bibfield  {title} {\enquote {\bibinfo {title} {Strong-coupling
  theory of charge-density-wave transitions},}\ }\href {\doibase
  10.1103/PhysRevLett.51.138} {\bibfield  {journal} {\bibinfo  {journal} {Phys.
  Rev. Lett.}\ }\textbf {\bibinfo {volume} {51}},\ \bibinfo {pages} {138--141}
  (\bibinfo {year} {1983})}\BibitemShut {NoStop}%
\bibitem [{\citenamefont {Johannes}\ and\ \citenamefont
  {Mazin}(2008)}]{Mazin:2008}%
  \BibitemOpen
  \bibfield  {author} {\bibinfo {author} {\bibfnamefont {M.~D.}\ \bibnamefont
  {Johannes}}\ and\ \bibinfo {author} {\bibfnamefont {I.~I.}\ \bibnamefont
  {Mazin}},\ }\bibfield  {title} {\enquote {\bibinfo {title} {Fermi surface
  nesting and the origin of charge density waves in metals},}\ }\href {\doibase
  10.1103/PhysRevB.77.165135} {\bibfield  {journal} {\bibinfo  {journal} {Phys.
  Rev. B}\ }\textbf {\bibinfo {volume} {77}},\ \bibinfo {pages} {165135}
  (\bibinfo {year} {2008})}\BibitemShut {NoStop}%
\bibitem [{\citenamefont {Zhu}\ \emph {et~al.}(2015)\citenamefont {Zhu},
  \citenamefont {Cao}, \citenamefont {Zhang}, \citenamefont {Plummer},\ and\
  \citenamefont {Guo}}]{Jiandong:2015}%
  \BibitemOpen
  \bibfield  {author} {\bibinfo {author} {\bibfnamefont {Xuetao}\ \bibnamefont
  {Zhu}}, \bibinfo {author} {\bibfnamefont {Yanwei}\ \bibnamefont {Cao}},
  \bibinfo {author} {\bibfnamefont {Jiandi}\ \bibnamefont {Zhang}}, \bibinfo
  {author} {\bibfnamefont {E.~W.}\ \bibnamefont {Plummer}}, \ and\ \bibinfo
  {author} {\bibfnamefont {Jiandong}\ \bibnamefont {Guo}},\ }\bibfield  {title}
  {\enquote {\bibinfo {title} {Classification of charge density waves based on
  their nature},}\ }\href {\doibase 10.1073/pnas.1424791112} {\bibfield
  {journal} {\bibinfo  {journal} {Proceedings of the National Academy of
  Sciences}\ }\textbf {\bibinfo {volume} {112}},\ \bibinfo {pages} {2367--2371}
  (\bibinfo {year} {2015})}\BibitemShut {NoStop}%
\bibitem [{\citenamefont {Tranquada}\ \emph {et~al.}(1995)\citenamefont
  {Tranquada}, \citenamefont {Sternlieb}, \citenamefont {Axe}, \citenamefont
  {Nakamura},\ and\ \citenamefont {Uchida}}]{Tranquada:1995}%
  \BibitemOpen
  \bibfield  {author} {\bibinfo {author} {\bibfnamefont {J.~M.}\ \bibnamefont
  {Tranquada}}, \bibinfo {author} {\bibfnamefont {B.~J.}\ \bibnamefont
  {Sternlieb}}, \bibinfo {author} {\bibfnamefont {J.~D.}\ \bibnamefont {Axe}},
  \bibinfo {author} {\bibfnamefont {Y.}~\bibnamefont {Nakamura}}, \ and\
  \bibinfo {author} {\bibfnamefont {S.}~\bibnamefont {Uchida}},\ }\bibfield
  {title} {\enquote {\bibinfo {title} {Evidence for stripe correlations of
  spins and holes in copper oxide superconductors},}\ }\href {\doibase
  10.1038/375561a0} {\bibfield  {journal} {\bibinfo  {journal} {Nature}\
  }\textbf {\bibinfo {volume} {375}},\ \bibinfo {pages} {561--563} (\bibinfo
  {year} {1995})}\BibitemShut {NoStop}%
\bibitem [{\citenamefont {Hoffman}\ \emph {et~al.}(2002)\citenamefont
  {Hoffman}, \citenamefont {Hudson}, \citenamefont {Lang}, \citenamefont
  {Madhavan}, \citenamefont {Eisaki}, \citenamefont {Uchida},\ and\
  \citenamefont {Davis}}]{Hoffman:2002}%
  \BibitemOpen
  \bibfield  {author} {\bibinfo {author} {\bibfnamefont {J.~E.}\ \bibnamefont
  {Hoffman}}, \bibinfo {author} {\bibfnamefont {E.~W.}\ \bibnamefont {Hudson}},
  \bibinfo {author} {\bibfnamefont {K.~M.}\ \bibnamefont {Lang}}, \bibinfo
  {author} {\bibfnamefont {V.}~\bibnamefont {Madhavan}}, \bibinfo {author}
  {\bibfnamefont {H.}~\bibnamefont {Eisaki}}, \bibinfo {author} {\bibfnamefont
  {S.}~\bibnamefont {Uchida}}, \ and\ \bibinfo {author} {\bibfnamefont {J.~C.}\
  \bibnamefont {Davis}},\ }\bibfield  {title} {\enquote {\bibinfo {title} {A
  four unit cell periodic pattern of quasi-particle states surrounding vortex
  cores in {B}i$_2${S}r$_2${C}a{C}u$_2${O}$_{8+\delta}$},}\ }\href {\doibase
  10.1126/science.1066974} {\bibfield  {journal} {\bibinfo  {journal}
  {Science}\ }\textbf {\bibinfo {volume} {295}},\ \bibinfo {pages} {466--469}
  (\bibinfo {year} {2002})}\BibitemShut {NoStop}%
\bibitem [{\citenamefont {Kivelson}\ \emph {et~al.}(2003)\citenamefont
  {Kivelson}, \citenamefont {Bindloss}, \citenamefont {Fradkin}, \citenamefont
  {Oganesyan}, \citenamefont {Tranquada}, \citenamefont {Kapitulnik},\ and\
  \citenamefont {Howald}}]{Kivelson:2003}%
  \BibitemOpen
  \bibfield  {author} {\bibinfo {author} {\bibfnamefont {S.~A.}\ \bibnamefont
  {Kivelson}}, \bibinfo {author} {\bibfnamefont {I.~P.}\ \bibnamefont
  {Bindloss}}, \bibinfo {author} {\bibfnamefont {E.}~\bibnamefont {Fradkin}},
  \bibinfo {author} {\bibfnamefont {V.}~\bibnamefont {Oganesyan}}, \bibinfo
  {author} {\bibfnamefont {J.~M.}\ \bibnamefont {Tranquada}}, \bibinfo {author}
  {\bibfnamefont {A.}~\bibnamefont {Kapitulnik}}, \ and\ \bibinfo {author}
  {\bibfnamefont {C.}~\bibnamefont {Howald}},\ }\bibfield  {title} {\enquote
  {\bibinfo {title} {How to detect fluctuating stripes in the high-temperature
  superconductors},}\ }\href {\doibase 10.1103/RevModPhys.75.1201} {\bibfield
  {journal} {\bibinfo  {journal} {Rev. Mod. Phys.}\ }\textbf {\bibinfo {volume}
  {75}},\ \bibinfo {pages} {1201--1241} (\bibinfo {year} {2003})}\BibitemShut
  {NoStop}%
\bibitem [{\citenamefont {Wise}\ \emph {et~al.}(2008)\citenamefont {Wise},
  \citenamefont {Boyer}, \citenamefont {Chatterjee}, \citenamefont {Kondo},
  \citenamefont {Takeuchi}, \citenamefont {Ikuta}, \citenamefont {Wang},\ and\
  \citenamefont {Hudson}}]{Wise:2008}%
  \BibitemOpen
  \bibfield  {author} {\bibinfo {author} {\bibfnamefont {W.~D.}\ \bibnamefont
  {Wise}}, \bibinfo {author} {\bibfnamefont {M.~C.}\ \bibnamefont {Boyer}},
  \bibinfo {author} {\bibfnamefont {Kamalesh}\ \bibnamefont {Chatterjee}},
  \bibinfo {author} {\bibfnamefont {Takeshi}\ \bibnamefont {Kondo}}, \bibinfo
  {author} {\bibfnamefont {T.}~\bibnamefont {Takeuchi}}, \bibinfo {author}
  {\bibfnamefont {H.}~\bibnamefont {Ikuta}}, \bibinfo {author} {\bibfnamefont
  {Yayu}\ \bibnamefont {Wang}}, \ and\ \bibinfo {author} {\bibfnamefont
  {E.~W.}\ \bibnamefont {Hudson}},\ }\bibfield  {title} {\enquote {\bibinfo
  {title} {Charge-density-wave origin of cuprate checkerboard visualized by
  scanning tunnelling microscopy},}\ }\href {\doibase 10.1038/nphys1021}
  {\bibfield  {journal} {\bibinfo  {journal} {Nature Physics}\ }\textbf
  {\bibinfo {volume} {4}},\ \bibinfo {pages} {696--699} (\bibinfo {year}
  {2008})}\BibitemShut {NoStop}%
\bibitem [{\citenamefont {Reznik}(2012)}]{Reznik:2012}%
  \BibitemOpen
  \bibfield  {author} {\bibinfo {author} {\bibfnamefont {D.}~\bibnamefont
  {Reznik}},\ }\bibfield  {title} {\enquote {\bibinfo {title} {Phonon anomalies
  and dynamic stripes},}\ }\href {\doibase
  https://doi.org/10.1016/j.physc.2012.01.024} {\bibfield  {journal} {\bibinfo
  {journal} {Physica C: Superconductivity}\ }\textbf {\bibinfo {volume}
  {481}},\ \bibinfo {pages} {75--92} (\bibinfo {year} {2012})},\ \bibinfo
  {note} {stripes and Electronic Liquid Crystals in Strongly Correlated
  Materials}\BibitemShut {NoStop}%
\bibitem [{\citenamefont {Blackburn}\ \emph {et~al.}(2013)\citenamefont
  {Blackburn}, \citenamefont {Chang}, \citenamefont {Said}, \citenamefont
  {Leu}, \citenamefont {Liang}, \citenamefont {Bonn}, \citenamefont {Hardy},
  \citenamefont {Forgan},\ and\ \citenamefont {Hayden}}]{Blackburn:2013}%
  \BibitemOpen
  \bibfield  {author} {\bibinfo {author} {\bibfnamefont {E.}~\bibnamefont
  {Blackburn}}, \bibinfo {author} {\bibfnamefont {J.}~\bibnamefont {Chang}},
  \bibinfo {author} {\bibfnamefont {A.~H.}\ \bibnamefont {Said}}, \bibinfo
  {author} {\bibfnamefont {B.~M.}\ \bibnamefont {Leu}}, \bibinfo {author}
  {\bibfnamefont {Ruixing}\ \bibnamefont {Liang}}, \bibinfo {author}
  {\bibfnamefont {D.~A.}\ \bibnamefont {Bonn}}, \bibinfo {author}
  {\bibfnamefont {W.~N.}\ \bibnamefont {Hardy}}, \bibinfo {author}
  {\bibfnamefont {E.~M.}\ \bibnamefont {Forgan}}, \ and\ \bibinfo {author}
  {\bibfnamefont {S.~M.}\ \bibnamefont {Hayden}},\ }\bibfield  {title}
  {\enquote {\bibinfo {title} {Inelastic x-ray study of phonon broadening and
  charge-density wave formation in ortho-ii-ordered
  {Y}{B}a$_2${C}u$_3${O}$_{6.54}$},}\ }\href {\doibase
  10.1103/PhysRevB.88.054506} {\bibfield  {journal} {\bibinfo  {journal} {Phys.
  Rev. B}\ }\textbf {\bibinfo {volume} {88}},\ \bibinfo {pages} {054506}
  (\bibinfo {year} {2013})}\BibitemShut {NoStop}%
\bibitem [{\citenamefont {Comin}\ \emph {et~al.}(2014)\citenamefont {Comin},
  \citenamefont {Frano}, \citenamefont {Yee}, \citenamefont {Yoshida},
  \citenamefont {Eisaki}, \citenamefont {Schierle}, \citenamefont {Weschke},
  \citenamefont {Sutarto}, \citenamefont {He}, \citenamefont {Soumyanarayanan},
  \citenamefont {He}, \citenamefont {Tacon}, \citenamefont {Elfimov},
  \citenamefont {Hoffman}, \citenamefont {Sawatzky}, \citenamefont {Keimer},\
  and\ \citenamefont {Damascelli}}]{Comin:2014}%
  \BibitemOpen
  \bibfield  {author} {\bibinfo {author} {\bibfnamefont {R.}~\bibnamefont
  {Comin}}, \bibinfo {author} {\bibfnamefont {A.}~\bibnamefont {Frano}},
  \bibinfo {author} {\bibfnamefont {M.~M.}\ \bibnamefont {Yee}}, \bibinfo
  {author} {\bibfnamefont {Y.}~\bibnamefont {Yoshida}}, \bibinfo {author}
  {\bibfnamefont {H.}~\bibnamefont {Eisaki}}, \bibinfo {author} {\bibfnamefont
  {E.}~\bibnamefont {Schierle}}, \bibinfo {author} {\bibfnamefont
  {E.}~\bibnamefont {Weschke}}, \bibinfo {author} {\bibfnamefont
  {R.}~\bibnamefont {Sutarto}}, \bibinfo {author} {\bibfnamefont
  {F.}~\bibnamefont {He}}, \bibinfo {author} {\bibfnamefont {A.}~\bibnamefont
  {Soumyanarayanan}}, \bibinfo {author} {\bibfnamefont {Yang}\ \bibnamefont
  {He}}, \bibinfo {author} {\bibfnamefont {M.~Le}\ \bibnamefont {Tacon}},
  \bibinfo {author} {\bibfnamefont {I.~S.}\ \bibnamefont {Elfimov}}, \bibinfo
  {author} {\bibfnamefont {Jennifer~E.}\ \bibnamefont {Hoffman}}, \bibinfo
  {author} {\bibfnamefont {G.~A.}\ \bibnamefont {Sawatzky}}, \bibinfo {author}
  {\bibfnamefont {B.}~\bibnamefont {Keimer}}, \ and\ \bibinfo {author}
  {\bibfnamefont {A.}~\bibnamefont {Damascelli}},\ }\bibfield  {title}
  {\enquote {\bibinfo {title} {Charge order driven by fermi-arc instability in
  {B}i$_2${S}r$_{2-x}${L}a$_x${C}u{O}$_{6+\delta}$},}\ }\href {\doibase
  10.1126/science.1242996} {\bibfield  {journal} {\bibinfo  {journal}
  {Science}\ }\textbf {\bibinfo {volume} {343}},\ \bibinfo {pages} {390--392}
  (\bibinfo {year} {2014})}\BibitemShut {NoStop}%
\bibitem [{\citenamefont {Le~Tacon}\ \emph {et~al.}(2014)\citenamefont
  {Le~Tacon}, \citenamefont {Bosak}, \citenamefont {Souliou}, \citenamefont
  {Dellea}, \citenamefont {Loew}, \citenamefont {Heid}, \citenamefont {Bohnen},
  \citenamefont {Ghiringhelli}, \citenamefont {Krisch},\ and\ \citenamefont
  {Keimer}}]{LeTacon:2014}%
  \BibitemOpen
  \bibfield  {author} {\bibinfo {author} {\bibfnamefont {M.}~\bibnamefont
  {Le~Tacon}}, \bibinfo {author} {\bibfnamefont {A.}~\bibnamefont {Bosak}},
  \bibinfo {author} {\bibfnamefont {S.~M.}\ \bibnamefont {Souliou}}, \bibinfo
  {author} {\bibfnamefont {G.}~\bibnamefont {Dellea}}, \bibinfo {author}
  {\bibfnamefont {T.}~\bibnamefont {Loew}}, \bibinfo {author} {\bibfnamefont
  {R.}~\bibnamefont {Heid}}, \bibinfo {author} {\bibfnamefont {K.-P.}\
  \bibnamefont {Bohnen}}, \bibinfo {author} {\bibfnamefont {G.}~\bibnamefont
  {Ghiringhelli}}, \bibinfo {author} {\bibfnamefont {M.}~\bibnamefont
  {Krisch}}, \ and\ \bibinfo {author} {\bibfnamefont {B.}~\bibnamefont
  {Keimer}},\ }\bibfield  {title} {\enquote {\bibinfo {title} {Inelastic x-ray
  scattering in {YB}a$_2${C}u$_3${O}$_{6.6}$ reveals giant phonon anomalies and
  elastic central peak due to charge-density-wave formation},}\ }\href
  {\doibase 10.1038/nphys2805} {\bibfield  {journal} {\bibinfo  {journal}
  {Nature Physics}\ }\textbf {\bibinfo {volume} {10}},\ \bibinfo {pages}
  {52--58} (\bibinfo {year} {2014})}\BibitemShut {NoStop}%
\bibitem [{\citenamefont {Fujita}\ \emph {et~al.}(2014)\citenamefont {Fujita},
  \citenamefont {Hamidian}, \citenamefont {Edkins}, \citenamefont {Kim},
  \citenamefont {Kohsaka}, \citenamefont {Azuma}, \citenamefont {Takano},
  \citenamefont {Takagi}, \citenamefont {Eisaki}, \citenamefont {ichi Uchida},
  \citenamefont {Allais}, \citenamefont {Lawler}, \citenamefont {Kim},
  \citenamefont {Sachdev},\ and\ \citenamefont {Davis}}]{Fujita:2014}%
  \BibitemOpen
  \bibfield  {author} {\bibinfo {author} {\bibfnamefont {Kazuhiro}\
  \bibnamefont {Fujita}}, \bibinfo {author} {\bibfnamefont {Mohammad~H.}\
  \bibnamefont {Hamidian}}, \bibinfo {author} {\bibfnamefont {Stephen~D.}\
  \bibnamefont {Edkins}}, \bibinfo {author} {\bibfnamefont {Chung~Koo}\
  \bibnamefont {Kim}}, \bibinfo {author} {\bibfnamefont {Yuhki}\ \bibnamefont
  {Kohsaka}}, \bibinfo {author} {\bibfnamefont {Masaki}\ \bibnamefont {Azuma}},
  \bibinfo {author} {\bibfnamefont {Mikio}\ \bibnamefont {Takano}}, \bibinfo
  {author} {\bibfnamefont {Hidenori}\ \bibnamefont {Takagi}}, \bibinfo {author}
  {\bibfnamefont {Hiroshi}\ \bibnamefont {Eisaki}}, \bibinfo {author}
  {\bibfnamefont {Shin}\ \bibnamefont {ichi Uchida}}, \bibinfo {author}
  {\bibfnamefont {Andrea}\ \bibnamefont {Allais}}, \bibinfo {author}
  {\bibfnamefont {Michael~J.}\ \bibnamefont {Lawler}}, \bibinfo {author}
  {\bibfnamefont {Eun-Ah}\ \bibnamefont {Kim}}, \bibinfo {author}
  {\bibfnamefont {Subir}\ \bibnamefont {Sachdev}}, \ and\ \bibinfo {author}
  {\bibfnamefont {J.~C.~S{\'{e}}amus}\ \bibnamefont {Davis}},\ }\bibfield
  {title} {\enquote {\bibinfo {title} {Direct phase-sensitive identification of
  a d-form factor density wave in underdoped cuprates},}\ }\href {\doibase
  10.1073/pnas.1406297111} {\bibfield  {journal} {\bibinfo  {journal}
  {Proceedings of the National Academy of Sciences}\ }\textbf {\bibinfo
  {volume} {111}} (\bibinfo {year} {2014}),\
  10.1073/pnas.1406297111}\BibitemShut {NoStop}%
\bibitem [{\citenamefont {Tranquada}\ \emph {et~al.}(1994)\citenamefont
  {Tranquada}, \citenamefont {Buttrey}, \citenamefont {Sachan},\ and\
  \citenamefont {Lorenzo}}]{Tranquada:1994}%
  \BibitemOpen
  \bibfield  {author} {\bibinfo {author} {\bibfnamefont {J.~M.}\ \bibnamefont
  {Tranquada}}, \bibinfo {author} {\bibfnamefont {D.~J.}\ \bibnamefont
  {Buttrey}}, \bibinfo {author} {\bibfnamefont {V.}~\bibnamefont {Sachan}}, \
  and\ \bibinfo {author} {\bibfnamefont {J.~E.}\ \bibnamefont {Lorenzo}},\
  }\bibfield  {title} {\enquote {\bibinfo {title} {Simultaneous ordering of
  holes and spins in {L}a$_2${N}i{O}$_{4.125}$},}\ }\href {\doibase
  10.1103/PhysRevLett.73.1003} {\bibfield  {journal} {\bibinfo  {journal}
  {Phys. Rev. Lett.}\ }\textbf {\bibinfo {volume} {73}},\ \bibinfo {pages}
  {1003--1006} (\bibinfo {year} {1994})}\BibitemShut {NoStop}%
\bibitem [{\citenamefont {Lee}\ and\ \citenamefont {Cheong}(1997)}]{Lee:1997}%
  \BibitemOpen
  \bibfield  {author} {\bibinfo {author} {\bibfnamefont {S.-H.}\ \bibnamefont
  {Lee}}\ and\ \bibinfo {author} {\bibfnamefont {S-W.}\ \bibnamefont
  {Cheong}},\ }\bibfield  {title} {\enquote {\bibinfo {title} {Melting of
  quasi-two-dimensional charge stripes in
  {L}a$_{5/3}${S}r$_{1/3}${N}i{O}$_{4}$},}\ }\href {\doibase
  10.1103/PhysRevLett.79.2514} {\bibfield  {journal} {\bibinfo  {journal}
  {Phys. Rev. Lett.}\ }\textbf {\bibinfo {volume} {79}},\ \bibinfo {pages}
  {2514--2517} (\bibinfo {year} {1997})}\BibitemShut {NoStop}%
\bibitem [{\citenamefont {Jiang}\ \emph {et~al.}(2021)\citenamefont {Jiang},
  \citenamefont {Yin}, \citenamefont {Denner}, \citenamefont {Shumiya},
  \citenamefont {Ortiz}, \citenamefont {Xu}, \citenamefont {Guguchia},
  \citenamefont {He}, \citenamefont {Hossain}, \citenamefont {Liu},
  \citenamefont {Ruff}, \citenamefont {Kautzsch}, \citenamefont {Zhang},
  \citenamefont {Chang}, \citenamefont {Belopolski}, \citenamefont {Zhang},
  \citenamefont {Cochran}, \citenamefont {Multer}, \citenamefont {Litskevich},
  \citenamefont {Cheng}, \citenamefont {Yang}, \citenamefont {Wang},
  \citenamefont {Thomale}, \citenamefont {Neupert}, \citenamefont {Wilson},\
  and\ \citenamefont {Hasan}}]{Jiang:2021}%
  \BibitemOpen
  \bibfield  {author} {\bibinfo {author} {\bibfnamefont {Yu-Xiao}\ \bibnamefont
  {Jiang}}, \bibinfo {author} {\bibfnamefont {Jia-Xin}\ \bibnamefont {Yin}},
  \bibinfo {author} {\bibfnamefont {M.~Michael}\ \bibnamefont {Denner}},
  \bibinfo {author} {\bibfnamefont {Nana}\ \bibnamefont {Shumiya}}, \bibinfo
  {author} {\bibfnamefont {Brenden~R.}\ \bibnamefont {Ortiz}}, \bibinfo
  {author} {\bibfnamefont {Gang}\ \bibnamefont {Xu}}, \bibinfo {author}
  {\bibfnamefont {Zurab}\ \bibnamefont {Guguchia}}, \bibinfo {author}
  {\bibfnamefont {Junyi}\ \bibnamefont {He}}, \bibinfo {author} {\bibfnamefont
  {Md~Shafayat}\ \bibnamefont {Hossain}}, \bibinfo {author} {\bibfnamefont
  {Xiaoxiong}\ \bibnamefont {Liu}}, \bibinfo {author} {\bibfnamefont {Jacob}\
  \bibnamefont {Ruff}}, \bibinfo {author} {\bibfnamefont {Linus}\ \bibnamefont
  {Kautzsch}}, \bibinfo {author} {\bibfnamefont {Songtian~S.}\ \bibnamefont
  {Zhang}}, \bibinfo {author} {\bibfnamefont {Guoqing}\ \bibnamefont {Chang}},
  \bibinfo {author} {\bibfnamefont {Ilya}\ \bibnamefont {Belopolski}}, \bibinfo
  {author} {\bibfnamefont {Qi}~\bibnamefont {Zhang}}, \bibinfo {author}
  {\bibfnamefont {Tyler~A.}\ \bibnamefont {Cochran}}, \bibinfo {author}
  {\bibfnamefont {Daniel}\ \bibnamefont {Multer}}, \bibinfo {author}
  {\bibfnamefont {Maksim}\ \bibnamefont {Litskevich}}, \bibinfo {author}
  {\bibfnamefont {Zi-Jia}\ \bibnamefont {Cheng}}, \bibinfo {author}
  {\bibfnamefont {Xian~P.}\ \bibnamefont {Yang}}, \bibinfo {author}
  {\bibfnamefont {Ziqiang}\ \bibnamefont {Wang}}, \bibinfo {author}
  {\bibfnamefont {Ronny}\ \bibnamefont {Thomale}}, \bibinfo {author}
  {\bibfnamefont {Titus}\ \bibnamefont {Neupert}}, \bibinfo {author}
  {\bibfnamefont {Stephen~D.}\ \bibnamefont {Wilson}}, \ and\ \bibinfo {author}
  {\bibfnamefont {M.~Zahid}\ \bibnamefont {Hasan}},\ }\bibfield  {title}
  {\enquote {\bibinfo {title} {Unconventional chiral charge order in kagome
  superconductor {KV}$_3${S}b$_5$},}\ }\href {\doibase
  10.1038/s41563-021-01034-y} {\bibfield  {journal} {\bibinfo  {journal}
  {Nature Materials}\ }\textbf {\bibinfo {volume} {20}},\ \bibinfo {pages}
  {1353--1357} (\bibinfo {year} {2021})}\BibitemShut {NoStop}%
\bibitem [{\citenamefont {Mielke}\ \emph {et~al.}(2022)\citenamefont {Mielke},
  \citenamefont {Das}, \citenamefont {Yin}, \citenamefont {Liu}, \citenamefont
  {Gupta}, \citenamefont {Jiang}, \citenamefont {Medarde}, \citenamefont {Wu},
  \citenamefont {Lei}, \citenamefont {Chang}, \citenamefont {Dai},
  \citenamefont {Si}, \citenamefont {Miao}, \citenamefont {Thomale},
  \citenamefont {Neupert}, \citenamefont {Shi}, \citenamefont {Khasanov},
  \citenamefont {Hasan}, \citenamefont {Luetkens},\ and\ \citenamefont
  {Guguchia}}]{Mielke:2022}%
  \BibitemOpen
  \bibfield  {author} {\bibinfo {author} {\bibfnamefont {C.}~\bibnamefont
  {Mielke}}, \bibinfo {author} {\bibfnamefont {D.}~\bibnamefont {Das}},
  \bibinfo {author} {\bibfnamefont {J.-X.}\ \bibnamefont {Yin}}, \bibinfo
  {author} {\bibfnamefont {H.}~\bibnamefont {Liu}}, \bibinfo {author}
  {\bibfnamefont {R.}~\bibnamefont {Gupta}}, \bibinfo {author} {\bibfnamefont
  {Y.-X.}\ \bibnamefont {Jiang}}, \bibinfo {author} {\bibfnamefont
  {M.}~\bibnamefont {Medarde}}, \bibinfo {author} {\bibfnamefont
  {X.}~\bibnamefont {Wu}}, \bibinfo {author} {\bibfnamefont {H.~C.}\
  \bibnamefont {Lei}}, \bibinfo {author} {\bibfnamefont {J.}~\bibnamefont
  {Chang}}, \bibinfo {author} {\bibfnamefont {Pengcheng}\ \bibnamefont {Dai}},
  \bibinfo {author} {\bibfnamefont {Q.}~\bibnamefont {Si}}, \bibinfo {author}
  {\bibfnamefont {H.}~\bibnamefont {Miao}}, \bibinfo {author} {\bibfnamefont
  {R.}~\bibnamefont {Thomale}}, \bibinfo {author} {\bibfnamefont
  {T.}~\bibnamefont {Neupert}}, \bibinfo {author} {\bibfnamefont
  {Y.}~\bibnamefont {Shi}}, \bibinfo {author} {\bibfnamefont {R.}~\bibnamefont
  {Khasanov}}, \bibinfo {author} {\bibfnamefont {M.~Z.}\ \bibnamefont {Hasan}},
  \bibinfo {author} {\bibfnamefont {H.}~\bibnamefont {Luetkens}}, \ and\
  \bibinfo {author} {\bibfnamefont {Z.}~\bibnamefont {Guguchia}},\ }\bibfield
  {title} {\enquote {\bibinfo {title} {Time-reversal symmetry-breaking charge
  order in a kagome superconductor},}\ }\href {\doibase
  10.1038/s41586-021-04327-z} {\bibfield  {journal} {\bibinfo  {journal}
  {Nature}\ }\textbf {\bibinfo {volume} {602}},\ \bibinfo {pages} {245--250}
  (\bibinfo {year} {2022})}\BibitemShut {NoStop}%
\bibitem [{\citenamefont {Ortiz}\ \emph {et~al.}(2019)\citenamefont {Ortiz},
  \citenamefont {Gomes}, \citenamefont {Morey}, \citenamefont {Winiarski},
  \citenamefont {Bordelon}, \citenamefont {Mangum}, \citenamefont {Oswald},
  \citenamefont {Rodriguez-Rivera}, \citenamefont {Neilson}, \citenamefont
  {Wilson}, \citenamefont {Ertekin}, \citenamefont {McQueen},\ and\
  \citenamefont {Toberer}}]{Ortiz:2019}%
  \BibitemOpen
  \bibfield  {author} {\bibinfo {author} {\bibfnamefont {Brenden~R.}\
  \bibnamefont {Ortiz}}, \bibinfo {author} {\bibfnamefont {L\'{\i}dia~C.}\
  \bibnamefont {Gomes}}, \bibinfo {author} {\bibfnamefont {Jennifer~R.}\
  \bibnamefont {Morey}}, \bibinfo {author} {\bibfnamefont {Michal}\
  \bibnamefont {Winiarski}}, \bibinfo {author} {\bibfnamefont {Mitchell}\
  \bibnamefont {Bordelon}}, \bibinfo {author} {\bibfnamefont {John~S.}\
  \bibnamefont {Mangum}}, \bibinfo {author} {\bibfnamefont {Iain W.~H.}\
  \bibnamefont {Oswald}}, \bibinfo {author} {\bibfnamefont {Jose~A.}\
  \bibnamefont {Rodriguez-Rivera}}, \bibinfo {author} {\bibfnamefont
  {James~R.}\ \bibnamefont {Neilson}}, \bibinfo {author} {\bibfnamefont
  {Stephen~D.}\ \bibnamefont {Wilson}}, \bibinfo {author} {\bibfnamefont
  {Elif}\ \bibnamefont {Ertekin}}, \bibinfo {author} {\bibfnamefont {Tyrel~M.}\
  \bibnamefont {McQueen}}, \ and\ \bibinfo {author} {\bibfnamefont {Eric~S.}\
  \bibnamefont {Toberer}},\ }\bibfield  {title} {\enquote {\bibinfo {title}
  {New kagome prototype materials: discovery of {K}{V}$_3${S}b$_5$,
  {R}b{V}$_3${S}b$_5$ and {C}s{V}$_3${S}b$_5$},}\ }\href {\doibase
  10.1103/PhysRevMaterials.3.094407} {\bibfield  {journal} {\bibinfo  {journal}
  {Phys. Rev. Mater.}\ }\textbf {\bibinfo {volume} {3}},\ \bibinfo {pages}
  {094407} (\bibinfo {year} {2019})}\BibitemShut {NoStop}%
\bibitem [{\citenamefont {Ortiz}\ \emph {et~al.}(2020)\citenamefont {Ortiz},
  \citenamefont {Teicher}, \citenamefont {Hu}, \citenamefont {Zuo},
  \citenamefont {Sarte}, \citenamefont {Schueller}, \citenamefont {Abeykoon},
  \citenamefont {Krogstad}, \citenamefont {Rosenkranz}, \citenamefont {Osborn},
  \citenamefont {Seshadri}, \citenamefont {Balents}, \citenamefont {He},\ and\
  \citenamefont {Wilson}}]{Stephen:2020}%
  \BibitemOpen
  \bibfield  {author} {\bibinfo {author} {\bibfnamefont {Brenden~R.}\
  \bibnamefont {Ortiz}}, \bibinfo {author} {\bibfnamefont {Samuel M.~L.}\
  \bibnamefont {Teicher}}, \bibinfo {author} {\bibfnamefont {Yong}\
  \bibnamefont {Hu}}, \bibinfo {author} {\bibfnamefont {Julia~L.}\ \bibnamefont
  {Zuo}}, \bibinfo {author} {\bibfnamefont {Paul~M.}\ \bibnamefont {Sarte}},
  \bibinfo {author} {\bibfnamefont {Emily~C.}\ \bibnamefont {Schueller}},
  \bibinfo {author} {\bibfnamefont {A.~M.~Milinda}\ \bibnamefont {Abeykoon}},
  \bibinfo {author} {\bibfnamefont {Matthew~J.}\ \bibnamefont {Krogstad}},
  \bibinfo {author} {\bibfnamefont {Stephan}\ \bibnamefont {Rosenkranz}},
  \bibinfo {author} {\bibfnamefont {Raymond}\ \bibnamefont {Osborn}}, \bibinfo
  {author} {\bibfnamefont {Ram}\ \bibnamefont {Seshadri}}, \bibinfo {author}
  {\bibfnamefont {Leon}\ \bibnamefont {Balents}}, \bibinfo {author}
  {\bibfnamefont {Junfeng}\ \bibnamefont {He}}, \ and\ \bibinfo {author}
  {\bibfnamefont {Stephen~D.}\ \bibnamefont {Wilson}},\ }\bibfield  {title}
  {\enquote {\bibinfo {title} {{C}s{V}$_3${S}b$_5$: A {Z}$_2$ topological
  kagome metal with a superconducting ground state},}\ }\href {\doibase
  10.1103/PhysRevLett.125.247002} {\bibfield  {journal} {\bibinfo  {journal}
  {Phys. Rev. Lett.}\ }\textbf {\bibinfo {volume} {125}},\ \bibinfo {pages}
  {247002} (\bibinfo {year} {2020})}\BibitemShut {NoStop}%
\bibitem [{\citenamefont {Yang}\ \emph {et~al.}(2020)\citenamefont {Yang},
  \citenamefont {Wang}, \citenamefont {Ortiz}, \citenamefont {Liu},
  \citenamefont {Gayles}, \citenamefont {Derunova}, \citenamefont
  {Gonzalez-Hernandez}, \citenamefont {Šmejkal}, \citenamefont {Chen},
  \citenamefont {Parkin}, \citenamefont {Wilson}, \citenamefont {Toberer},
  \citenamefont {McQueen},\ and\ \citenamefont {Ali}}]{Mazhar:2020}%
  \BibitemOpen
  \bibfield  {author} {\bibinfo {author} {\bibfnamefont {Shuo-Ying}\
  \bibnamefont {Yang}}, \bibinfo {author} {\bibfnamefont {Yaojia}\ \bibnamefont
  {Wang}}, \bibinfo {author} {\bibfnamefont {Brenden~R.}\ \bibnamefont
  {Ortiz}}, \bibinfo {author} {\bibfnamefont {Defa}\ \bibnamefont {Liu}},
  \bibinfo {author} {\bibfnamefont {Jacob}\ \bibnamefont {Gayles}}, \bibinfo
  {author} {\bibfnamefont {Elena}\ \bibnamefont {Derunova}}, \bibinfo {author}
  {\bibfnamefont {Rafael}\ \bibnamefont {Gonzalez-Hernandez}}, \bibinfo
  {author} {\bibfnamefont {Libor}\ \bibnamefont {Šmejkal}}, \bibinfo {author}
  {\bibfnamefont {Yulin}\ \bibnamefont {Chen}}, \bibinfo {author}
  {\bibfnamefont {Stuart S.~P.}\ \bibnamefont {Parkin}}, \bibinfo {author}
  {\bibfnamefont {Stephen~D.}\ \bibnamefont {Wilson}}, \bibinfo {author}
  {\bibfnamefont {Eric~S.}\ \bibnamefont {Toberer}}, \bibinfo {author}
  {\bibfnamefont {Tyrel}\ \bibnamefont {McQueen}}, \ and\ \bibinfo {author}
  {\bibfnamefont {Mazhar~N.}\ \bibnamefont {Ali}},\ }\bibfield  {title}
  {\enquote {\bibinfo {title} {Giant, unconventional anomalous hall effect in
  the metallic frustrated magnet candidate, {KV}$_3${S}b$_5$},}\ }\href
  {\doibase 10.1126/sciadv.abb6003} {\bibfield  {journal} {\bibinfo  {journal}
  {Science Advances}\ }\textbf {\bibinfo {volume} {6}},\ \bibinfo {pages}
  {eabb6003} (\bibinfo {year} {2020})}\BibitemShut {NoStop}%
\bibitem [{\citenamefont {Feng}\ \emph {et~al.}(2021)\citenamefont {Feng},
  \citenamefont {Jiang}, \citenamefont {Wang},\ and\ \citenamefont
  {Hu}}]{Jiangping:2021}%
  \BibitemOpen
  \bibfield  {author} {\bibinfo {author} {\bibfnamefont {Xilin}\ \bibnamefont
  {Feng}}, \bibinfo {author} {\bibfnamefont {Kun}\ \bibnamefont {Jiang}},
  \bibinfo {author} {\bibfnamefont {Ziqiang}\ \bibnamefont {Wang}}, \ and\
  \bibinfo {author} {\bibfnamefont {Jiangping}\ \bibnamefont {Hu}},\ }\bibfield
   {title} {\enquote {\bibinfo {title} {Chiral flux phase in the kagome
  superconductor {AV}$_3${S}b$_5$},}\ }\href {\doibase
  https://doi.org/10.1016/j.scib.2021.04.043} {\bibfield  {journal} {\bibinfo
  {journal} {Science Bulletin}\ }\textbf {\bibinfo {volume} {66}},\ \bibinfo
  {pages} {1384--1388} (\bibinfo {year} {2021})}\BibitemShut {NoStop}%
\bibitem [{\citenamefont {Kiesel}\ \emph {et~al.}(2013)\citenamefont {Kiesel},
  \citenamefont {Platt},\ and\ \citenamefont {Thomale}}]{Thomale:2013}%
  \BibitemOpen
  \bibfield  {author} {\bibinfo {author} {\bibfnamefont {Maximilian~L.}\
  \bibnamefont {Kiesel}}, \bibinfo {author} {\bibfnamefont {Christian}\
  \bibnamefont {Platt}}, \ and\ \bibinfo {author} {\bibfnamefont {Ronny}\
  \bibnamefont {Thomale}},\ }\bibfield  {title} {\enquote {\bibinfo {title}
  {Unconventional fermi surface instabilities in the kagome hubbard model},}\
  }\href {\doibase 10.1103/PhysRevLett.110.126405} {\bibfield  {journal}
  {\bibinfo  {journal} {Phys. Rev. Lett.}\ }\textbf {\bibinfo {volume} {110}},\
  \bibinfo {pages} {126405} (\bibinfo {year} {2013})}\BibitemShut {NoStop}%
\bibitem [{\citenamefont {Wang}\ \emph {et~al.}(2013)\citenamefont {Wang},
  \citenamefont {Li}, \citenamefont {Xiang},\ and\ \citenamefont
  {Wang}}]{Qianghua:2013}%
  \BibitemOpen
  \bibfield  {author} {\bibinfo {author} {\bibfnamefont {Wan-Sheng}\
  \bibnamefont {Wang}}, \bibinfo {author} {\bibfnamefont {Zheng-Zhao}\
  \bibnamefont {Li}}, \bibinfo {author} {\bibfnamefont {Yuan-Yuan}\
  \bibnamefont {Xiang}}, \ and\ \bibinfo {author} {\bibfnamefont {Qiang-Hua}\
  \bibnamefont {Wang}},\ }\bibfield  {title} {\enquote {\bibinfo {title}
  {Competing electronic orders on kagome lattices at van hove filling},}\
  }\href {\doibase 10.1103/PhysRevB.87.115135} {\bibfield  {journal} {\bibinfo
  {journal} {Phys. Rev. B}\ }\textbf {\bibinfo {volume} {87}},\ \bibinfo
  {pages} {115135} (\bibinfo {year} {2013})}\BibitemShut {NoStop}%
\bibitem [{\citenamefont {Tan}\ \emph {et~al.}(2021)\citenamefont {Tan},
  \citenamefont {Liu}, \citenamefont {Wang},\ and\ \citenamefont
  {Yan}}]{Binghai:2021}%
  \BibitemOpen
  \bibfield  {author} {\bibinfo {author} {\bibfnamefont {Hengxin}\ \bibnamefont
  {Tan}}, \bibinfo {author} {\bibfnamefont {Yizhou}\ \bibnamefont {Liu}},
  \bibinfo {author} {\bibfnamefont {Ziqiang}\ \bibnamefont {Wang}}, \ and\
  \bibinfo {author} {\bibfnamefont {Binghai}\ \bibnamefont {Yan}},\ }\bibfield
  {title} {\enquote {\bibinfo {title} {Charge density waves and electronic
  properties of superconducting kagome metals},}\ }\href {\doibase
  10.1103/PhysRevLett.127.046401} {\bibfield  {journal} {\bibinfo  {journal}
  {Phys. Rev. Lett.}\ }\textbf {\bibinfo {volume} {127}},\ \bibinfo {pages}
  {046401} (\bibinfo {year} {2021})}\BibitemShut {NoStop}%
\bibitem [{\citenamefont {Li}\ \emph {et~al.}(2021)\citenamefont {Li},
  \citenamefont {Zhang}, \citenamefont {Yilmaz}, \citenamefont {Pai},
  \citenamefont {Marvinney}, \citenamefont {Said}, \citenamefont {Yin},
  \citenamefont {Gong}, \citenamefont {Tu}, \citenamefont {Vescovo},
  \citenamefont {Nelson}, \citenamefont {Moore}, \citenamefont {Murakami},
  \citenamefont {Lei}, \citenamefont {Lee}, \citenamefont {Lawrie},\ and\
  \citenamefont {Miao}}]{Miao:2021}%
  \BibitemOpen
  \bibfield  {author} {\bibinfo {author} {\bibfnamefont {Haoxiang}\
  \bibnamefont {Li}}, \bibinfo {author} {\bibfnamefont {T.~T.}\ \bibnamefont
  {Zhang}}, \bibinfo {author} {\bibfnamefont {T.}~\bibnamefont {Yilmaz}},
  \bibinfo {author} {\bibfnamefont {Y.~Y.}\ \bibnamefont {Pai}}, \bibinfo
  {author} {\bibfnamefont {C.~E.}\ \bibnamefont {Marvinney}}, \bibinfo {author}
  {\bibfnamefont {A.}~\bibnamefont {Said}}, \bibinfo {author} {\bibfnamefont
  {Q.~W.}\ \bibnamefont {Yin}}, \bibinfo {author} {\bibfnamefont {C.~S.}\
  \bibnamefont {Gong}}, \bibinfo {author} {\bibfnamefont {Z.~J.}\ \bibnamefont
  {Tu}}, \bibinfo {author} {\bibfnamefont {E.}~\bibnamefont {Vescovo}},
  \bibinfo {author} {\bibfnamefont {C.~S.}\ \bibnamefont {Nelson}}, \bibinfo
  {author} {\bibfnamefont {R.~G.}\ \bibnamefont {Moore}}, \bibinfo {author}
  {\bibfnamefont {S.}~\bibnamefont {Murakami}}, \bibinfo {author}
  {\bibfnamefont {H.~C.}\ \bibnamefont {Lei}}, \bibinfo {author} {\bibfnamefont
  {H.~N.}\ \bibnamefont {Lee}}, \bibinfo {author} {\bibfnamefont {B.~J.}\
  \bibnamefont {Lawrie}}, \ and\ \bibinfo {author} {\bibfnamefont
  {H.}~\bibnamefont {Miao}},\ }\bibfield  {title} {\enquote {\bibinfo {title}
  {Observation of unconventional charge density wave without acoustic phonon
  anomaly in kagome superconductors {AV}$_3${S}b$_5$ ({A}={R}b, {C}s)},}\
  }\href {\doibase 10.1103/PhysRevX.11.031050} {\bibfield  {journal} {\bibinfo
  {journal} {Phys. Rev. X}\ }\textbf {\bibinfo {volume} {11}},\ \bibinfo
  {pages} {031050} (\bibinfo {year} {2021})}\BibitemShut {NoStop}%
\bibitem [{\citenamefont {Denner}\ \emph {et~al.}(2021)\citenamefont {Denner},
  \citenamefont {Thomale},\ and\ \citenamefont {Neupert}}]{Denner:2021}%
  \BibitemOpen
  \bibfield  {author} {\bibinfo {author} {\bibfnamefont {M.~Michael}\
  \bibnamefont {Denner}}, \bibinfo {author} {\bibfnamefont {Ronny}\
  \bibnamefont {Thomale}}, \ and\ \bibinfo {author} {\bibfnamefont {Titus}\
  \bibnamefont {Neupert}},\ }\bibfield  {title} {\enquote {\bibinfo {title}
  {Analysis of charge order in the kagome metal {AV}$_3${S}b$_5$ ({A}={K},
  {R}b, {C}s)},}\ }\href {\doibase 10.1103/PhysRevLett.127.217601} {\bibfield
  {journal} {\bibinfo  {journal} {Phys. Rev. Lett.}\ }\textbf {\bibinfo
  {volume} {127}},\ \bibinfo {pages} {217601} (\bibinfo {year}
  {2021})}\BibitemShut {NoStop}%
\bibitem [{\citenamefont {Lin}\ and\ \citenamefont
  {Nandkishore}(2021)}]{Rahul:2021}%
  \BibitemOpen
  \bibfield  {author} {\bibinfo {author} {\bibfnamefont {Yu-Ping}\ \bibnamefont
  {Lin}}\ and\ \bibinfo {author} {\bibfnamefont {Rahul~M.}\ \bibnamefont
  {Nandkishore}},\ }\bibfield  {title} {\enquote {\bibinfo {title} {Complex
  charge density waves at van hove singularity on hexagonal lattices:
  Haldane-model phase diagram and potential realization in the kagome metals
  {AV}$_{3}${S}b$_{5}$ ({A}={K}, {R}b, {C}s)},}\ }\href {\doibase
  10.1103/PhysRevB.104.045122} {\bibfield  {journal} {\bibinfo  {journal}
  {Phys. Rev. B}\ }\textbf {\bibinfo {volume} {104}},\ \bibinfo {pages}
  {045122} (\bibinfo {year} {2021})}\BibitemShut {NoStop}%
\bibitem [{\citenamefont {Park}\ \emph {et~al.}(2021)\citenamefont {Park},
  \citenamefont {Ye},\ and\ \citenamefont {Balents}}]{Balents:2021}%
  \BibitemOpen
  \bibfield  {author} {\bibinfo {author} {\bibfnamefont {Takamori}\
  \bibnamefont {Park}}, \bibinfo {author} {\bibfnamefont {Mengxing}\
  \bibnamefont {Ye}}, \ and\ \bibinfo {author} {\bibfnamefont {Leon}\
  \bibnamefont {Balents}},\ }\bibfield  {title} {\enquote {\bibinfo {title}
  {Electronic instabilities of kagome metals: Saddle points and landau
  theory},}\ }\href {\doibase 10.1103/PhysRevB.104.035142} {\bibfield
  {journal} {\bibinfo  {journal} {Phys. Rev. B}\ }\textbf {\bibinfo {volume}
  {104}},\ \bibinfo {pages} {035142} (\bibinfo {year} {2021})}\BibitemShut
  {NoStop}%
\bibitem [{\citenamefont {Christensen}\ \emph {et~al.}(2021)\citenamefont
  {Christensen}, \citenamefont {Birol}, \citenamefont {Andersen},\ and\
  \citenamefont {Fernandes}}]{Rafael:2021}%
  \BibitemOpen
  \bibfield  {author} {\bibinfo {author} {\bibfnamefont {Morten~H.}\
  \bibnamefont {Christensen}}, \bibinfo {author} {\bibfnamefont {Turan}\
  \bibnamefont {Birol}}, \bibinfo {author} {\bibfnamefont {Brian~M.}\
  \bibnamefont {Andersen}}, \ and\ \bibinfo {author} {\bibfnamefont
  {Rafael~M.}\ \bibnamefont {Fernandes}},\ }\bibfield  {title} {\enquote
  {\bibinfo {title} {Theory of the charge density wave in {AV}$_3${S}b$_5$
  kagome metals},}\ }\href {\doibase 10.1103/PhysRevB.104.214513} {\bibfield
  {journal} {\bibinfo  {journal} {Phys. Rev. B}\ }\textbf {\bibinfo {volume}
  {104}},\ \bibinfo {pages} {214513} (\bibinfo {year} {2021})}\BibitemShut
  {NoStop}%
\bibitem [{\citenamefont {Zhao}\ \emph {et~al.}(2021)\citenamefont {Zhao},
  \citenamefont {Li}, \citenamefont {Ortiz}, \citenamefont {Teicher},
  \citenamefont {Park}, \citenamefont {Ye}, \citenamefont {Wang}, \citenamefont
  {Balents}, \citenamefont {Wilson},\ and\ \citenamefont
  {Zeljkovic}}]{Zhao:2021}%
  \BibitemOpen
  \bibfield  {author} {\bibinfo {author} {\bibfnamefont {He}~\bibnamefont
  {Zhao}}, \bibinfo {author} {\bibfnamefont {Hong}\ \bibnamefont {Li}},
  \bibinfo {author} {\bibfnamefont {Brenden~R.}\ \bibnamefont {Ortiz}},
  \bibinfo {author} {\bibfnamefont {Samuel M.~L.}\ \bibnamefont {Teicher}},
  \bibinfo {author} {\bibfnamefont {Takamori}\ \bibnamefont {Park}}, \bibinfo
  {author} {\bibfnamefont {Mengxing}\ \bibnamefont {Ye}}, \bibinfo {author}
  {\bibfnamefont {Ziqiang}\ \bibnamefont {Wang}}, \bibinfo {author}
  {\bibfnamefont {Leon}\ \bibnamefont {Balents}}, \bibinfo {author}
  {\bibfnamefont {Stephen~D.}\ \bibnamefont {Wilson}}, \ and\ \bibinfo {author}
  {\bibfnamefont {Ilija}\ \bibnamefont {Zeljkovic}},\ }\bibfield  {title}
  {\enquote {\bibinfo {title} {Cascade of correlated electron states in the
  kagome superconductor {C}s{V}$_3${S}b$_5$},}\ }\href {\doibase
  10.1038/s41586-021-03946-w} {\bibfield  {journal} {\bibinfo  {journal}
  {Nature}\ }\textbf {\bibinfo {volume} {599}},\ \bibinfo {pages} {216--221}
  (\bibinfo {year} {2021})}\BibitemShut {NoStop}%
\bibitem [{\citenamefont {Jiang}\ \emph {et~al.}(2022)\citenamefont {Jiang},
  \citenamefont {Wu}, \citenamefont {Yin}, \citenamefont {Wang}, \citenamefont
  {Hasan}, \citenamefont {Wilson}, \citenamefont {Chen},\ and\ \citenamefont
  {Hu}}]{JiangKun:2022}%
  \BibitemOpen
  \bibfield  {author} {\bibinfo {author} {\bibfnamefont {Kun}\ \bibnamefont
  {Jiang}}, \bibinfo {author} {\bibfnamefont {Tao}\ \bibnamefont {Wu}},
  \bibinfo {author} {\bibfnamefont {Jia-Xin}\ \bibnamefont {Yin}}, \bibinfo
  {author} {\bibfnamefont {Zhenyu}\ \bibnamefont {Wang}}, \bibinfo {author}
  {\bibfnamefont {M~Zahid}\ \bibnamefont {Hasan}}, \bibinfo {author}
  {\bibfnamefont {Stephen~D}\ \bibnamefont {Wilson}}, \bibinfo {author}
  {\bibfnamefont {Xianhui}\ \bibnamefont {Chen}}, \ and\ \bibinfo {author}
  {\bibfnamefont {Jiangping}\ \bibnamefont {Hu}},\ }\bibfield  {title}
  {\enquote {\bibinfo {title} {{Kagome superconductors {AV}$_3${S}b$_5$
  ({A}={K}, {R}b, {C}s)}},}\ }\href {\doibase 10.1093/nsr/nwac199} {\bibfield
  {journal} {\bibinfo  {journal} {National Science Review}\ } (\bibinfo {year}
  {2022}),\ 10.1093/nsr/nwac199},\ \bibinfo {note} {nwac199}\BibitemShut
  {NoStop}%
\bibitem [{\citenamefont {Nie}\ \emph {et~al.}(2022)\citenamefont {Nie},
  \citenamefont {Sun}, \citenamefont {Ma}, \citenamefont {Song}, \citenamefont
  {Zheng}, \citenamefont {Liang}, \citenamefont {Wu}, \citenamefont {Yu},
  \citenamefont {Li}, \citenamefont {Shan}, \citenamefont {Zhao}, \citenamefont
  {Li}, \citenamefont {Kang}, \citenamefont {Wu}, \citenamefont {Zhou},
  \citenamefont {Liu}, \citenamefont {Xiang}, \citenamefont {Ying},
  \citenamefont {Wang}, \citenamefont {Wu},\ and\ \citenamefont
  {Chen}}]{Nie:2022}%
  \BibitemOpen
  \bibfield  {author} {\bibinfo {author} {\bibfnamefont {Linpeng}\ \bibnamefont
  {Nie}}, \bibinfo {author} {\bibfnamefont {Kuanglv}\ \bibnamefont {Sun}},
  \bibinfo {author} {\bibfnamefont {Wanru}\ \bibnamefont {Ma}}, \bibinfo
  {author} {\bibfnamefont {Dianwu}\ \bibnamefont {Song}}, \bibinfo {author}
  {\bibfnamefont {Lixuan}\ \bibnamefont {Zheng}}, \bibinfo {author}
  {\bibfnamefont {Zuowei}\ \bibnamefont {Liang}}, \bibinfo {author}
  {\bibfnamefont {Ping}\ \bibnamefont {Wu}}, \bibinfo {author} {\bibfnamefont
  {Fanghang}\ \bibnamefont {Yu}}, \bibinfo {author} {\bibfnamefont {Jian}\
  \bibnamefont {Li}}, \bibinfo {author} {\bibfnamefont {Min}\ \bibnamefont
  {Shan}}, \bibinfo {author} {\bibfnamefont {Dan}\ \bibnamefont {Zhao}},
  \bibinfo {author} {\bibfnamefont {Shunjiao}\ \bibnamefont {Li}}, \bibinfo
  {author} {\bibfnamefont {Baolei}\ \bibnamefont {Kang}}, \bibinfo {author}
  {\bibfnamefont {Zhimian}\ \bibnamefont {Wu}}, \bibinfo {author}
  {\bibfnamefont {Yanbing}\ \bibnamefont {Zhou}}, \bibinfo {author}
  {\bibfnamefont {Kai}\ \bibnamefont {Liu}}, \bibinfo {author} {\bibfnamefont
  {Ziji}\ \bibnamefont {Xiang}}, \bibinfo {author} {\bibfnamefont {Jianjun}\
  \bibnamefont {Ying}}, \bibinfo {author} {\bibfnamefont {Zhenyu}\ \bibnamefont
  {Wang}}, \bibinfo {author} {\bibfnamefont {Tao}\ \bibnamefont {Wu}}, \ and\
  \bibinfo {author} {\bibfnamefont {Xianhui}\ \bibnamefont {Chen}},\ }\bibfield
   {title} {\enquote {\bibinfo {title} {Charge-density-wave-driven electronic
  nematicity in a kagome superconductor},}\ }\href {\doibase
  10.1038/s41586-022-04493-8} {\bibfield  {journal} {\bibinfo  {journal}
  {Nature}\ }\textbf {\bibinfo {volume} {604}},\ \bibinfo {pages} {59--64}
  (\bibinfo {year} {2022})}\BibitemShut {NoStop}%
\bibitem [{\citenamefont {Syôzi}(1951)}]{syozi:1951}%
  \BibitemOpen
  \bibfield  {author} {\bibinfo {author} {\bibfnamefont {Itiro}\ \bibnamefont
  {Syôzi}},\ }\bibfield  {title} {\enquote {\bibinfo {title} {{Statistics of
  Kagomé Lattice}},}\ }\href {\doibase 10.1143/ptp/6.3.306} {\bibfield
  {journal} {\bibinfo  {journal} {Progress of Theoretical Physics}\ }\textbf
  {\bibinfo {volume} {6}},\ \bibinfo {pages} {306--308} (\bibinfo {year}
  {1951})}\BibitemShut {NoStop}%
\bibitem [{\citenamefont {Sachdev}(1992)}]{Sachdev:1992}%
  \BibitemOpen
  \bibfield  {author} {\bibinfo {author} {\bibfnamefont {Subir}\ \bibnamefont
  {Sachdev}},\ }\bibfield  {title} {\enquote {\bibinfo {title} {Kagome- and
  triangular-lattice {H}eisenberg antiferromagnets: Ordering from quantum
  fluctuations and quantum-disordered ground states with unconfined bosonic
  spinons},}\ }\href {\doibase 10.1103/PhysRevB.45.12377} {\bibfield  {journal}
  {\bibinfo  {journal} {Phys. Rev. B}\ }\textbf {\bibinfo {volume} {45}},\
  \bibinfo {pages} {12377--12396} (\bibinfo {year} {1992})}\BibitemShut
  {NoStop}%
\bibitem [{\citenamefont {Norman}(2016)}]{Norman:2016}%
  \BibitemOpen
  \bibfield  {author} {\bibinfo {author} {\bibfnamefont {M.~R.}\ \bibnamefont
  {Norman}},\ }\bibfield  {title} {\enquote {\bibinfo {title} {Colloquium:
  Herbertsmithite and the search for the quantum spin liquid},}\ }\href
  {\doibase 10.1103/RevModPhys.88.041002} {\bibfield  {journal} {\bibinfo
  {journal} {Rev. Mod. Phys.}\ }\textbf {\bibinfo {volume} {88}},\ \bibinfo
  {pages} {041002} (\bibinfo {year} {2016})}\BibitemShut {NoStop}%
\bibitem [{\citenamefont {Tang}\ \emph {et~al.}(2011)\citenamefont {Tang},
  \citenamefont {Mei},\ and\ \citenamefont {Wen}}]{wenxg:2011}%
  \BibitemOpen
  \bibfield  {author} {\bibinfo {author} {\bibfnamefont {Evelyn}\ \bibnamefont
  {Tang}}, \bibinfo {author} {\bibfnamefont {Jia-Wei}\ \bibnamefont {Mei}}, \
  and\ \bibinfo {author} {\bibfnamefont {Xiao-Gang}\ \bibnamefont {Wen}},\
  }\bibfield  {title} {\enquote {\bibinfo {title} {High-temperature fractional
  quantum {H}all states},}\ }\href {\doibase 10.1103/PhysRevLett.106.236802}
  {\bibfield  {journal} {\bibinfo  {journal} {Phys. Rev. Lett.}\ }\textbf
  {\bibinfo {volume} {106}},\ \bibinfo {pages} {236802} (\bibinfo {year}
  {2011})}\BibitemShut {NoStop}%
\bibitem [{\citenamefont {Yin}\ \emph {et~al.}(2018)\citenamefont {Yin},
  \citenamefont {Zhang}, \citenamefont {Li}, \citenamefont {Jiang},
  \citenamefont {Chang}, \citenamefont {Zhang}, \citenamefont {Lian},
  \citenamefont {Xiang}, \citenamefont {Belopolski}, \citenamefont {Zheng},
  \citenamefont {Cochran}, \citenamefont {Xu}, \citenamefont {Bian},
  \citenamefont {Liu}, \citenamefont {Chang}, \citenamefont {Lin},
  \citenamefont {Lu}, \citenamefont {Wang}, \citenamefont {Jia}, \citenamefont
  {Wang},\ and\ \citenamefont {Hasan}}]{Yin:2018}%
  \BibitemOpen
  \bibfield  {author} {\bibinfo {author} {\bibfnamefont {Jia-Xin}\ \bibnamefont
  {Yin}}, \bibinfo {author} {\bibfnamefont {Songtian~S.}\ \bibnamefont
  {Zhang}}, \bibinfo {author} {\bibfnamefont {Hang}\ \bibnamefont {Li}},
  \bibinfo {author} {\bibfnamefont {Kun}\ \bibnamefont {Jiang}}, \bibinfo
  {author} {\bibfnamefont {Guoqing}\ \bibnamefont {Chang}}, \bibinfo {author}
  {\bibfnamefont {Bingjing}\ \bibnamefont {Zhang}}, \bibinfo {author}
  {\bibfnamefont {Biao}\ \bibnamefont {Lian}}, \bibinfo {author} {\bibfnamefont
  {Cheng}\ \bibnamefont {Xiang}}, \bibinfo {author} {\bibfnamefont {Ilya}\
  \bibnamefont {Belopolski}}, \bibinfo {author} {\bibfnamefont {Hao}\
  \bibnamefont {Zheng}}, \bibinfo {author} {\bibfnamefont {Tyler~A.}\
  \bibnamefont {Cochran}}, \bibinfo {author} {\bibfnamefont {Su-Yang}\
  \bibnamefont {Xu}}, \bibinfo {author} {\bibfnamefont {Guang}\ \bibnamefont
  {Bian}}, \bibinfo {author} {\bibfnamefont {Kai}\ \bibnamefont {Liu}},
  \bibinfo {author} {\bibfnamefont {Tay-Rong}\ \bibnamefont {Chang}}, \bibinfo
  {author} {\bibfnamefont {Hsin}\ \bibnamefont {Lin}}, \bibinfo {author}
  {\bibfnamefont {Zhong-Yi}\ \bibnamefont {Lu}}, \bibinfo {author}
  {\bibfnamefont {Ziqiang}\ \bibnamefont {Wang}}, \bibinfo {author}
  {\bibfnamefont {Shuang}\ \bibnamefont {Jia}}, \bibinfo {author}
  {\bibfnamefont {Wenhong}\ \bibnamefont {Wang}}, \ and\ \bibinfo {author}
  {\bibfnamefont {M.~Zahid}\ \bibnamefont {Hasan}},\ }\bibfield  {title}
  {\enquote {\bibinfo {title} {Giant and anisotropic many-body spin--orbit
  tunability in a strongly correlated kagome magnet},}\ }\href {\doibase
  10.1038/s41586-018-0502-7} {\bibfield  {journal} {\bibinfo  {journal}
  {Nature}\ }\textbf {\bibinfo {volume} {562}},\ \bibinfo {pages} {91--95}
  (\bibinfo {year} {2018})}\BibitemShut {NoStop}%
\bibitem [{\citenamefont {Yin}\ \emph {et~al.}(2020{\natexlab{a}})\citenamefont
  {Yin}, \citenamefont {Shumiya}, \citenamefont {Mardanya}, \citenamefont
  {Wang}, \citenamefont {Zhang}, \citenamefont {Tien}, \citenamefont {Multer},
  \citenamefont {Jiang}, \citenamefont {Cheng}, \citenamefont {Yao},
  \citenamefont {Wu}, \citenamefont {Wu}, \citenamefont {Deng}, \citenamefont
  {Ye}, \citenamefont {He}, \citenamefont {Chang}, \citenamefont {Liu},
  \citenamefont {Jiang}, \citenamefont {Wang}, \citenamefont {Neupert},
  \citenamefont {Agarwal}, \citenamefont {Chang}, \citenamefont {Chu},
  \citenamefont {Lei},\ and\ \citenamefont {Hasan}}]{Yin:2020}%
  \BibitemOpen
  \bibfield  {author} {\bibinfo {author} {\bibfnamefont {J.-X.}\ \bibnamefont
  {Yin}}, \bibinfo {author} {\bibfnamefont {Nana}\ \bibnamefont {Shumiya}},
  \bibinfo {author} {\bibfnamefont {Sougata}\ \bibnamefont {Mardanya}},
  \bibinfo {author} {\bibfnamefont {Qi}~\bibnamefont {Wang}}, \bibinfo {author}
  {\bibfnamefont {Songtian~S.}\ \bibnamefont {Zhang}}, \bibinfo {author}
  {\bibfnamefont {Hung-Ju}\ \bibnamefont {Tien}}, \bibinfo {author}
  {\bibfnamefont {Daniel}\ \bibnamefont {Multer}}, \bibinfo {author}
  {\bibfnamefont {Yuxiao}\ \bibnamefont {Jiang}}, \bibinfo {author}
  {\bibfnamefont {Guangming}\ \bibnamefont {Cheng}}, \bibinfo {author}
  {\bibfnamefont {Nan}\ \bibnamefont {Yao}}, \bibinfo {author} {\bibfnamefont
  {Shangfei}\ \bibnamefont {Wu}}, \bibinfo {author} {\bibfnamefont {Desheng}\
  \bibnamefont {Wu}}, \bibinfo {author} {\bibfnamefont {Liangzi}\ \bibnamefont
  {Deng}}, \bibinfo {author} {\bibfnamefont {Zhipeng}\ \bibnamefont {Ye}},
  \bibinfo {author} {\bibfnamefont {Rui}\ \bibnamefont {He}}, \bibinfo {author}
  {\bibfnamefont {Guoqing}\ \bibnamefont {Chang}}, \bibinfo {author}
  {\bibfnamefont {Zhonghao}\ \bibnamefont {Liu}}, \bibinfo {author}
  {\bibfnamefont {Kun}\ \bibnamefont {Jiang}}, \bibinfo {author} {\bibfnamefont
  {Ziqiang}\ \bibnamefont {Wang}}, \bibinfo {author} {\bibfnamefont {Titus}\
  \bibnamefont {Neupert}}, \bibinfo {author} {\bibfnamefont {Amit}\
  \bibnamefont {Agarwal}}, \bibinfo {author} {\bibfnamefont {Tay-Rong}\
  \bibnamefont {Chang}}, \bibinfo {author} {\bibfnamefont {Ching-Wu}\
  \bibnamefont {Chu}}, \bibinfo {author} {\bibfnamefont {Hechang}\ \bibnamefont
  {Lei}}, \ and\ \bibinfo {author} {\bibfnamefont {M.~Zahid}\ \bibnamefont
  {Hasan}},\ }\bibfield  {title} {\enquote {\bibinfo {title} {Fermion--boson
  many-body interplay in a frustrated kagome paramagnet},}\ }\href {\doibase
  10.1038/s41467-020-17464-2} {\bibfield  {journal} {\bibinfo  {journal}
  {Nature Communications}\ }\textbf {\bibinfo {volume} {11}},\ \bibinfo {pages}
  {4003} (\bibinfo {year} {2020}{\natexlab{a}})}\BibitemShut {NoStop}%
\bibitem [{\citenamefont {Liu}\ \emph {et~al.}(2020)\citenamefont {Liu},
  \citenamefont {Li}, \citenamefont {Wang}, \citenamefont {Wang}, \citenamefont
  {Wen}, \citenamefont {Jiang}, \citenamefont {Lu}, \citenamefont {Yan},
  \citenamefont {Huang}, \citenamefont {Shen}, \citenamefont {Yin},
  \citenamefont {Wang}, \citenamefont {Yin}, \citenamefont {Lei},\ and\
  \citenamefont {Wang}}]{Liu:2020}%
  \BibitemOpen
  \bibfield  {author} {\bibinfo {author} {\bibfnamefont {Zhonghao}\
  \bibnamefont {Liu}}, \bibinfo {author} {\bibfnamefont {Man}\ \bibnamefont
  {Li}}, \bibinfo {author} {\bibfnamefont {Qi}~\bibnamefont {Wang}}, \bibinfo
  {author} {\bibfnamefont {Guangwei}\ \bibnamefont {Wang}}, \bibinfo {author}
  {\bibfnamefont {Chenhaoping}\ \bibnamefont {Wen}}, \bibinfo {author}
  {\bibfnamefont {Kun}\ \bibnamefont {Jiang}}, \bibinfo {author} {\bibfnamefont
  {Xiangle}\ \bibnamefont {Lu}}, \bibinfo {author} {\bibfnamefont {Shichao}\
  \bibnamefont {Yan}}, \bibinfo {author} {\bibfnamefont {Yaobo}\ \bibnamefont
  {Huang}}, \bibinfo {author} {\bibfnamefont {Dawei}\ \bibnamefont {Shen}},
  \bibinfo {author} {\bibfnamefont {Jia-Xin}\ \bibnamefont {Yin}}, \bibinfo
  {author} {\bibfnamefont {Ziqiang}\ \bibnamefont {Wang}}, \bibinfo {author}
  {\bibfnamefont {Zhiping}\ \bibnamefont {Yin}}, \bibinfo {author}
  {\bibfnamefont {Hechang}\ \bibnamefont {Lei}}, \ and\ \bibinfo {author}
  {\bibfnamefont {Shancai}\ \bibnamefont {Wang}},\ }\bibfield  {title}
  {\enquote {\bibinfo {title} {Orbital-selective dirac fermions and extremely
  flat bands in frustrated kagome-lattice metal {C}o{S}n},}\ }\href {\doibase
  10.1038/s41467-020-17462-4} {\bibfield  {journal} {\bibinfo  {journal}
  {Nature Communications}\ }\textbf {\bibinfo {volume} {11}},\ \bibinfo {pages}
  {4002} (\bibinfo {year} {2020})}\BibitemShut {NoStop}%
\bibitem [{\citenamefont {Kang}\ \emph {et~al.}(2020)\citenamefont {Kang},
  \citenamefont {Ye}, \citenamefont {Fang}, \citenamefont {You}, \citenamefont
  {Levitan}, \citenamefont {Han}, \citenamefont {Facio}, \citenamefont
  {Jozwiak}, \citenamefont {Bostwick}, \citenamefont {Rotenberg}, \citenamefont
  {Chan}, \citenamefont {McDonald}, \citenamefont {Graf}, \citenamefont
  {Kaznatcheev}, \citenamefont {Vescovo}, \citenamefont {Bell}, \citenamefont
  {Kaxiras}, \citenamefont {van~den Brink}, \citenamefont {Richter},
  \citenamefont {Prasad~Ghimire}, \citenamefont {Checkelsky},\ and\
  \citenamefont {Comin}}]{Kang:2020}%
  \BibitemOpen
  \bibfield  {author} {\bibinfo {author} {\bibfnamefont {Mingu}\ \bibnamefont
  {Kang}}, \bibinfo {author} {\bibfnamefont {Linda}\ \bibnamefont {Ye}},
  \bibinfo {author} {\bibfnamefont {Shiang}\ \bibnamefont {Fang}}, \bibinfo
  {author} {\bibfnamefont {Jhih-Shih}\ \bibnamefont {You}}, \bibinfo {author}
  {\bibfnamefont {Abe}\ \bibnamefont {Levitan}}, \bibinfo {author}
  {\bibfnamefont {Minyong}\ \bibnamefont {Han}}, \bibinfo {author}
  {\bibfnamefont {Jorge~I.}\ \bibnamefont {Facio}}, \bibinfo {author}
  {\bibfnamefont {Chris}\ \bibnamefont {Jozwiak}}, \bibinfo {author}
  {\bibfnamefont {Aaron}\ \bibnamefont {Bostwick}}, \bibinfo {author}
  {\bibfnamefont {Eli}\ \bibnamefont {Rotenberg}}, \bibinfo {author}
  {\bibfnamefont {Mun~K.}\ \bibnamefont {Chan}}, \bibinfo {author}
  {\bibfnamefont {Ross~D.}\ \bibnamefont {McDonald}}, \bibinfo {author}
  {\bibfnamefont {David}\ \bibnamefont {Graf}}, \bibinfo {author}
  {\bibfnamefont {Konstantine}\ \bibnamefont {Kaznatcheev}}, \bibinfo {author}
  {\bibfnamefont {Elio}\ \bibnamefont {Vescovo}}, \bibinfo {author}
  {\bibfnamefont {David~C.}\ \bibnamefont {Bell}}, \bibinfo {author}
  {\bibfnamefont {Efthimios}\ \bibnamefont {Kaxiras}}, \bibinfo {author}
  {\bibfnamefont {Jeroen}\ \bibnamefont {van~den Brink}}, \bibinfo {author}
  {\bibfnamefont {Manuel}\ \bibnamefont {Richter}}, \bibinfo {author}
  {\bibfnamefont {Madhav}\ \bibnamefont {Prasad~Ghimire}}, \bibinfo {author}
  {\bibfnamefont {Joseph~G.}\ \bibnamefont {Checkelsky}}, \ and\ \bibinfo
  {author} {\bibfnamefont {Riccardo}\ \bibnamefont {Comin}},\ }\bibfield
  {title} {\enquote {\bibinfo {title} {Dirac fermions and flat bands in the
  ideal kagome metal {F}e{S}n},}\ }\href {\doibase 10.1038/s41563-019-0531-0}
  {\bibfield  {journal} {\bibinfo  {journal} {Nature Materials}\ }\textbf
  {\bibinfo {volume} {19}},\ \bibinfo {pages} {163--169} (\bibinfo {year}
  {2020})}\BibitemShut {NoStop}%
\bibitem [{\citenamefont {Wang}(2023)}]{Yilin:2023}%
  \BibitemOpen
  \bibfield  {author} {\bibinfo {author} {\bibfnamefont {Yilin}\ \bibnamefont
  {Wang}},\ }\href@noop {} {\enquote {\bibinfo {title} {Electronic correlation
  effects on stabilizing a perfect kagome lattice and ferromagnetic fluctuation
  in {L}a{R}u$_3${S}i$_2$},}\ } (\bibinfo {year} {2023}),\ \Eprint
  {http://arxiv.org/abs/arXiv:2303.12273} {arXiv:2303.12273} \BibitemShut
  {NoStop}%
\bibitem [{\citenamefont {Tasaki}(1992)}]{Tasaki:1992}%
  \BibitemOpen
  \bibfield  {author} {\bibinfo {author} {\bibfnamefont {Hal}\ \bibnamefont
  {Tasaki}},\ }\bibfield  {title} {\enquote {\bibinfo {title} {Ferromagnetism
  in the {H}ubbard models with degenerate single-electron ground states},}\
  }\href {\doibase 10.1103/PhysRevLett.69.1608} {\bibfield  {journal} {\bibinfo
   {journal} {Phys. Rev. Lett.}\ }\textbf {\bibinfo {volume} {69}},\ \bibinfo
  {pages} {1608--1611} (\bibinfo {year} {1992})}\BibitemShut {NoStop}%
\bibitem [{\citenamefont {Lin}\ \emph {et~al.}(2018)\citenamefont {Lin},
  \citenamefont {Choi}, \citenamefont {Zhang}, \citenamefont {Qin},
  \citenamefont {Yi}, \citenamefont {Wang}, \citenamefont {Li}, \citenamefont
  {Wang}, \citenamefont {Zhang}, \citenamefont {Sun}, \citenamefont {Wei},
  \citenamefont {Zhang}, \citenamefont {Guo}, \citenamefont {Lu}, \citenamefont
  {Cho}, \citenamefont {Zeng},\ and\ \citenamefont {Zhang}}]{Zhenyu:2018}%
  \BibitemOpen
  \bibfield  {author} {\bibinfo {author} {\bibfnamefont {Zhiyong}\ \bibnamefont
  {Lin}}, \bibinfo {author} {\bibfnamefont {Jin-Ho}\ \bibnamefont {Choi}},
  \bibinfo {author} {\bibfnamefont {Qiang}\ \bibnamefont {Zhang}}, \bibinfo
  {author} {\bibfnamefont {Wei}\ \bibnamefont {Qin}}, \bibinfo {author}
  {\bibfnamefont {Seho}\ \bibnamefont {Yi}}, \bibinfo {author} {\bibfnamefont
  {Pengdong}\ \bibnamefont {Wang}}, \bibinfo {author} {\bibfnamefont {Lin}\
  \bibnamefont {Li}}, \bibinfo {author} {\bibfnamefont {Yifan}\ \bibnamefont
  {Wang}}, \bibinfo {author} {\bibfnamefont {Hui}\ \bibnamefont {Zhang}},
  \bibinfo {author} {\bibfnamefont {Zhe}\ \bibnamefont {Sun}}, \bibinfo
  {author} {\bibfnamefont {Laiming}\ \bibnamefont {Wei}}, \bibinfo {author}
  {\bibfnamefont {Shengbai}\ \bibnamefont {Zhang}}, \bibinfo {author}
  {\bibfnamefont {Tengfei}\ \bibnamefont {Guo}}, \bibinfo {author}
  {\bibfnamefont {Qingyou}\ \bibnamefont {Lu}}, \bibinfo {author}
  {\bibfnamefont {Jun-Hyung}\ \bibnamefont {Cho}}, \bibinfo {author}
  {\bibfnamefont {Changgan}\ \bibnamefont {Zeng}}, \ and\ \bibinfo {author}
  {\bibfnamefont {Zhenyu}\ \bibnamefont {Zhang}},\ }\bibfield  {title}
  {\enquote {\bibinfo {title} {Flatbands and emergent ferromagnetic ordering in
  {F}e$_{3}${S}n$_{2}$ kagome lattices},}\ }\href {\doibase
  10.1103/PhysRevLett.121.096401} {\bibfield  {journal} {\bibinfo  {journal}
  {Phys. Rev. Lett.}\ }\textbf {\bibinfo {volume} {121}},\ \bibinfo {pages}
  {096401} (\bibinfo {year} {2018})}\BibitemShut {NoStop}%
\bibitem [{\citenamefont {Yin}\ \emph {et~al.}(2019)\citenamefont {Yin},
  \citenamefont {Zhang}, \citenamefont {Chang}, \citenamefont {Wang},
  \citenamefont {Tsirkin}, \citenamefont {Guguchia}, \citenamefont {Lian},
  \citenamefont {Zhou}, \citenamefont {Jiang}, \citenamefont {Belopolski},
  \citenamefont {Shumiya}, \citenamefont {Multer}, \citenamefont {Litskevich},
  \citenamefont {Cochran}, \citenamefont {Lin}, \citenamefont {Wang},
  \citenamefont {Neupert}, \citenamefont {Jia}, \citenamefont {Lei},\ and\
  \citenamefont {Hasan}}]{Yin:2019}%
  \BibitemOpen
  \bibfield  {author} {\bibinfo {author} {\bibfnamefont {Jia-Xin}\ \bibnamefont
  {Yin}}, \bibinfo {author} {\bibfnamefont {Songtian~S.}\ \bibnamefont
  {Zhang}}, \bibinfo {author} {\bibfnamefont {Guoqing}\ \bibnamefont {Chang}},
  \bibinfo {author} {\bibfnamefont {Qi}~\bibnamefont {Wang}}, \bibinfo {author}
  {\bibfnamefont {Stepan~S.}\ \bibnamefont {Tsirkin}}, \bibinfo {author}
  {\bibfnamefont {Zurab}\ \bibnamefont {Guguchia}}, \bibinfo {author}
  {\bibfnamefont {Biao}\ \bibnamefont {Lian}}, \bibinfo {author} {\bibfnamefont
  {Huibin}\ \bibnamefont {Zhou}}, \bibinfo {author} {\bibfnamefont {Kun}\
  \bibnamefont {Jiang}}, \bibinfo {author} {\bibfnamefont {Ilya}\ \bibnamefont
  {Belopolski}}, \bibinfo {author} {\bibfnamefont {Nana}\ \bibnamefont
  {Shumiya}}, \bibinfo {author} {\bibfnamefont {Daniel}\ \bibnamefont
  {Multer}}, \bibinfo {author} {\bibfnamefont {Maksim}\ \bibnamefont
  {Litskevich}}, \bibinfo {author} {\bibfnamefont {Tyler~A.}\ \bibnamefont
  {Cochran}}, \bibinfo {author} {\bibfnamefont {Hsin}\ \bibnamefont {Lin}},
  \bibinfo {author} {\bibfnamefont {Ziqiang}\ \bibnamefont {Wang}}, \bibinfo
  {author} {\bibfnamefont {Titus}\ \bibnamefont {Neupert}}, \bibinfo {author}
  {\bibfnamefont {Shuang}\ \bibnamefont {Jia}}, \bibinfo {author}
  {\bibfnamefont {Hechang}\ \bibnamefont {Lei}}, \ and\ \bibinfo {author}
  {\bibfnamefont {M.~Zahid}\ \bibnamefont {Hasan}},\ }\bibfield  {title}
  {\enquote {\bibinfo {title} {Negative flat band magnetism in a
  spin--orbit-coupled correlated kagome magnet},}\ }\href {\doibase
  10.1038/s41567-019-0426-7} {\bibfield  {journal} {\bibinfo  {journal} {Nature
  Physics}\ }\textbf {\bibinfo {volume} {15}},\ \bibinfo {pages} {443--448}
  (\bibinfo {year} {2019})}\BibitemShut {NoStop}%
\bibitem [{\citenamefont {Yin}\ \emph {et~al.}(2020{\natexlab{b}})\citenamefont
  {Yin}, \citenamefont {Ma}, \citenamefont {Cochran}, \citenamefont {Xu},
  \citenamefont {Zhang}, \citenamefont {Tien}, \citenamefont {Shumiya},
  \citenamefont {Cheng}, \citenamefont {Jiang}, \citenamefont {Lian},
  \citenamefont {Song}, \citenamefont {Chang}, \citenamefont {Belopolski},
  \citenamefont {Multer}, \citenamefont {Litskevich}, \citenamefont {Cheng},
  \citenamefont {Yang}, \citenamefont {Swidler}, \citenamefont {Zhou},
  \citenamefont {Lin}, \citenamefont {Neupert}, \citenamefont {Wang},
  \citenamefont {Yao}, \citenamefont {Chang}, \citenamefont {Jia},\ and\
  \citenamefont {Zahid~Hasan}}]{Yin:2020b}%
  \BibitemOpen
  \bibfield  {author} {\bibinfo {author} {\bibfnamefont {Jia-Xin}\ \bibnamefont
  {Yin}}, \bibinfo {author} {\bibfnamefont {Wenlong}\ \bibnamefont {Ma}},
  \bibinfo {author} {\bibfnamefont {Tyler~A.}\ \bibnamefont {Cochran}},
  \bibinfo {author} {\bibfnamefont {Xitong}\ \bibnamefont {Xu}}, \bibinfo
  {author} {\bibfnamefont {Songtian~S.}\ \bibnamefont {Zhang}}, \bibinfo
  {author} {\bibfnamefont {Hung-Ju}\ \bibnamefont {Tien}}, \bibinfo {author}
  {\bibfnamefont {Nana}\ \bibnamefont {Shumiya}}, \bibinfo {author}
  {\bibfnamefont {Guangming}\ \bibnamefont {Cheng}}, \bibinfo {author}
  {\bibfnamefont {Kun}\ \bibnamefont {Jiang}}, \bibinfo {author} {\bibfnamefont
  {Biao}\ \bibnamefont {Lian}}, \bibinfo {author} {\bibfnamefont {Zhida}\
  \bibnamefont {Song}}, \bibinfo {author} {\bibfnamefont {Guoqing}\
  \bibnamefont {Chang}}, \bibinfo {author} {\bibfnamefont {Ilya}\ \bibnamefont
  {Belopolski}}, \bibinfo {author} {\bibfnamefont {Daniel}\ \bibnamefont
  {Multer}}, \bibinfo {author} {\bibfnamefont {Maksim}\ \bibnamefont
  {Litskevich}}, \bibinfo {author} {\bibfnamefont {Zi-Jia}\ \bibnamefont
  {Cheng}}, \bibinfo {author} {\bibfnamefont {Xian~P.}\ \bibnamefont {Yang}},
  \bibinfo {author} {\bibfnamefont {Bianca}\ \bibnamefont {Swidler}}, \bibinfo
  {author} {\bibfnamefont {Huibin}\ \bibnamefont {Zhou}}, \bibinfo {author}
  {\bibfnamefont {Hsin}\ \bibnamefont {Lin}}, \bibinfo {author} {\bibfnamefont
  {Titus}\ \bibnamefont {Neupert}}, \bibinfo {author} {\bibfnamefont {Ziqiang}\
  \bibnamefont {Wang}}, \bibinfo {author} {\bibfnamefont {Nan}\ \bibnamefont
  {Yao}}, \bibinfo {author} {\bibfnamefont {Tay-Rong}\ \bibnamefont {Chang}},
  \bibinfo {author} {\bibfnamefont {Shuang}\ \bibnamefont {Jia}}, \ and\
  \bibinfo {author} {\bibfnamefont {M.}~\bibnamefont {Zahid~Hasan}},\
  }\bibfield  {title} {\enquote {\bibinfo {title} {Quantum-limit chern
  topological magnetism in {T}b{M}n$_6${S}n$_6$},}\ }\href {\doibase
  10.1038/s41586-020-2482-7} {\bibfield  {journal} {\bibinfo  {journal}
  {Nature}\ }\textbf {\bibinfo {volume} {583}},\ \bibinfo {pages} {533--536}
  (\bibinfo {year} {2020}{\natexlab{b}})}\BibitemShut {NoStop}%
\bibitem [{\citenamefont {Zhang}\ \emph {et~al.}(2021)\citenamefont {Zhang},
  \citenamefont {Yilmaz}, \citenamefont {Meier}, \citenamefont {Pai},
  \citenamefont {Lapano}, \citenamefont {Li}, \citenamefont {Kaznatcheev},
  \citenamefont {Vescovo}, \citenamefont {Huon}, \citenamefont {Brahlek},
  \citenamefont {Ward}, \citenamefont {Lawrie}, \citenamefont {Moore},
  \citenamefont {Lee}, \citenamefont {Wang}, \citenamefont {Miao},\ and\
  \citenamefont {Sales}}]{Miao:2021b}%
  \BibitemOpen
  \bibfield  {author} {\bibinfo {author} {\bibfnamefont {Jie}\ \bibnamefont
  {Zhang}}, \bibinfo {author} {\bibfnamefont {T}~\bibnamefont {Yilmaz}},
  \bibinfo {author} {\bibfnamefont {J~W~R}\ \bibnamefont {Meier}}, \bibinfo
  {author} {\bibfnamefont {J~Y}\ \bibnamefont {Pai}}, \bibinfo {author}
  {\bibfnamefont {J}~\bibnamefont {Lapano}}, \bibinfo {author} {\bibfnamefont
  {H~X}\ \bibnamefont {Li}}, \bibinfo {author} {\bibfnamefont {K}~\bibnamefont
  {Kaznatcheev}}, \bibinfo {author} {\bibfnamefont {E}~\bibnamefont {Vescovo}},
  \bibinfo {author} {\bibfnamefont {A}~\bibnamefont {Huon}}, \bibinfo {author}
  {\bibfnamefont {M}~\bibnamefont {Brahlek}}, \bibinfo {author} {\bibfnamefont
  {T~Z}\ \bibnamefont {Ward}}, \bibinfo {author} {\bibfnamefont
  {B}~\bibnamefont {Lawrie}}, \bibinfo {author} {\bibfnamefont {R~G}\
  \bibnamefont {Moore}}, \bibinfo {author} {\bibfnamefont {H~N}\ \bibnamefont
  {Lee}}, \bibinfo {author} {\bibfnamefont {Y~L}\ \bibnamefont {Wang}},
  \bibinfo {author} {\bibfnamefont {H}~\bibnamefont {Miao}}, \ and\ \bibinfo
  {author} {\bibfnamefont {B}~\bibnamefont {Sales}},\ }\bibfield  {title}
  {\enquote {\bibinfo {title} {Flat band induced negative magnetoresistance in
  {Multi-Orbital} kagome metal},}\ }\href@noop {} {\  (\bibinfo {year}
  {2021})},\ \Eprint {http://arxiv.org/abs/2105.08888} {arXiv:2105.08888
  [cond-mat.str-el]} \BibitemShut {NoStop}%
\bibitem [{\citenamefont {Huang}\ \emph {et~al.}(2022)\citenamefont {Huang},
  \citenamefont {Zheng}, \citenamefont {Lin}, \citenamefont {Guo},
  \citenamefont {Wang}, \citenamefont {Zhang}, \citenamefont {Zhang},
  \citenamefont {Sun}, \citenamefont {Wang}, \citenamefont {Weng},
  \citenamefont {Li}, \citenamefont {Wu}, \citenamefont {Chen},\ and\
  \citenamefont {Zeng}}]{Changgan:2022}%
  \BibitemOpen
  \bibfield  {author} {\bibinfo {author} {\bibfnamefont {Hao}\ \bibnamefont
  {Huang}}, \bibinfo {author} {\bibfnamefont {Lixuan}\ \bibnamefont {Zheng}},
  \bibinfo {author} {\bibfnamefont {Zhiyong}\ \bibnamefont {Lin}}, \bibinfo
  {author} {\bibfnamefont {Xu}~\bibnamefont {Guo}}, \bibinfo {author}
  {\bibfnamefont {Sheng}\ \bibnamefont {Wang}}, \bibinfo {author}
  {\bibfnamefont {Shuai}\ \bibnamefont {Zhang}}, \bibinfo {author}
  {\bibfnamefont {Chi}\ \bibnamefont {Zhang}}, \bibinfo {author} {\bibfnamefont
  {Zhe}\ \bibnamefont {Sun}}, \bibinfo {author} {\bibfnamefont {Zhengfei}\
  \bibnamefont {Wang}}, \bibinfo {author} {\bibfnamefont {Hongming}\
  \bibnamefont {Weng}}, \bibinfo {author} {\bibfnamefont {Lin}\ \bibnamefont
  {Li}}, \bibinfo {author} {\bibfnamefont {Tao}\ \bibnamefont {Wu}}, \bibinfo
  {author} {\bibfnamefont {Xianhui}\ \bibnamefont {Chen}}, \ and\ \bibinfo
  {author} {\bibfnamefont {Changgan}\ \bibnamefont {Zeng}},\ }\bibfield
  {title} {\enquote {\bibinfo {title} {Flat-band-induced anomalous anisotropic
  charge transport and orbital magnetism in kagome metal {C}o{S}n},}\ }\href
  {\doibase 10.1103/PhysRevLett.128.096601} {\bibfield  {journal} {\bibinfo
  {journal} {Phys. Rev. Lett.}\ }\textbf {\bibinfo {volume} {128}},\ \bibinfo
  {pages} {096601} (\bibinfo {year} {2022})}\BibitemShut {NoStop}%
\bibitem [{\citenamefont {Ye}\ \emph {et~al.}(2018)\citenamefont {Ye},
  \citenamefont {Kang}, \citenamefont {Liu}, \citenamefont {von Cube},
  \citenamefont {Wicker}, \citenamefont {Suzuki}, \citenamefont {Jozwiak},
  \citenamefont {Bostwick}, \citenamefont {Rotenberg}, \citenamefont {Bell},
  \citenamefont {Fu}, \citenamefont {Comin},\ and\ \citenamefont
  {Checkelsky}}]{Ye:2018}%
  \BibitemOpen
  \bibfield  {author} {\bibinfo {author} {\bibfnamefont {Linda}\ \bibnamefont
  {Ye}}, \bibinfo {author} {\bibfnamefont {Mingu}\ \bibnamefont {Kang}},
  \bibinfo {author} {\bibfnamefont {Junwei}\ \bibnamefont {Liu}}, \bibinfo
  {author} {\bibfnamefont {Felix}\ \bibnamefont {von Cube}}, \bibinfo {author}
  {\bibfnamefont {Christina~R.}\ \bibnamefont {Wicker}}, \bibinfo {author}
  {\bibfnamefont {Takehito}\ \bibnamefont {Suzuki}}, \bibinfo {author}
  {\bibfnamefont {Chris}\ \bibnamefont {Jozwiak}}, \bibinfo {author}
  {\bibfnamefont {Aaron}\ \bibnamefont {Bostwick}}, \bibinfo {author}
  {\bibfnamefont {Eli}\ \bibnamefont {Rotenberg}}, \bibinfo {author}
  {\bibfnamefont {David~C.}\ \bibnamefont {Bell}}, \bibinfo {author}
  {\bibfnamefont {Liang}\ \bibnamefont {Fu}}, \bibinfo {author} {\bibfnamefont
  {Riccardo}\ \bibnamefont {Comin}}, \ and\ \bibinfo {author} {\bibfnamefont
  {Joseph~G.}\ \bibnamefont {Checkelsky}},\ }\bibfield  {title} {\enquote
  {\bibinfo {title} {Massive dirac fermions in a ferromagnetic kagome metal},}\
  }\href {\doibase 10.1038/nature25987} {\bibfield  {journal} {\bibinfo
  {journal} {Nature}\ }\textbf {\bibinfo {volume} {555}},\ \bibinfo {pages}
  {638--642} (\bibinfo {year} {2018})}\BibitemShut {NoStop}%
\bibitem [{\citenamefont {Liang}\ \emph {et~al.}(2021)\citenamefont {Liang},
  \citenamefont {Hou}, \citenamefont {Zhang}, \citenamefont {Ma}, \citenamefont
  {Wu}, \citenamefont {Zhang}, \citenamefont {Yu}, \citenamefont {Ying},
  \citenamefont {Jiang}, \citenamefont {Shan}, \citenamefont {Wang},\ and\
  \citenamefont {Chen}}]{Liang:2021}%
  \BibitemOpen
  \bibfield  {author} {\bibinfo {author} {\bibfnamefont {Zuowei}\ \bibnamefont
  {Liang}}, \bibinfo {author} {\bibfnamefont {Xingyuan}\ \bibnamefont {Hou}},
  \bibinfo {author} {\bibfnamefont {Fan}\ \bibnamefont {Zhang}}, \bibinfo
  {author} {\bibfnamefont {Wanru}\ \bibnamefont {Ma}}, \bibinfo {author}
  {\bibfnamefont {Ping}\ \bibnamefont {Wu}}, \bibinfo {author} {\bibfnamefont
  {Zongyuan}\ \bibnamefont {Zhang}}, \bibinfo {author} {\bibfnamefont
  {Fanghang}\ \bibnamefont {Yu}}, \bibinfo {author} {\bibfnamefont {J.-J.}\
  \bibnamefont {Ying}}, \bibinfo {author} {\bibfnamefont {Kun}\ \bibnamefont
  {Jiang}}, \bibinfo {author} {\bibfnamefont {Lei}\ \bibnamefont {Shan}},
  \bibinfo {author} {\bibfnamefont {Zhenyu}\ \bibnamefont {Wang}}, \ and\
  \bibinfo {author} {\bibfnamefont {X.-H.}\ \bibnamefont {Chen}},\ }\bibfield
  {title} {\enquote {\bibinfo {title} {Three-dimensional charge density wave
  and surface-dependent vortex-core states in a kagome superconductor
  {C}s{V}$_3${S}b$_5$},}\ }\href {\doibase 10.1103/PhysRevX.11.031026}
  {\bibfield  {journal} {\bibinfo  {journal} {Phys. Rev. X}\ }\textbf {\bibinfo
  {volume} {11}},\ \bibinfo {pages} {031026} (\bibinfo {year}
  {2021})}\BibitemShut {NoStop}%
\bibitem [{\citenamefont {Zhou}\ and\ \citenamefont
  {Wang}(2022)}]{Ziqiang:2022}%
  \BibitemOpen
  \bibfield  {author} {\bibinfo {author} {\bibfnamefont {Sen}\ \bibnamefont
  {Zhou}}\ and\ \bibinfo {author} {\bibfnamefont {Ziqiang}\ \bibnamefont
  {Wang}},\ }\bibfield  {title} {\enquote {\bibinfo {title} {Chern fermi
  pocket, topological pair density wave, and charge-4e and charge-6e
  superconductivity in kagom{\'e} superconductors},}\ }\href {\doibase
  10.1038/s41467-022-34832-2} {\bibfield  {journal} {\bibinfo  {journal}
  {Nature Communications}\ }\textbf {\bibinfo {volume} {13}},\ \bibinfo {pages}
  {7288} (\bibinfo {year} {2022})}\BibitemShut {NoStop}%
\bibitem [{\citenamefont {Ye}\ \emph {et~al.}(2022)\citenamefont {Ye},
  \citenamefont {Luo}, \citenamefont {Yin}, \citenamefont {Hasan},\ and\
  \citenamefont {Xu}}]{GangXu:2022}%
  \BibitemOpen
  \bibfield  {author} {\bibinfo {author} {\bibfnamefont {Zijin}\ \bibnamefont
  {Ye}}, \bibinfo {author} {\bibfnamefont {Aiyun}\ \bibnamefont {Luo}},
  \bibinfo {author} {\bibfnamefont {Jia-Xin}\ \bibnamefont {Yin}}, \bibinfo
  {author} {\bibfnamefont {M.~Zahid}\ \bibnamefont {Hasan}}, \ and\ \bibinfo
  {author} {\bibfnamefont {Gang}\ \bibnamefont {Xu}},\ }\bibfield  {title}
  {\enquote {\bibinfo {title} {Structural instability and charge modulations in
  the kagome superconductor {AV}$_3${S}b$_5$},}\ }\href {\doibase
  10.1103/PhysRevB.105.245121} {\bibfield  {journal} {\bibinfo  {journal}
  {Phys. Rev. B}\ }\textbf {\bibinfo {volume} {105}},\ \bibinfo {pages}
  {245121} (\bibinfo {year} {2022})}\BibitemShut {NoStop}%
\bibitem [{\citenamefont {Xie}\ \emph {et~al.}(2022)\citenamefont {Xie},
  \citenamefont {Li}, \citenamefont {Bourges}, \citenamefont {Ivanov},
  \citenamefont {Ye}, \citenamefont {Yin}, \citenamefont {Hasan}, \citenamefont
  {Luo}, \citenamefont {Yao}, \citenamefont {Wang}, \citenamefont {Xu},\ and\
  \citenamefont {Dai}}]{Pengcheng:2022}%
  \BibitemOpen
  \bibfield  {author} {\bibinfo {author} {\bibfnamefont {Yaofeng}\ \bibnamefont
  {Xie}}, \bibinfo {author} {\bibfnamefont {Yongkai}\ \bibnamefont {Li}},
  \bibinfo {author} {\bibfnamefont {Philippe}\ \bibnamefont {Bourges}},
  \bibinfo {author} {\bibfnamefont {Alexandre}\ \bibnamefont {Ivanov}},
  \bibinfo {author} {\bibfnamefont {Zijin}\ \bibnamefont {Ye}}, \bibinfo
  {author} {\bibfnamefont {Jia-Xin}\ \bibnamefont {Yin}}, \bibinfo {author}
  {\bibfnamefont {M.~Zahid}\ \bibnamefont {Hasan}}, \bibinfo {author}
  {\bibfnamefont {Aiyun}\ \bibnamefont {Luo}}, \bibinfo {author} {\bibfnamefont
  {Yugui}\ \bibnamefont {Yao}}, \bibinfo {author} {\bibfnamefont {Zhiwei}\
  \bibnamefont {Wang}}, \bibinfo {author} {\bibfnamefont {Gang}\ \bibnamefont
  {Xu}}, \ and\ \bibinfo {author} {\bibfnamefont {Pengcheng}\ \bibnamefont
  {Dai}},\ }\bibfield  {title} {\enquote {\bibinfo {title} {Electron-phonon
  coupling in the charge density wave state of {C}s{V}$_3${S}b$_5$},}\ }\href
  {\doibase 10.1103/PhysRevB.105.L140501} {\bibfield  {journal} {\bibinfo
  {journal} {Phys. Rev. B}\ }\textbf {\bibinfo {volume} {105}},\ \bibinfo
  {pages} {L140501} (\bibinfo {year} {2022})}\BibitemShut {NoStop}%
\bibitem [{\citenamefont {Liu}\ \emph {et~al.}(2022)\citenamefont {Liu},
  \citenamefont {Ma}, \citenamefont {He}, \citenamefont {Li}, \citenamefont
  {Tan}, \citenamefont {Liu}, \citenamefont {Xu}, \citenamefont {Tang},
  \citenamefont {Watanabe}, \citenamefont {Taniguchi}, \citenamefont {Gao},
  \citenamefont {Dai}, \citenamefont {Wen}, \citenamefont {Yan},\ and\
  \citenamefont {Xi}}]{Liu:2022}%
  \BibitemOpen
  \bibfield  {author} {\bibinfo {author} {\bibfnamefont {Gan}\ \bibnamefont
  {Liu}}, \bibinfo {author} {\bibfnamefont {Xinran}\ \bibnamefont {Ma}},
  \bibinfo {author} {\bibfnamefont {Kuanyu}\ \bibnamefont {He}}, \bibinfo
  {author} {\bibfnamefont {Qing}\ \bibnamefont {Li}}, \bibinfo {author}
  {\bibfnamefont {Hengxin}\ \bibnamefont {Tan}}, \bibinfo {author}
  {\bibfnamefont {Yizhou}\ \bibnamefont {Liu}}, \bibinfo {author}
  {\bibfnamefont {Jie}\ \bibnamefont {Xu}}, \bibinfo {author} {\bibfnamefont
  {Wenna}\ \bibnamefont {Tang}}, \bibinfo {author} {\bibfnamefont {Kenji}\
  \bibnamefont {Watanabe}}, \bibinfo {author} {\bibfnamefont {Takashi}\
  \bibnamefont {Taniguchi}}, \bibinfo {author} {\bibfnamefont {Libo}\
  \bibnamefont {Gao}}, \bibinfo {author} {\bibfnamefont {Yaomin}\ \bibnamefont
  {Dai}}, \bibinfo {author} {\bibfnamefont {Hai-Hu}\ \bibnamefont {Wen}},
  \bibinfo {author} {\bibfnamefont {Binghai}\ \bibnamefont {Yan}}, \ and\
  \bibinfo {author} {\bibfnamefont {Xiaoxiang}\ \bibnamefont {Xi}},\ }\bibfield
   {title} {\enquote {\bibinfo {title} {Observation of anomalous amplitude
  modes in the kagome metal {C}s{V}$_3${S}b$_5$},}\ }\href {\doibase
  10.1038/s41467-022-31162-1} {\bibfield  {journal} {\bibinfo  {journal}
  {Nature Communications}\ }\textbf {\bibinfo {volume} {13}},\ \bibinfo {pages}
  {3461} (\bibinfo {year} {2022})}\BibitemShut {NoStop}%
\bibitem [{\citenamefont {Zhong}\ \emph {et~al.}(2022)\citenamefont {Zhong},
  \citenamefont {Li}, \citenamefont {Liu}, \citenamefont {Dong}, \citenamefont
  {Aido}, \citenamefont {Arai}, \citenamefont {Li}, \citenamefont {Zhang},
  \citenamefont {Shi}, \citenamefont {Wang}, \citenamefont {Shin},
  \citenamefont {Lee}, \citenamefont {Miao}, \citenamefont {Kondo},\ and\
  \citenamefont {Okazaki}}]{Okazaki:2022}%
  \BibitemOpen
  \bibfield  {author} {\bibinfo {author} {\bibfnamefont {Yigui}\ \bibnamefont
  {Zhong}}, \bibinfo {author} {\bibfnamefont {Shaozhi}\ \bibnamefont {Li}},
  \bibinfo {author} {\bibfnamefont {Hongxiong}\ \bibnamefont {Liu}}, \bibinfo
  {author} {\bibfnamefont {Yuyang}\ \bibnamefont {Dong}}, \bibinfo {author}
  {\bibfnamefont {Kohei}\ \bibnamefont {Aido}}, \bibinfo {author}
  {\bibfnamefont {Yosuke}\ \bibnamefont {Arai}}, \bibinfo {author}
  {\bibfnamefont {Haoxiang}\ \bibnamefont {Li}}, \bibinfo {author}
  {\bibfnamefont {Weilu}\ \bibnamefont {Zhang}}, \bibinfo {author}
  {\bibfnamefont {Youguo}\ \bibnamefont {Shi}}, \bibinfo {author}
  {\bibfnamefont {Ziqiang}\ \bibnamefont {Wang}}, \bibinfo {author}
  {\bibfnamefont {Shik}\ \bibnamefont {Shin}}, \bibinfo {author} {\bibfnamefont
  {H.~N.}\ \bibnamefont {Lee}}, \bibinfo {author} {\bibfnamefont
  {H.}~\bibnamefont {Miao}}, \bibinfo {author} {\bibfnamefont {Takeshi}\
  \bibnamefont {Kondo}}, \ and\ \bibinfo {author} {\bibfnamefont {Kozo}\
  \bibnamefont {Okazaki}},\ }\href@noop {} {\enquote {\bibinfo {title} {Testing
  electron-phonon coupling for the superconductivity in kagome metal
  $\rm{CsV_3Sb_5}$},}\ } (\bibinfo {year} {2022}),\ \Eprint
  {http://arxiv.org/abs/arXiv:2207.02407} {arXiv:2207.02407} \BibitemShut
  {NoStop}%
\bibitem [{\citenamefont {Teng}\ \emph {et~al.}(2022)\citenamefont {Teng},
  \citenamefont {Chen}, \citenamefont {Ye}, \citenamefont {Rosenberg},
  \citenamefont {Liu}, \citenamefont {Yin}, \citenamefont {Jiang},
  \citenamefont {Oh}, \citenamefont {Hasan}, \citenamefont {Neubauer},
  \citenamefont {Gao}, \citenamefont {Xie}, \citenamefont {Hashimoto},
  \citenamefont {Lu}, \citenamefont {Jozwiak}, \citenamefont {Bostwick},
  \citenamefont {Rotenberg}, \citenamefont {Birgeneau}, \citenamefont {Chu},
  \citenamefont {Yi},\ and\ \citenamefont {Dai}}]{Teng:2022}%
  \BibitemOpen
  \bibfield  {author} {\bibinfo {author} {\bibfnamefont {Xiaokun}\ \bibnamefont
  {Teng}}, \bibinfo {author} {\bibfnamefont {Lebing}\ \bibnamefont {Chen}},
  \bibinfo {author} {\bibfnamefont {Feng}\ \bibnamefont {Ye}}, \bibinfo
  {author} {\bibfnamefont {Elliott}\ \bibnamefont {Rosenberg}}, \bibinfo
  {author} {\bibfnamefont {Zhaoyu}\ \bibnamefont {Liu}}, \bibinfo {author}
  {\bibfnamefont {Jia-Xin}\ \bibnamefont {Yin}}, \bibinfo {author}
  {\bibfnamefont {Yu-Xiao}\ \bibnamefont {Jiang}}, \bibinfo {author}
  {\bibfnamefont {Ji~Seop}\ \bibnamefont {Oh}}, \bibinfo {author}
  {\bibfnamefont {M.~Zahid}\ \bibnamefont {Hasan}}, \bibinfo {author}
  {\bibfnamefont {Kelly~J.}\ \bibnamefont {Neubauer}}, \bibinfo {author}
  {\bibfnamefont {Bin}\ \bibnamefont {Gao}}, \bibinfo {author} {\bibfnamefont
  {Yaofeng}\ \bibnamefont {Xie}}, \bibinfo {author} {\bibfnamefont {Makoto}\
  \bibnamefont {Hashimoto}}, \bibinfo {author} {\bibfnamefont {Donghui}\
  \bibnamefont {Lu}}, \bibinfo {author} {\bibfnamefont {Chris}\ \bibnamefont
  {Jozwiak}}, \bibinfo {author} {\bibfnamefont {Aaron}\ \bibnamefont
  {Bostwick}}, \bibinfo {author} {\bibfnamefont {Eli}\ \bibnamefont
  {Rotenberg}}, \bibinfo {author} {\bibfnamefont {Robert~J.}\ \bibnamefont
  {Birgeneau}}, \bibinfo {author} {\bibfnamefont {Jiun-Haw}\ \bibnamefont
  {Chu}}, \bibinfo {author} {\bibfnamefont {Ming}\ \bibnamefont {Yi}}, \ and\
  \bibinfo {author} {\bibfnamefont {Pengcheng}\ \bibnamefont {Dai}},\
  }\bibfield  {title} {\enquote {\bibinfo {title} {Discovery of charge density
  wave in a kagome lattice antiferromagnet},}\ }\href {\doibase
  10.1038/s41586-022-05034-z} {\bibfield  {journal} {\bibinfo  {journal}
  {Nature}\ }\textbf {\bibinfo {volume} {609}},\ \bibinfo {pages} {490--495}
  (\bibinfo {year} {2022})}\BibitemShut {NoStop}%
\bibitem [{\citenamefont {Yin}\ \emph {et~al.}(2022)\citenamefont {Yin},
  \citenamefont {Jiang}, \citenamefont {Teng}, \citenamefont {Hossain},
  \citenamefont {Mardanya}, \citenamefont {Chang}, \citenamefont {Ye},
  \citenamefont {Xu}, \citenamefont {Denner}, \citenamefont {Neupert},
  \citenamefont {Lienhard}, \citenamefont {Deng}, \citenamefont {Setty},
  \citenamefont {Si}, \citenamefont {Chang}, \citenamefont {Guguchia},
  \citenamefont {Gao}, \citenamefont {Shumiya}, \citenamefont {Zhang},
  \citenamefont {Cochran}, \citenamefont {Multer}, \citenamefont {Yi},
  \citenamefont {Dai},\ and\ \citenamefont {Hasan}}]{YinJiaXin:2022}%
  \BibitemOpen
  \bibfield  {author} {\bibinfo {author} {\bibfnamefont {Jia-Xin}\ \bibnamefont
  {Yin}}, \bibinfo {author} {\bibfnamefont {Yu-Xiao}\ \bibnamefont {Jiang}},
  \bibinfo {author} {\bibfnamefont {Xiaokun}\ \bibnamefont {Teng}}, \bibinfo
  {author} {\bibfnamefont {Md.~Shafayat}\ \bibnamefont {Hossain}}, \bibinfo
  {author} {\bibfnamefont {Sougata}\ \bibnamefont {Mardanya}}, \bibinfo
  {author} {\bibfnamefont {Tay-Rong}\ \bibnamefont {Chang}}, \bibinfo {author}
  {\bibfnamefont {Zijin}\ \bibnamefont {Ye}}, \bibinfo {author} {\bibfnamefont
  {Gang}\ \bibnamefont {Xu}}, \bibinfo {author} {\bibfnamefont {M.~Michael}\
  \bibnamefont {Denner}}, \bibinfo {author} {\bibfnamefont {Titus}\
  \bibnamefont {Neupert}}, \bibinfo {author} {\bibfnamefont {Benjamin}\
  \bibnamefont {Lienhard}}, \bibinfo {author} {\bibfnamefont {Han-Bin}\
  \bibnamefont {Deng}}, \bibinfo {author} {\bibfnamefont {Chandan}\
  \bibnamefont {Setty}}, \bibinfo {author} {\bibfnamefont {Qimiao}\
  \bibnamefont {Si}}, \bibinfo {author} {\bibfnamefont {Guoqing}\ \bibnamefont
  {Chang}}, \bibinfo {author} {\bibfnamefont {Zurab}\ \bibnamefont {Guguchia}},
  \bibinfo {author} {\bibfnamefont {Bin}\ \bibnamefont {Gao}}, \bibinfo
  {author} {\bibfnamefont {Nana}\ \bibnamefont {Shumiya}}, \bibinfo {author}
  {\bibfnamefont {Qi}~\bibnamefont {Zhang}}, \bibinfo {author} {\bibfnamefont
  {Tyler~A.}\ \bibnamefont {Cochran}}, \bibinfo {author} {\bibfnamefont
  {Daniel}\ \bibnamefont {Multer}}, \bibinfo {author} {\bibfnamefont {Ming}\
  \bibnamefont {Yi}}, \bibinfo {author} {\bibfnamefont {Pengcheng}\
  \bibnamefont {Dai}}, \ and\ \bibinfo {author} {\bibfnamefont {M.~Zahid}\
  \bibnamefont {Hasan}},\ }\bibfield  {title} {\enquote {\bibinfo {title}
  {Discovery of charge order and corresponding edge state in kagome magnet
  {F}e{G}e},}\ }\href {\doibase 10.1103/PhysRevLett.129.166401} {\bibfield
  {journal} {\bibinfo  {journal} {Phys. Rev. Lett.}\ }\textbf {\bibinfo
  {volume} {129}},\ \bibinfo {pages} {166401} (\bibinfo {year}
  {2022})}\BibitemShut {NoStop}%
\bibitem [{\citenamefont {Miao}\ \emph {et~al.}(2022)\citenamefont {Miao},
  \citenamefont {Zhang}, \citenamefont {Li}, \citenamefont {Fabbris},
  \citenamefont {Said}, \citenamefont {Tartaglia}, \citenamefont {Yilmaz},
  \citenamefont {Vescovo}, \citenamefont {Yin}, \citenamefont {Murakami},
  \citenamefont {Feng}, \citenamefont {Jiang}, \citenamefont {Wu},
  \citenamefont {Wang}, \citenamefont {Okamoto}, \citenamefont {Wang},\ and\
  \citenamefont {Lee}}]{Miao:2022}%
  \BibitemOpen
  \bibfield  {author} {\bibinfo {author} {\bibfnamefont {H.}~\bibnamefont
  {Miao}}, \bibinfo {author} {\bibfnamefont {T.~T.}\ \bibnamefont {Zhang}},
  \bibinfo {author} {\bibfnamefont {H.~X.}\ \bibnamefont {Li}}, \bibinfo
  {author} {\bibfnamefont {G.}~\bibnamefont {Fabbris}}, \bibinfo {author}
  {\bibfnamefont {A.~H.}\ \bibnamefont {Said}}, \bibinfo {author}
  {\bibfnamefont {R.}~\bibnamefont {Tartaglia}}, \bibinfo {author}
  {\bibfnamefont {T.}~\bibnamefont {Yilmaz}}, \bibinfo {author} {\bibfnamefont
  {E.}~\bibnamefont {Vescovo}}, \bibinfo {author} {\bibfnamefont {J.~X.}\
  \bibnamefont {Yin}}, \bibinfo {author} {\bibfnamefont {S.}~\bibnamefont
  {Murakami}}, \bibinfo {author} {\bibfnamefont {L.~X.}\ \bibnamefont {Feng}},
  \bibinfo {author} {\bibfnamefont {K.}~\bibnamefont {Jiang}}, \bibinfo
  {author} {\bibfnamefont {X.~L.}\ \bibnamefont {Wu}}, \bibinfo {author}
  {\bibfnamefont {A.~F.}\ \bibnamefont {Wang}}, \bibinfo {author}
  {\bibfnamefont {S.}~\bibnamefont {Okamoto}}, \bibinfo {author} {\bibfnamefont
  {Y.~L.}\ \bibnamefont {Wang}}, \ and\ \bibinfo {author} {\bibfnamefont
  {H.~N.}\ \bibnamefont {Lee}},\ }\href@noop {} {\enquote {\bibinfo {title}
  {Spin-phonon coupling driven charge density wave in a kagome magnet},}\ }
  (\bibinfo {year} {2022}),\ \Eprint {http://arxiv.org/abs/arXiv:2210.06359}
  {arXiv:2210.06359} \BibitemShut {NoStop}%
\bibitem [{\citenamefont {Yu}\ \emph {et~al.}(2021)\citenamefont {Yu},
  \citenamefont {Wu}, \citenamefont {Wang}, \citenamefont {Lei}, \citenamefont
  {Zhuo}, \citenamefont {Ying},\ and\ \citenamefont {Chen}}]{Yingjianjun:2021}%
  \BibitemOpen
  \bibfield  {author} {\bibinfo {author} {\bibfnamefont {F.~H.}\ \bibnamefont
  {Yu}}, \bibinfo {author} {\bibfnamefont {T.}~\bibnamefont {Wu}}, \bibinfo
  {author} {\bibfnamefont {Z.~Y.}\ \bibnamefont {Wang}}, \bibinfo {author}
  {\bibfnamefont {B.}~\bibnamefont {Lei}}, \bibinfo {author} {\bibfnamefont
  {W.~Z.}\ \bibnamefont {Zhuo}}, \bibinfo {author} {\bibfnamefont {J.~J.}\
  \bibnamefont {Ying}}, \ and\ \bibinfo {author} {\bibfnamefont {X.~H.}\
  \bibnamefont {Chen}},\ }\bibfield  {title} {\enquote {\bibinfo {title}
  {Concurrence of anomalous hall effect and charge density wave in a
  superconducting topological kagome metal},}\ }\href {\doibase
  10.1103/PhysRevB.104.L041103} {\bibfield  {journal} {\bibinfo  {journal}
  {Phys. Rev. B}\ }\textbf {\bibinfo {volume} {104}},\ \bibinfo {pages}
  {L041103} (\bibinfo {year} {2021})}\BibitemShut {NoStop}%
\bibitem [{\citenamefont {Teng}\ \emph {et~al.}(2023)\citenamefont {Teng},
  \citenamefont {Oh}, \citenamefont {Tan}, \citenamefont {Chen}, \citenamefont
  {Huang}, \citenamefont {Gao}, \citenamefont {Yin}, \citenamefont {Chu},
  \citenamefont {Hashimoto}, \citenamefont {Lu}, \citenamefont {Jozwiak},
  \citenamefont {Bostwick}, \citenamefont {Rotenberg}, \citenamefont
  {Granroth}, \citenamefont {Yan}, \citenamefont {Birgeneau}, \citenamefont
  {Dai},\ and\ \citenamefont {Yi}}]{Teng:2023}%
  \BibitemOpen
  \bibfield  {author} {\bibinfo {author} {\bibfnamefont {Xiaokun}\ \bibnamefont
  {Teng}}, \bibinfo {author} {\bibfnamefont {Ji~Seop}\ \bibnamefont {Oh}},
  \bibinfo {author} {\bibfnamefont {Hengxin}\ \bibnamefont {Tan}}, \bibinfo
  {author} {\bibfnamefont {Lebing}\ \bibnamefont {Chen}}, \bibinfo {author}
  {\bibfnamefont {Jianwei}\ \bibnamefont {Huang}}, \bibinfo {author}
  {\bibfnamefont {Bin}\ \bibnamefont {Gao}}, \bibinfo {author} {\bibfnamefont
  {Jia-Xin}\ \bibnamefont {Yin}}, \bibinfo {author} {\bibfnamefont {Jiun-Haw}\
  \bibnamefont {Chu}}, \bibinfo {author} {\bibfnamefont {Makoto}\ \bibnamefont
  {Hashimoto}}, \bibinfo {author} {\bibfnamefont {Donghui}\ \bibnamefont {Lu}},
  \bibinfo {author} {\bibfnamefont {Chris}\ \bibnamefont {Jozwiak}}, \bibinfo
  {author} {\bibfnamefont {Aaron}\ \bibnamefont {Bostwick}}, \bibinfo {author}
  {\bibfnamefont {Eli}\ \bibnamefont {Rotenberg}}, \bibinfo {author}
  {\bibfnamefont {Garrett~E.}\ \bibnamefont {Granroth}}, \bibinfo {author}
  {\bibfnamefont {Binghai}\ \bibnamefont {Yan}}, \bibinfo {author}
  {\bibfnamefont {Robert~J.}\ \bibnamefont {Birgeneau}}, \bibinfo {author}
  {\bibfnamefont {Pengcheng}\ \bibnamefont {Dai}}, \ and\ \bibinfo {author}
  {\bibfnamefont {Ming}\ \bibnamefont {Yi}},\ }\bibfield  {title} {\enquote
  {\bibinfo {title} {Magnetism and charge density wave order in kagome
  {F}e{G}e},}\ }\href {\doibase 10.1038/s41567-023-01985-w} {\bibfield
  {journal} {\bibinfo  {journal} {Nature Physics}\ } (\bibinfo {year} {2023}),\
  10.1038/s41567-023-01985-w}\BibitemShut {NoStop}%
\bibitem [{\citenamefont {Setty}\ \emph {et~al.}(2022)\citenamefont {Setty},
  \citenamefont {Lane}, \citenamefont {Chen}, \citenamefont {Hu}, \citenamefont
  {Zhu},\ and\ \citenamefont {Si}}]{Setty:2022}%
  \BibitemOpen
  \bibfield  {author} {\bibinfo {author} {\bibfnamefont {Chandan}\ \bibnamefont
  {Setty}}, \bibinfo {author} {\bibfnamefont {Christopher~A.}\ \bibnamefont
  {Lane}}, \bibinfo {author} {\bibfnamefont {Lei}\ \bibnamefont {Chen}},
  \bibinfo {author} {\bibfnamefont {Haoyu}\ \bibnamefont {Hu}}, \bibinfo
  {author} {\bibfnamefont {Jian-Xin}\ \bibnamefont {Zhu}}, \ and\ \bibinfo
  {author} {\bibfnamefont {Qimiao}\ \bibnamefont {Si}},\ }\href@noop {}
  {\enquote {\bibinfo {title} {Electron correlations and charge density wave in
  the topological kagome metal {F}e{G}e},}\ } (\bibinfo {year} {2022}),\
  \Eprint {http://arxiv.org/abs/arXiv:2203.01930} {arXiv:2203.01930}
  \BibitemShut {NoStop}%
\bibitem [{\citenamefont {Shao}\ \emph {et~al.}(2022)\citenamefont {Shao},
  \citenamefont {Yin}, \citenamefont {Belopolski}, \citenamefont {You},
  \citenamefont {Hou}, \citenamefont {Chen}, \citenamefont {Jiang},
  \citenamefont {Hossain}, \citenamefont {Yahyavi}, \citenamefont {Hsu},
  \citenamefont {Feng}, \citenamefont {Bansil}, \citenamefont {Hasan},\ and\
  \citenamefont {Chang}}]{Guoqing:2022}%
  \BibitemOpen
  \bibfield  {author} {\bibinfo {author} {\bibfnamefont {Sen}\ \bibnamefont
  {Shao}}, \bibinfo {author} {\bibfnamefont {Jia-Xin}\ \bibnamefont {Yin}},
  \bibinfo {author} {\bibfnamefont {Ilya}\ \bibnamefont {Belopolski}}, \bibinfo
  {author} {\bibfnamefont {Jing-Yang}\ \bibnamefont {You}}, \bibinfo {author}
  {\bibfnamefont {Tao}\ \bibnamefont {Hou}}, \bibinfo {author} {\bibfnamefont
  {Hongyu}\ \bibnamefont {Chen}}, \bibinfo {author} {\bibfnamefont {Yu-Xiao}\
  \bibnamefont {Jiang}}, \bibinfo {author} {\bibfnamefont {Md~Shafayat}\
  \bibnamefont {Hossain}}, \bibinfo {author} {\bibfnamefont {Mohammad}\
  \bibnamefont {Yahyavi}}, \bibinfo {author} {\bibfnamefont {Chia-Hsiu}\
  \bibnamefont {Hsu}}, \bibinfo {author} {\bibfnamefont {Yuanping}\
  \bibnamefont {Feng}}, \bibinfo {author} {\bibfnamefont {Arun}\ \bibnamefont
  {Bansil}}, \bibinfo {author} {\bibfnamefont {M.~Zahid}\ \bibnamefont
  {Hasan}}, \ and\ \bibinfo {author} {\bibfnamefont {Guoqing}\ \bibnamefont
  {Chang}},\ }\href@noop {} {\enquote {\bibinfo {title} {Charge density wave
  interaction in a kagome-honeycomb antiferromagnet},}\ } (\bibinfo {year}
  {2022}),\ \Eprint {http://arxiv.org/abs/arXiv:2206.12033} {arXiv:2206.12033}
  \BibitemShut {NoStop}%
\bibitem [{\citenamefont {Zhou}\ \emph {et~al.}(2022)\citenamefont {Zhou},
  \citenamefont {Yan}, \citenamefont {Fan}, \citenamefont {Wang},\ and\
  \citenamefont {Wan}}]{Xiangang:2022}%
  \BibitemOpen
  \bibfield  {author} {\bibinfo {author} {\bibfnamefont {Hanjing}\ \bibnamefont
  {Zhou}}, \bibinfo {author} {\bibfnamefont {Songsong}\ \bibnamefont {Yan}},
  \bibinfo {author} {\bibfnamefont {Dongze}\ \bibnamefont {Fan}}, \bibinfo
  {author} {\bibfnamefont {Di}~\bibnamefont {Wang}}, \ and\ \bibinfo {author}
  {\bibfnamefont {Xiangang}\ \bibnamefont {Wan}},\ }\href@noop {} {\enquote
  {\bibinfo {title} {Magnetic interactions and possible structural distortion
  in kagome {F}e{G}e from first-principles study and symmetry analysis},}\ }
  (\bibinfo {year} {2022}),\ \Eprint {http://arxiv.org/abs/arXiv:2211.15545}
  {arXiv:2211.15545} \BibitemShut {NoStop}%
\bibitem [{\citenamefont {Wu}\ \emph {et~al.}(2023)\citenamefont {Wu},
  \citenamefont {Hu}, \citenamefont {Wang},\ and\ \citenamefont
  {Wan}}]{Xiangang:2023}%
  \BibitemOpen
  \bibfield  {author} {\bibinfo {author} {\bibfnamefont {Lin}\ \bibnamefont
  {Wu}}, \bibinfo {author} {\bibfnamefont {Yating}\ \bibnamefont {Hu}},
  \bibinfo {author} {\bibfnamefont {Di}~\bibnamefont {Wang}}, \ and\ \bibinfo
  {author} {\bibfnamefont {Xiangang}\ \bibnamefont {Wan}},\ }\href@noop {}
  {\enquote {\bibinfo {title} {Novel three-dimensional fermi surface and
  electron-correlation-induced charge density wave in {F}e{G}e},}\ } (\bibinfo
  {year} {2023}),\ \Eprint {http://arxiv.org/abs/arXiv:2302.03622}
  {arXiv:2302.03622} \BibitemShut {NoStop}%
\bibitem [{\citenamefont {Ma}\ \emph {et~al.}(2023)\citenamefont {Ma},
  \citenamefont {Yin}, \citenamefont {Hasan},\ and\ \citenamefont
  {Liu}}]{Jianpeng:2023}%
  \BibitemOpen
  \bibfield  {author} {\bibinfo {author} {\bibfnamefont {Hai-Yang}\
  \bibnamefont {Ma}}, \bibinfo {author} {\bibfnamefont {Jia-Xin}\ \bibnamefont
  {Yin}}, \bibinfo {author} {\bibfnamefont {M.~Zahid}\ \bibnamefont {Hasan}}, \
  and\ \bibinfo {author} {\bibfnamefont {Jianpeng}\ \bibnamefont {Liu}},\
  }\href@noop {} {\enquote {\bibinfo {title} {Theory for charge density wave
  and orbital-flux state in antiferromagnetic kagome metal {F}e{G}e},}\ }
  (\bibinfo {year} {2023}),\ \Eprint {http://arxiv.org/abs/arXiv:2303.02824}
  {arXiv:2303.02824} \BibitemShut {NoStop}%
\bibitem [{\citenamefont {Lichtenstein}\ \emph {et~al.}(2001)\citenamefont
  {Lichtenstein}, \citenamefont {Katsnelson},\ and\ \citenamefont
  {Kotliar}}]{lichtenstein:2001}%
  \BibitemOpen
  \bibfield  {author} {\bibinfo {author} {\bibfnamefont {A.~I.}\ \bibnamefont
  {Lichtenstein}}, \bibinfo {author} {\bibfnamefont {M.~I.}\ \bibnamefont
  {Katsnelson}}, \ and\ \bibinfo {author} {\bibfnamefont {G.}~\bibnamefont
  {Kotliar}},\ }\bibfield  {title} {\enquote {\bibinfo {title}
  {Finite-temperature magnetism of transition metals: An ab initio dynamical
  mean-field theory},}\ }\href {\doibase 10.1103/PhysRevLett.87.067205}
  {\bibfield  {journal} {\bibinfo  {journal} {Phys. Rev. Lett.}\ }\textbf
  {\bibinfo {volume} {87}},\ \bibinfo {pages} {067205} (\bibinfo {year}
  {2001})}\BibitemShut {NoStop}%
\bibitem [{\citenamefont {Kotliar}\ \emph {et~al.}(2006)\citenamefont
  {Kotliar}, \citenamefont {Savrasov}, \citenamefont {Haule}, \citenamefont
  {Oudovenko}, \citenamefont {Parcollet},\ and\ \citenamefont
  {Marianetti}}]{kotliar:2006}%
  \BibitemOpen
  \bibfield  {author} {\bibinfo {author} {\bibfnamefont {G.}~\bibnamefont
  {Kotliar}}, \bibinfo {author} {\bibfnamefont {S.~Y.}\ \bibnamefont
  {Savrasov}}, \bibinfo {author} {\bibfnamefont {K.}~\bibnamefont {Haule}},
  \bibinfo {author} {\bibfnamefont {V.~S.}\ \bibnamefont {Oudovenko}}, \bibinfo
  {author} {\bibfnamefont {O.}~\bibnamefont {Parcollet}}, \ and\ \bibinfo
  {author} {\bibfnamefont {C.~A.}\ \bibnamefont {Marianetti}},\ }\bibfield
  {title} {\enquote {\bibinfo {title} {Electronic structure calculations with
  dynamical mean-field theory},}\ }\href {\doibase 10.1103/RevModPhys.78.865}
  {\bibfield  {journal} {\bibinfo  {journal} {Rev. Mod. Phys.}\ }\textbf
  {\bibinfo {volume} {78}},\ \bibinfo {pages} {865--951} (\bibinfo {year}
  {2006})}\BibitemShut {NoStop}%
\bibitem [{\citenamefont {Kresse}\ and\ \citenamefont
  {Furthm\"uller}(1996)}]{kresse:1996}%
  \BibitemOpen
  \bibfield  {author} {\bibinfo {author} {\bibfnamefont {G.}~\bibnamefont
  {Kresse}}\ and\ \bibinfo {author} {\bibfnamefont {J.}~\bibnamefont
  {Furthm\"uller}},\ }\bibfield  {title} {\enquote {\bibinfo {title} {Efficient
  iterative schemes for \textit{ab initio} total-energy calculations using a
  plane-wave basis set},}\ }\href {\doibase 10.1103/PhysRevB.54.11169}
  {\bibfield  {journal} {\bibinfo  {journal} {Phys. Rev. B}\ }\textbf {\bibinfo
  {volume} {54}},\ \bibinfo {pages} {11169--11186} (\bibinfo {year}
  {1996})}\BibitemShut {NoStop}%
\bibitem [{\citenamefont {Bl\"ochl}(1994)}]{blochl:1994}%
  \BibitemOpen
  \bibfield  {author} {\bibinfo {author} {\bibfnamefont {P.~E.}\ \bibnamefont
  {Bl\"ochl}},\ }\bibfield  {title} {\enquote {\bibinfo {title} {Projector
  augmented-wave method},}\ }\href {\doibase 10.1103/PhysRevB.50.17953}
  {\bibfield  {journal} {\bibinfo  {journal} {Phys. Rev. B}\ }\textbf {\bibinfo
  {volume} {50}},\ \bibinfo {pages} {17953--17979} (\bibinfo {year}
  {1994})}\BibitemShut {NoStop}%
\bibitem [{\citenamefont {Perdew}\ \emph {et~al.}(1996)\citenamefont {Perdew},
  \citenamefont {Burke},\ and\ \citenamefont {Ernzerhof}}]{perdew:1996}%
  \BibitemOpen
  \bibfield  {author} {\bibinfo {author} {\bibfnamefont {John~P.}\ \bibnamefont
  {Perdew}}, \bibinfo {author} {\bibfnamefont {Kieron}\ \bibnamefont {Burke}},
  \ and\ \bibinfo {author} {\bibfnamefont {Matthias}\ \bibnamefont
  {Ernzerhof}},\ }\bibfield  {title} {\enquote {\bibinfo {title} {Generalized
  gradient approximation made simple},}\ }\href {\doibase
  10.1103/PhysRevLett.77.3865} {\bibfield  {journal} {\bibinfo  {journal}
  {Phys. Rev. Lett.}\ }\textbf {\bibinfo {volume} {77}},\ \bibinfo {pages}
  {3865--3868} (\bibinfo {year} {1996})}\BibitemShut {NoStop}%
\bibitem [{\citenamefont {Meier}\ \emph {et~al.}(2020)\citenamefont {Meier},
  \citenamefont {Du}, \citenamefont {Okamoto}, \citenamefont {Mohanta},
  \citenamefont {May}, \citenamefont {McGuire}, \citenamefont {Bridges},
  \citenamefont {Samolyuk},\ and\ \citenamefont {Sales}}]{Brian:2020}%
  \BibitemOpen
  \bibfield  {author} {\bibinfo {author} {\bibfnamefont {William~R.}\
  \bibnamefont {Meier}}, \bibinfo {author} {\bibfnamefont {Mao-Hua}\
  \bibnamefont {Du}}, \bibinfo {author} {\bibfnamefont {Satoshi}\ \bibnamefont
  {Okamoto}}, \bibinfo {author} {\bibfnamefont {Narayan}\ \bibnamefont
  {Mohanta}}, \bibinfo {author} {\bibfnamefont {Andrew~F.}\ \bibnamefont
  {May}}, \bibinfo {author} {\bibfnamefont {Michael~A.}\ \bibnamefont
  {McGuire}}, \bibinfo {author} {\bibfnamefont {Craig~A.}\ \bibnamefont
  {Bridges}}, \bibinfo {author} {\bibfnamefont {German~D.}\ \bibnamefont
  {Samolyuk}}, \ and\ \bibinfo {author} {\bibfnamefont {Brian~C.}\ \bibnamefont
  {Sales}},\ }\bibfield  {title} {\enquote {\bibinfo {title} {Flat bands in the
  {C}o{S}n-type compounds},}\ }\href {\doibase 10.1103/PhysRevB.102.075148}
  {\bibfield  {journal} {\bibinfo  {journal} {Phys. Rev. B}\ }\textbf {\bibinfo
  {volume} {102}},\ \bibinfo {pages} {075148} (\bibinfo {year}
  {2020})}\BibitemShut {NoStop}%
\bibitem [{\citenamefont {Haule}\ \emph {et~al.}(2010)\citenamefont {Haule},
  \citenamefont {Yee},\ and\ \citenamefont {Kim}}]{Haule:2010}%
  \BibitemOpen
  \bibfield  {author} {\bibinfo {author} {\bibfnamefont {Kristjan}\
  \bibnamefont {Haule}}, \bibinfo {author} {\bibfnamefont {Chuck-Hou}\
  \bibnamefont {Yee}}, \ and\ \bibinfo {author} {\bibfnamefont {Kyoo}\
  \bibnamefont {Kim}},\ }\bibfield  {title} {\enquote {\bibinfo {title}
  {Dynamical mean-field theory within the full-potential methods: Electronic
  structure of {C}e{I}r{I}n$_{5}$, {C}e{C}o{I}n$_{5}$, and
  {C}e{R}h{I}n$_{5}$},}\ }\href {\doibase 10.1103/PhysRevB.81.195107}
  {\bibfield  {journal} {\bibinfo  {journal} {Phys. Rev. B}\ }\textbf {\bibinfo
  {volume} {81}},\ \bibinfo {pages} {195107} (\bibinfo {year}
  {2010})}\BibitemShut {NoStop}%
\bibitem [{\citenamefont {Haule}\ and\ \citenamefont
  {Birol}(2015)}]{Haule:2015free}%
  \BibitemOpen
  \bibfield  {author} {\bibinfo {author} {\bibfnamefont {Kristjan}\
  \bibnamefont {Haule}}\ and\ \bibinfo {author} {\bibfnamefont {Turan}\
  \bibnamefont {Birol}},\ }\bibfield  {title} {\enquote {\bibinfo {title} {Free
  energy from stationary implementation of the $\mathrm{DFT}+\mathrm{DMFT}$
  functional},}\ }\href {\doibase 10.1103/PhysRevLett.115.256402} {\bibfield
  {journal} {\bibinfo  {journal} {Phys. Rev. Lett.}\ }\textbf {\bibinfo
  {volume} {115}},\ \bibinfo {pages} {256402} (\bibinfo {year}
  {2015})}\BibitemShut {NoStop}%
\bibitem [{\citenamefont {Blaha}\ \emph {et~al.}(2020)\citenamefont {Blaha},
  \citenamefont {Schwarz}, \citenamefont {Tran}, \citenamefont {Laskowski},
  \citenamefont {Madsen},\ and\ \citenamefont {Marks}}]{Blaha:2020}%
  \BibitemOpen
  \bibfield  {author} {\bibinfo {author} {\bibfnamefont {Peter}\ \bibnamefont
  {Blaha}}, \bibinfo {author} {\bibfnamefont {Karlheinz}\ \bibnamefont
  {Schwarz}}, \bibinfo {author} {\bibfnamefont {Fabien}\ \bibnamefont {Tran}},
  \bibinfo {author} {\bibfnamefont {Robert}\ \bibnamefont {Laskowski}},
  \bibinfo {author} {\bibfnamefont {Georg K.~H.}\ \bibnamefont {Madsen}}, \
  and\ \bibinfo {author} {\bibfnamefont {Laurence~D.}\ \bibnamefont {Marks}},\
  }\bibfield  {title} {\enquote {\bibinfo {title} {{WIEN}2k: An {APW}+lo
  program for calculating the properties of solids},}\ }\href {\doibase
  10.1063/1.5143061} {\bibfield  {journal} {\bibinfo  {journal} {The Journal of
  Chemical Physics}\ }\textbf {\bibinfo {volume} {152}},\ \bibinfo {pages}
  {074101} (\bibinfo {year} {2020})}\BibitemShut {NoStop}%
\bibitem [{sup()}]{suppl}%
  \BibitemOpen
  \href@noop {} {\bibinfo  {journal} {See Supporting Information at [url] for:
  (1) Computational details, (2) DFT calculations of the $2\times 2\times 2$
  superstructure of FeSn. The Supplemental Material includes
  Refs.~\cite{Dudarev:1998,Gull:2011,Haule:2016force,tutor:MnO}}\ }\BibitemShut
  {NoStop}%
\bibitem [{\citenamefont {Herrera}\ and\ \citenamefont
  {Naumis}(2020)}]{Herrera:2020}%
  \BibitemOpen
\bibfield  {journal} {  }\bibfield  {author} {\bibinfo {author} {\bibfnamefont
  {Sa\'ul~A.}\ \bibnamefont {Herrera}}\ and\ \bibinfo {author} {\bibfnamefont
  {Gerardo~G.}\ \bibnamefont {Naumis}},\ }\bibfield  {title} {\enquote
  {\bibinfo {title} {Electronic and optical conductivity of kekul\'e-patterned
  graphene: Intravalley and intervalley transport},}\ }\href {\doibase
  10.1103/PhysRevB.101.205413} {\bibfield  {journal} {\bibinfo  {journal}
  {Phys. Rev. B}\ }\textbf {\bibinfo {volume} {101}},\ \bibinfo {pages}
  {205413} (\bibinfo {year} {2020})}\BibitemShut {NoStop}%
\bibitem [{\citenamefont {Chen}\ \emph {et~al.}(2023)\citenamefont {Chen},
  \citenamefont {Wu}, \citenamefont {Yin}, \citenamefont {Zhang}, \citenamefont
  {Wang}, \citenamefont {Li}, \citenamefont {Li}, \citenamefont {Wang},
  \citenamefont {Wang}, \citenamefont {Yan},\ and\ \citenamefont
  {Feng}}]{Ziyuan:2023}%
  \BibitemOpen
  \bibfield  {author} {\bibinfo {author} {\bibfnamefont {Ziyuan}\ \bibnamefont
  {Chen}}, \bibinfo {author} {\bibfnamefont {Xueliang}\ \bibnamefont {Wu}},
  \bibinfo {author} {\bibfnamefont {Ruotong}\ \bibnamefont {Yin}}, \bibinfo
  {author} {\bibfnamefont {Jiakang}\ \bibnamefont {Zhang}}, \bibinfo {author}
  {\bibfnamefont {Shiyuan}\ \bibnamefont {Wang}}, \bibinfo {author}
  {\bibfnamefont {Yuanji}\ \bibnamefont {Li}}, \bibinfo {author} {\bibfnamefont
  {Mingzhe}\ \bibnamefont {Li}}, \bibinfo {author} {\bibfnamefont {Aifeng}\
  \bibnamefont {Wang}}, \bibinfo {author} {\bibfnamefont {Yilin}\ \bibnamefont
  {Wang}}, \bibinfo {author} {\bibfnamefont {Ya-Jun}\ \bibnamefont {Yan}}, \
  and\ \bibinfo {author} {\bibfnamefont {Dong-Lai}\ \bibnamefont {Feng}},\
  }\href@noop {} {\enquote {\bibinfo {title} {Charge density wave with strong
  quantum phase fluctuations in kagome magnet {F}e{G}e},}\ } (\bibinfo {year}
  {2023}),\ \Eprint {http://arxiv.org/abs/arXiv:2302.04490} {arXiv:2302.04490}
  \BibitemShut {NoStop}%
\bibitem [{\citenamefont {Yin}\ \emph {et~al.}(2011)\citenamefont {Yin},
  \citenamefont {Haule},\ and\ \citenamefont {Kotliar}}]{YinZP:2011}%
  \BibitemOpen
  \bibfield  {author} {\bibinfo {author} {\bibfnamefont {Z.~P.}\ \bibnamefont
  {Yin}}, \bibinfo {author} {\bibfnamefont {K.}~\bibnamefont {Haule}}, \ and\
  \bibinfo {author} {\bibfnamefont {G.}~\bibnamefont {Kotliar}},\ }\bibfield
  {title} {\enquote {\bibinfo {title} {Kinetic frustration and the nature of
  the magnetic and paramagnetic states in iron pnictides and
  iron chalcogenides},}\ }\href {\doibase 10.1038/nmat3120} {\bibfield
  {journal} {\bibinfo  {journal} {Nature Materials}\ }\textbf {\bibinfo
  {volume} {10}},\ \bibinfo {pages} {932--935} (\bibinfo {year}
  {2011})}\BibitemShut {NoStop}%
\bibitem [{\citenamefont {Ieki}\ \emph {et~al.}(2014)\citenamefont {Ieki},
  \citenamefont {Nakayama}, \citenamefont {Miyata}, \citenamefont {Sato},
  \citenamefont {Miao}, \citenamefont {Xu}, \citenamefont {Wang}, \citenamefont
  {Zhang}, \citenamefont {Qian}, \citenamefont {Richard}, \citenamefont {Xu},
  \citenamefont {Wen}, \citenamefont {Gu}, \citenamefont {Luo}, \citenamefont
  {Wen}, \citenamefont {Ding},\ and\ \citenamefont {Takahashi}}]{Ieki:2014}%
  \BibitemOpen
  \bibfield  {author} {\bibinfo {author} {\bibfnamefont {E.}~\bibnamefont
  {Ieki}}, \bibinfo {author} {\bibfnamefont {K.}~\bibnamefont {Nakayama}},
  \bibinfo {author} {\bibfnamefont {Y.}~\bibnamefont {Miyata}}, \bibinfo
  {author} {\bibfnamefont {T.}~\bibnamefont {Sato}}, \bibinfo {author}
  {\bibfnamefont {H.}~\bibnamefont {Miao}}, \bibinfo {author} {\bibfnamefont
  {N.}~\bibnamefont {Xu}}, \bibinfo {author} {\bibfnamefont {X.-P.}\
  \bibnamefont {Wang}}, \bibinfo {author} {\bibfnamefont {P.}~\bibnamefont
  {Zhang}}, \bibinfo {author} {\bibfnamefont {T.}~\bibnamefont {Qian}},
  \bibinfo {author} {\bibfnamefont {P.}~\bibnamefont {Richard}}, \bibinfo
  {author} {\bibfnamefont {Z.-J.}\ \bibnamefont {Xu}}, \bibinfo {author}
  {\bibfnamefont {J.~S.}\ \bibnamefont {Wen}}, \bibinfo {author} {\bibfnamefont
  {G.~D.}\ \bibnamefont {Gu}}, \bibinfo {author} {\bibfnamefont {H.~Q.}\
  \bibnamefont {Luo}}, \bibinfo {author} {\bibfnamefont {H.-H.}\ \bibnamefont
  {Wen}}, \bibinfo {author} {\bibfnamefont {H.}~\bibnamefont {Ding}}, \ and\
  \bibinfo {author} {\bibfnamefont {T.}~\bibnamefont {Takahashi}},\ }\bibfield
  {title} {\enquote {\bibinfo {title} {Evolution from incoherent to coherent
  electronic states and its implications for superconductivity in
  {F}e{T}e$_{1-x}${S}e$_x$},}\ }\href {\doibase 10.1103/PhysRevB.89.140506}
  {\bibfield  {journal} {\bibinfo  {journal} {Phys. Rev. B}\ }\textbf {\bibinfo
  {volume} {89}},\ \bibinfo {pages} {140506} (\bibinfo {year}
  {2014})}\BibitemShut {NoStop}%
\bibitem [{\citenamefont {Haule}\ and\ \citenamefont
  {Pascut}(2016)}]{Haule:2016force}%
  \BibitemOpen
  \bibfield  {author} {\bibinfo {author} {\bibfnamefont {Kristjan}\
  \bibnamefont {Haule}}\ and\ \bibinfo {author} {\bibfnamefont {Gheorghe~L.}\
  \bibnamefont {Pascut}},\ }\bibfield  {title} {\enquote {\bibinfo {title}
  {Forces for structural optimizations in correlated materials within a
  {DFT}+embedded {DMFT} functional approach},}\ }\href {\doibase
  10.1103/PhysRevB.94.195146} {\bibfield  {journal} {\bibinfo  {journal} {Phys.
  Rev. B}\ }\textbf {\bibinfo {volume} {94}},\ \bibinfo {pages} {195146}
  (\bibinfo {year} {2016})}\BibitemShut {NoStop}%
\bibitem [{\citenamefont {Lin}\ \emph {et~al.}(2020)\citenamefont {Lin},
  \citenamefont {Wang}, \citenamefont {Wang}, \citenamefont {Yi}, \citenamefont
  {Li}, \citenamefont {Zhang}, \citenamefont {Wang}, \citenamefont {Wang},
  \citenamefont {Huang}, \citenamefont {Sun}, \citenamefont {Huang},
  \citenamefont {Shen}, \citenamefont {Feng}, \citenamefont {Sun},
  \citenamefont {Cho}, \citenamefont {Zeng},\ and\ \citenamefont
  {Zhang}}]{Zhenyu:2020}%
  \BibitemOpen
  \bibfield  {author} {\bibinfo {author} {\bibfnamefont {Zhiyong}\ \bibnamefont
  {Lin}}, \bibinfo {author} {\bibfnamefont {Chongze}\ \bibnamefont {Wang}},
  \bibinfo {author} {\bibfnamefont {Pengdong}\ \bibnamefont {Wang}}, \bibinfo
  {author} {\bibfnamefont {Seho}\ \bibnamefont {Yi}}, \bibinfo {author}
  {\bibfnamefont {Lin}\ \bibnamefont {Li}}, \bibinfo {author} {\bibfnamefont
  {Qiang}\ \bibnamefont {Zhang}}, \bibinfo {author} {\bibfnamefont {Yifan}\
  \bibnamefont {Wang}}, \bibinfo {author} {\bibfnamefont {Zhongyi}\
  \bibnamefont {Wang}}, \bibinfo {author} {\bibfnamefont {Hao}\ \bibnamefont
  {Huang}}, \bibinfo {author} {\bibfnamefont {Yan}\ \bibnamefont {Sun}},
  \bibinfo {author} {\bibfnamefont {Yaobo}\ \bibnamefont {Huang}}, \bibinfo
  {author} {\bibfnamefont {Dawei}\ \bibnamefont {Shen}}, \bibinfo {author}
  {\bibfnamefont {Donglai}\ \bibnamefont {Feng}}, \bibinfo {author}
  {\bibfnamefont {Zhe}\ \bibnamefont {Sun}}, \bibinfo {author} {\bibfnamefont
  {Jun-Hyung}\ \bibnamefont {Cho}}, \bibinfo {author} {\bibfnamefont
  {Changgan}\ \bibnamefont {Zeng}}, \ and\ \bibinfo {author} {\bibfnamefont
  {Zhenyu}\ \bibnamefont {Zhang}},\ }\bibfield  {title} {\enquote {\bibinfo
  {title} {Dirac fermions in antiferromagnetic {F}e{S}n kagome lattices with
  combined space inversion and time-reversal symmetry},}\ }\href {\doibase
  10.1103/PhysRevB.102.155103} {\bibfield  {journal} {\bibinfo  {journal}
  {Phys. Rev. B}\ }\textbf {\bibinfo {volume} {102}},\ \bibinfo {pages}
  {155103} (\bibinfo {year} {2020})}\BibitemShut {NoStop}%
\bibitem [{\citenamefont {Sales}\ \emph {et~al.}(2019)\citenamefont {Sales},
  \citenamefont {Yan}, \citenamefont {Meier}, \citenamefont {Christianson},
  \citenamefont {Okamoto},\ and\ \citenamefont {McGuire}}]{Brian:2019}%
  \BibitemOpen
  \bibfield  {author} {\bibinfo {author} {\bibfnamefont {Brian~C.}\
  \bibnamefont {Sales}}, \bibinfo {author} {\bibfnamefont {Jiaqiang}\
  \bibnamefont {Yan}}, \bibinfo {author} {\bibfnamefont {William~R.}\
  \bibnamefont {Meier}}, \bibinfo {author} {\bibfnamefont {Andrew~D.}\
  \bibnamefont {Christianson}}, \bibinfo {author} {\bibfnamefont {Satoshi}\
  \bibnamefont {Okamoto}}, \ and\ \bibinfo {author} {\bibfnamefont
  {Michael~A.}\ \bibnamefont {McGuire}},\ }\bibfield  {title} {\enquote
  {\bibinfo {title} {Electronic, magnetic, and thermodynamic properties of the
  kagome layer compound {F}e{S}n},}\ }\href {\doibase
  10.1103/PhysRevMaterials.3.114203} {\bibfield  {journal} {\bibinfo  {journal}
  {Phys. Rev. Mater.}\ }\textbf {\bibinfo {volume} {3}},\ \bibinfo {pages}
  {114203} (\bibinfo {year} {2019})}\BibitemShut {NoStop}%
\bibitem [{\citenamefont {Dudarev}\ \emph {et~al.}(1998)\citenamefont
  {Dudarev}, \citenamefont {Botton}, \citenamefont {Savrasov}, \citenamefont
  {Humphreys},\ and\ \citenamefont {Sutton}}]{Dudarev:1998}%
  \BibitemOpen
  \bibfield  {author} {\bibinfo {author} {\bibfnamefont {S.~L.}\ \bibnamefont
  {Dudarev}}, \bibinfo {author} {\bibfnamefont {G.~A.}\ \bibnamefont {Botton}},
  \bibinfo {author} {\bibfnamefont {S.~Y.}\ \bibnamefont {Savrasov}}, \bibinfo
  {author} {\bibfnamefont {C.~J.}\ \bibnamefont {Humphreys}}, \ and\ \bibinfo
  {author} {\bibfnamefont {A.~P.}\ \bibnamefont {Sutton}},\ }\bibfield  {title}
  {\enquote {\bibinfo {title} {Electron-energy-loss spectra and the structural
  stability of nickel oxide: An {LSDA}+{U} study},}\ }\href {\doibase
  10.1103/PhysRevB.57.1505} {\bibfield  {journal} {\bibinfo  {journal} {Phys.
  Rev. B}\ }\textbf {\bibinfo {volume} {57}},\ \bibinfo {pages} {1505--1509}
  (\bibinfo {year} {1998})}\BibitemShut {NoStop}%
\bibitem [{\citenamefont {Gull}\ \emph {et~al.}(2011)\citenamefont {Gull},
  \citenamefont {Millis}, \citenamefont {Lichtenstein}, \citenamefont
  {Rubtsov}, \citenamefont {Troyer},\ and\ \citenamefont {Werner}}]{Gull:2011}%
  \BibitemOpen
  \bibfield  {author} {\bibinfo {author} {\bibfnamefont {Emanuel}\ \bibnamefont
  {Gull}}, \bibinfo {author} {\bibfnamefont {Andrew~J.}\ \bibnamefont
  {Millis}}, \bibinfo {author} {\bibfnamefont {Alexander~I.}\ \bibnamefont
  {Lichtenstein}}, \bibinfo {author} {\bibfnamefont {Alexey~N.}\ \bibnamefont
  {Rubtsov}}, \bibinfo {author} {\bibfnamefont {Matthias}\ \bibnamefont
  {Troyer}}, \ and\ \bibinfo {author} {\bibfnamefont {Philipp}\ \bibnamefont
  {Werner}},\ }\bibfield  {title} {\enquote {\bibinfo {title} {Continuous-time
  monte carlo methods for quantum impurity models},}\ }\href {\doibase
  10.1103/RevModPhys.83.349} {\bibfield  {journal} {\bibinfo  {journal} {Rev.
  Mod. Phys.}\ }\textbf {\bibinfo {volume} {83}},\ \bibinfo {pages} {349--404}
  (\bibinfo {year} {2011})}\BibitemShut {NoStop}%
\bibitem [{tut()}]{tutor:MnO}%
  \BibitemOpen
  \href@noop {} {}\bibinfo {howpublished}
  {\url{http://hauleweb.rutgers.edu/tutorials/Tutorial1a.html}}\BibitemShut
  {NoStop}%
\end{thebibliography}%


\begin{thebibliography}{7}%
\makeatletter
\providecommand \@ifxundefined [1]{%
 \@ifx{#1\undefined}
}%
\providecommand \@ifnum [1]{%
 \ifnum #1\expandafter \@firstoftwo
 \else \expandafter \@secondoftwo
 \fi
}%
\providecommand \@ifx [1]{%
 \ifx #1\expandafter \@firstoftwo
 \else \expandafter \@secondoftwo
 \fi
}%
\providecommand \natexlab [1]{#1}%
\providecommand \enquote  [1]{``#1''}%
\providecommand \bibnamefont  [1]{#1}%
\providecommand \bibfnamefont [1]{#1}%
\providecommand \citenamefont [1]{#1}%
\providecommand \href@noop [0]{\@secondoftwo}%
\providecommand \href [0]{\begingroup \@sanitize@url \@href}%
\providecommand \@href[1]{\@@startlink{#1}\@@href}%
\providecommand \@@href[1]{\endgroup#1\@@endlink}%
\providecommand \@sanitize@url [0]{\catcode `\\12\catcode `\$12\catcode
  `\&12\catcode `\#12\catcode `\^12\catcode `\_12\catcode `\%12\relax}%
\providecommand \@@startlink[1]{}%
\providecommand \@@endlink[0]{}%
\providecommand \url  [0]{\begingroup\@sanitize@url \@url }%
\providecommand \@url [1]{\endgroup\@href {#1}{\urlprefix }}%
\providecommand \urlprefix  [0]{URL }%
\providecommand \Eprint [0]{\href }%
\providecommand \doibase [0]{http://dx.doi.org/}%
\providecommand \selectlanguage [0]{\@gobble}%
\providecommand \bibinfo  [0]{\@secondoftwo}%
\providecommand \bibfield  [0]{\@secondoftwo}%
\providecommand \translation [1]{[#1]}%
\providecommand \BibitemOpen [0]{}%
\providecommand \bibitemStop [0]{}%
\providecommand \bibitemNoStop [0]{.\EOS\space}%
\providecommand \EOS [0]{\spacefactor3000\relax}%
\providecommand \BibitemShut  [1]{\csname bibitem#1\endcsname}%
\let\auto@bib@innerbib\@empty
\bibitem [{\citenamefont {Dudarev}\ \emph {et~al.}(1998)\citenamefont
  {Dudarev}, \citenamefont {Botton}, \citenamefont {Savrasov}, \citenamefont
  {Humphreys},\ and\ \citenamefont {Sutton}}]{Dudarev:1998}%
  \BibitemOpen
  \bibfield  {author} {\bibinfo {author} {\bibfnamefont {S.~L.}\ \bibnamefont
  {Dudarev}}, \bibinfo {author} {\bibfnamefont {G.~A.}\ \bibnamefont {Botton}},
  \bibinfo {author} {\bibfnamefont {S.~Y.}\ \bibnamefont {Savrasov}}, \bibinfo
  {author} {\bibfnamefont {C.~J.}\ \bibnamefont {Humphreys}}, \ and\ \bibinfo
  {author} {\bibfnamefont {A.~P.}\ \bibnamefont {Sutton}},\ }\bibfield  {title}
  {\enquote {\bibinfo {title} {Electron-energy-loss spectra and the structural
  stability of nickel oxide: An {LSDA}+{U} study},}\ }\href {\doibase
  10.1103/PhysRevB.57.1505} {\bibfield  {journal} {\bibinfo  {journal} {Phys.
  Rev. B}\ }\textbf {\bibinfo {volume} {57}},\ \bibinfo {pages} {1505--1509}
  (\bibinfo {year} {1998})}\BibitemShut {NoStop}%
\bibitem [{\citenamefont {Gull}\ \emph {et~al.}(2011)\citenamefont {Gull},
  \citenamefont {Millis}, \citenamefont {Lichtenstein}, \citenamefont
  {Rubtsov}, \citenamefont {Troyer},\ and\ \citenamefont {Werner}}]{Gull:2011}%
  \BibitemOpen
  \bibfield  {author} {\bibinfo {author} {\bibfnamefont {Emanuel}\ \bibnamefont
  {Gull}}, \bibinfo {author} {\bibfnamefont {Andrew~J.}\ \bibnamefont
  {Millis}}, \bibinfo {author} {\bibfnamefont {Alexander~I.}\ \bibnamefont
  {Lichtenstein}}, \bibinfo {author} {\bibfnamefont {Alexey~N.}\ \bibnamefont
  {Rubtsov}}, \bibinfo {author} {\bibfnamefont {Matthias}\ \bibnamefont
  {Troyer}}, \ and\ \bibinfo {author} {\bibfnamefont {Philipp}\ \bibnamefont
  {Werner}},\ }\bibfield  {title} {\enquote {\bibinfo {title} {Continuous-time
  monte carlo methods for quantum impurity models},}\ }\href {\doibase
  10.1103/RevModPhys.83.349} {\bibfield  {journal} {\bibinfo  {journal} {Rev.
  Mod. Phys.}\ }\textbf {\bibinfo {volume} {83}},\ \bibinfo {pages} {349--404}
  (\bibinfo {year} {2011})}\BibitemShut {NoStop}%
\bibitem [{\citenamefont {Haule}\ and\ \citenamefont
  {Pascut}(2016)}]{Haule:2016force}%
  \BibitemOpen
  \bibfield  {author} {\bibinfo {author} {\bibfnamefont {Kristjan}\
  \bibnamefont {Haule}}\ and\ \bibinfo {author} {\bibfnamefont {Gheorghe~L.}\
  \bibnamefont {Pascut}},\ }\bibfield  {title} {\enquote {\bibinfo {title}
  {Forces for structural optimizations in correlated materials within a
  {DFT}+embedded {DMFT} functional approach},}\ }\href {\doibase
  10.1103/PhysRevB.94.195146} {\bibfield  {journal} {\bibinfo  {journal} {Phys.
  Rev. B}\ }\textbf {\bibinfo {volume} {94}},\ \bibinfo {pages} {195146}
  (\bibinfo {year} {2016})}\BibitemShut {NoStop}%
\bibitem [{\citenamefont {Haule}\ and\ \citenamefont
  {Birol}(2015)}]{Haule:2015free}%
  \BibitemOpen
  \bibfield  {author} {\bibinfo {author} {\bibfnamefont {Kristjan}\
  \bibnamefont {Haule}}\ and\ \bibinfo {author} {\bibfnamefont {Turan}\
  \bibnamefont {Birol}},\ }\bibfield  {title} {\enquote {\bibinfo {title} {Free
  energy from stationary implementation of the $\mathrm{DFT}+\mathrm{DMFT}$
  functional},}\ }\href {\doibase 10.1103/PhysRevLett.115.256402} {\bibfield
  {journal} {\bibinfo  {journal} {Phys. Rev. Lett.}\ }\textbf {\bibinfo
  {volume} {115}},\ \bibinfo {pages} {256402} (\bibinfo {year}
  {2015})}\BibitemShut {NoStop}%
\bibitem [{tut()}]{tutor:MnO}%
  \BibitemOpen
  \href@noop {} {}\bibinfo {howpublished}
  {\url{http://hauleweb.rutgers.edu/tutorials/Tutorial1a.html}}\BibitemShut
  {NoStop}%
\bibitem [{\citenamefont {Meier}\ \emph {et~al.}(2020)\citenamefont {Meier},
  \citenamefont {Du}, \citenamefont {Okamoto}, \citenamefont {Mohanta},
  \citenamefont {May}, \citenamefont {McGuire}, \citenamefont {Bridges},
  \citenamefont {Samolyuk},\ and\ \citenamefont {Sales}}]{Brian:2020}%
  \BibitemOpen
  \bibfield  {author} {\bibinfo {author} {\bibfnamefont {William~R.}\
  \bibnamefont {Meier}}, \bibinfo {author} {\bibfnamefont {Mao-Hua}\
  \bibnamefont {Du}}, \bibinfo {author} {\bibfnamefont {Satoshi}\ \bibnamefont
  {Okamoto}}, \bibinfo {author} {\bibfnamefont {Narayan}\ \bibnamefont
  {Mohanta}}, \bibinfo {author} {\bibfnamefont {Andrew~F.}\ \bibnamefont
  {May}}, \bibinfo {author} {\bibfnamefont {Michael~A.}\ \bibnamefont
  {McGuire}}, \bibinfo {author} {\bibfnamefont {Craig~A.}\ \bibnamefont
  {Bridges}}, \bibinfo {author} {\bibfnamefont {German~D.}\ \bibnamefont
  {Samolyuk}}, \ and\ \bibinfo {author} {\bibfnamefont {Brian~C.}\ \bibnamefont
  {Sales}},\ }\bibfield  {title} {\enquote {\bibinfo {title} {Flat bands in the
  {C}o{S}n-type compounds},}\ }\href {\doibase 10.1103/PhysRevB.102.075148}
  {\bibfield  {journal} {\bibinfo  {journal} {Phys. Rev. B}\ }\textbf {\bibinfo
  {volume} {102}},\ \bibinfo {pages} {075148} (\bibinfo {year}
  {2020})}\BibitemShut {NoStop}%
\bibitem [{\citenamefont {Sales}\ \emph {et~al.}(2019)\citenamefont {Sales},
  \citenamefont {Yan}, \citenamefont {Meier}, \citenamefont {Christianson},
  \citenamefont {Okamoto},\ and\ \citenamefont {McGuire}}]{Brian:2019}%
  \BibitemOpen
  \bibfield  {author} {\bibinfo {author} {\bibfnamefont {Brian~C.}\
  \bibnamefont {Sales}}, \bibinfo {author} {\bibfnamefont {Jiaqiang}\
  \bibnamefont {Yan}}, \bibinfo {author} {\bibfnamefont {William~R.}\
  \bibnamefont {Meier}}, \bibinfo {author} {\bibfnamefont {Andrew~D.}\
  \bibnamefont {Christianson}}, \bibinfo {author} {\bibfnamefont {Satoshi}\
  \bibnamefont {Okamoto}}, \ and\ \bibinfo {author} {\bibfnamefont
  {Michael~A.}\ \bibnamefont {McGuire}},\ }\bibfield  {title} {\enquote
  {\bibinfo {title} {Electronic, magnetic, and thermodynamic properties of the
  kagome layer compound {F}e{S}n},}\ }\href {\doibase
  10.1103/PhysRevMaterials.3.114203} {\bibfield  {journal} {\bibinfo  {journal}
  {Phys. Rev. Mater.}\ }\textbf {\bibinfo {volume} {3}},\ \bibinfo {pages}
  {114203} (\bibinfo {year} {2019})}\BibitemShut {NoStop}%
\end{thebibliography}%

\end{document}


\title{Supplementary Materials for ``Enhanced Spin-polarization via Partial Ge1-dimerization as the Driving Force of the 2$\times$2$\times$2 CDW in FeGe''}
\author{Yilin Wang}     
\affiliation{School of Future Technology, University of Science and Technology of China, Hefei, Anhui 230026, China} 

\date{\today}

\maketitle

\section{Computational Details}
In VASP calculations, the energy cutoff of the plane-wave basis is set to be 500 eV. $\Gamma$-centered $K$-point grid of $16\times 16\times 10$, $10\times 10\times 10$, $8\times 8\times 10$ and $6\times 6\times 10$ are used for the superstructures of $1\times 1\times 2$, $\sqrt{3}\times \sqrt{3}\times 2$, $2\times 2\times 2$ and $\sqrt{5}\times \sqrt{5}\times 2$, respectively. The criterion of total energy convergence (EDIFF) is set to be $10^{-8}$ eV. For the case of ``S1'', superstructures are construed by dimerizing one pair of Ge1-sites in adjacent layers along $c$-axis, with a dimerization strength $d=d_{\text{Ge1-Ge1}}^0-d_{\text{Ge1-Ge1}}$, where $d_{\text{Ge1-Ge1}}^0$ and $d_{\text{Ge1-Ge1}}$ are the Ge1-Ge1 bond lengths before and after dimerization, respectively (see Fig.1 in the main text). For the case of ``S2'', starting with the structures used in ``S1'' and fixing the positions of the dimerized Ge1-Ge1 pair, the internal atomic positions of all other sites are relaxed, until the force of each atom is smaller than 1 meV/\AA\: (selective dynamics). For DFT+$U$ calculations, the simplified (rotationally invariant) approach introduced by Dudarev \textit{et al.}~\cite{Dudarev:1998} is used (LDAUTYPE=2), which is parameterized by Hubbard $U$.

In DFT+DMFT calculations, the exchange-correlation functional of local density approximation (LDA) is chosen in the DFT part. We choose a wide hybridization energy window from -10 to 10 eV with respect to the Fermi level. All five Fe-$3d$ orbitals are considered as correlated ones and a local Coulomb interaction Hamiltonian with Ising form is applied. The local Anderson impurity model is solved by the continuous time quantum Monte Carlo solver~\cite{Gull:2011}. We use a ``nominal'' double counting scheme with a nominal Fe-$3d$ occupancy of 6.0, which is close to the value given by DFT calculation. The self-energy on real frequency $\Sigma(\omega)$ is obtained by the analytical continuation method of maximum entropy. The mass-enhancement is then computed by $\frac{m^{\star}}{m^{DFT}}=1-\frac{\partial{Re \Sigma(\omega)}}{\partial{\omega}}|_{\omega=0}$. We follow the method introduced by Haule \textit{et al.}~\cite{Haule:2016force} to perform structure relaxation in the framework of DFT+DMFT. All the calculations are preformed at $T=290$ K. Following Ref.~\cite{Haule:2015free}, we use the Yukawa representation of the screened Coulomb interaction, in which there is an unique relationship between $U$ and $J_H$. If $U$ is specified, $J_H$ is uniquely determined by a code in EDMFTF~\cite{tutor:MnO}. The $U$ is chosen to be 3.9 eV, which correctly produces the ordered magnetic moment of the AFM phase of FeGe (about 1.5 $\mu_B$/Fe). For AFM calculations, a constant self-energy which breaks the spin degeneracy is provided in the first step of DFT+DMFT self-consistency. For the WIEN2k calculations, a $12 \times 12 \times 13$ $K$-point grid is used for the $2\times 2\times 2$ superstructure. The RMT values for Fe, and Ge are 2.34 and 2.27, respectively, and $R_{\text{mt}}*K_{\text{max}}$ is 7.0. 

The spin-orbit coupling is not included in all the calculations, since it is very small for Fe and Ge ions and will not change the conclusions of the present work.

\section{DFT results for F\lowercase{e}S\lowercase{n}}

\begin{figure*}
        \centering
        \includegraphics[width=0.9\textwidth]{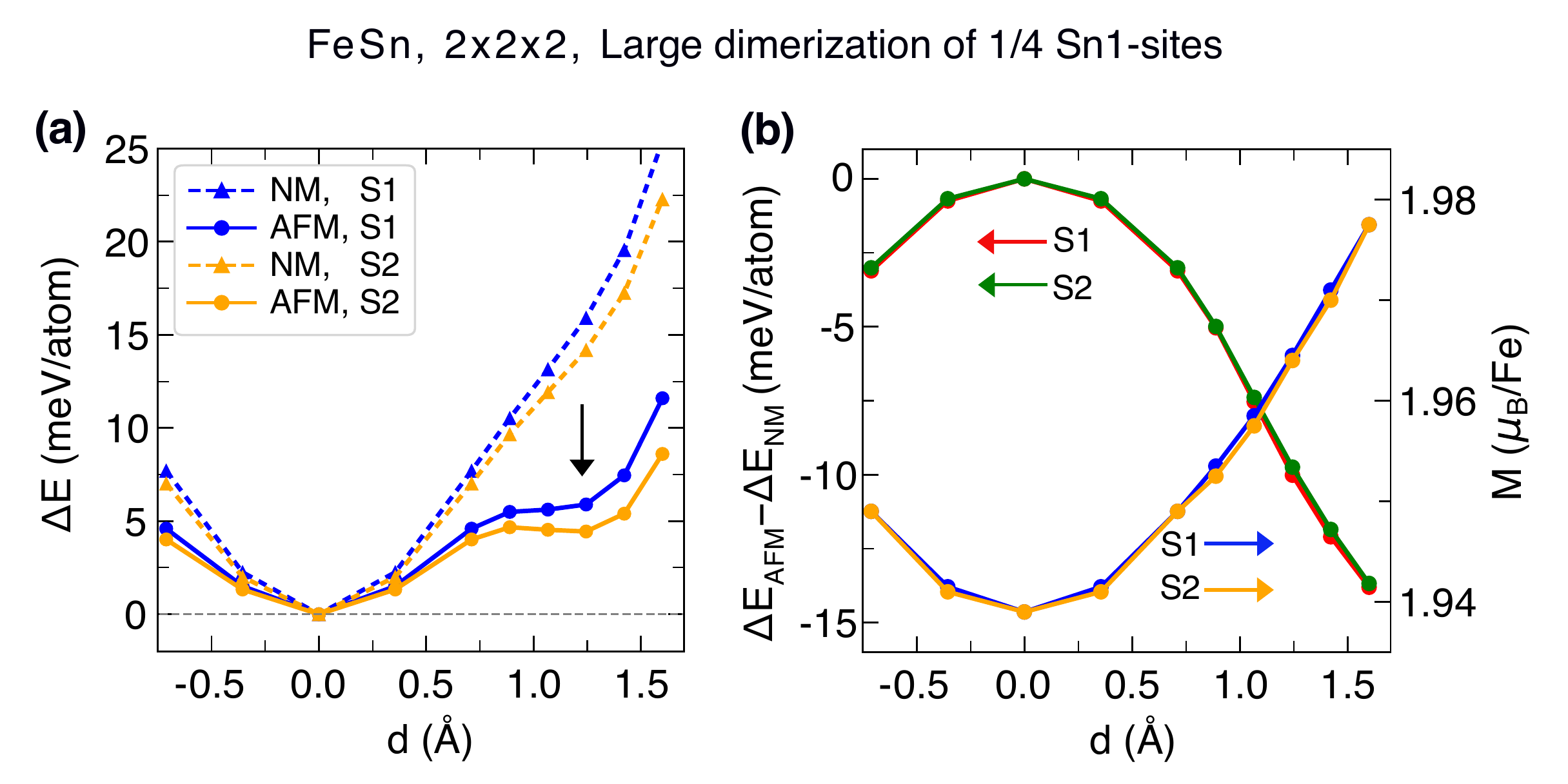}
        \caption{Analogous to Fig.2 in the main text, but calculated by DFT for FeSn with experimental lattice parameters $a=5.2765$ \AA, $c=4.4443$\AA. }
        \label{fig:fesn}
\end{figure*}

It is noted that the electronic band structures and magnetic orders of FeGe are very similar to its sister compound FeSn, but no CDW has been observed experimentally in FeSn.  We also perform DFT calculations for FeSn with the experimental lattice parameters $a=5.2765$ \AA\: and $c=4.4443$ \AA~\cite{Brian:2020}. The results of $2\times 2\times 2$ superstructure are shown in Fig.~\ref{fig:fesn}(a). The DFT method yields an ordered magnetic moment of about 1.94 $\mu_B$/Fe at $d=0$, which is also close to the value ($\sim$1.85 $\mu_B$/Fe) from neutron scattering experiment~\cite{Brian:2019}. Therefore, the DFT calculation without Hubbard $U$ is also applicable to the AFM states of FeSn. Although we also find similar behavior of enhanced spin-polarization via large partial Sn1-dimerization in FeSn, the local energy minimum around $d=1.5$ \AA\: is 5 meV/atom higher than that at $d=0$ and is far from becoming a true global minimum, consistent with the experiments. Further DFT+$U$ calculations show that it becomes a global minimum only at large $U$ ($>$ 2.2 eV), but that will yield a very large ordered magnetic moments of about 2.6 $\mu_B$/Fe, much larger than the experimental value. One possible reason may originate from that Sn has a larger atomic radius than Ge, which also leads to much larger lattice parameters in FeSn than FeGe, i.e., 5.8\% larger in $a$ and 9.8\% larger in $c$. The larger atomic radius of Sn and larger crystal volume may cause that it has to pay for more structural distortion energy by dimerizing Sn1-sites, such that the magnetic energy saving never wins in FeSn. Indeed, as shown by the dashed curves in Fig.~\ref{fig:fesn}(a), $\Delta E$ keeps increasing with $|d|$ without slowing down the rate when the dimerization strength $d$ is around 1.0 \AA $\sim$ 1.5 \AA, in sharp contrast to FeGe.

\bibliography{suppl}